\documentstyle[preprint,aps,eqsecnum,epsf,rotate]{revtex}

\begin{document}
\draft
\tighten


\title{Superfluid Bosons and Flux Liquids: Disorder, \\
        Thermal Fluctuations, and Finite--Size Effects}

\author{Uwe C. T\"auber \footnote
        {E--mail: {\it tauber@thphys.ox.ac.uk} , \ 
        Tel.: +44--1865--273963 , \ FAX: +44--1865--273947}}
\address{Department of Physics -- Theoretical Physics, 
         University of Oxford, \\ 1 Keble Road, Oxford OX1 3NP, U.K.}

\author{David R. Nelson}
\address{Lyman Laboratory of Physics, Harvard University, \\
         Cambridge, Massachusetts 02138, U.S.A.} 

\date{\today}
\maketitle


\begin{abstract}
The influence of different types of disorder (both uncorrelated and
correlated) on the superfluid properties of a weakly interacting or
dilute Bose gas, as well as on the corresponding quantities for
flux line liquids in high--temperature superconductors at low magnetic
fields are reviewed, investigated and compared. We exploit the formal
analogy between superfluid bosons and the statistical mechanics of
directed lines, and explore the influence of the different "imaginary
time" boundary conditions appropriate for a flux line liquid. For
superfluids, we discuss the density and momentum correlations, the
condensate fraction, and the normal--fluid density as function of
temperature for two-- and three--dimensional systems subject to a
space-- and time--dependent random potential as well as conventional
point--, line--, and plane--like defects. In the case of vortex
liquids subject to point disorder, twin boundaries, screw
dislocations, and various configurations of columnar damage tracks, we
calculate the corresponding quantities, namely density and tilt
correlations, the ``boson'' order parameter, and the tilt modulus. The
finite--size corrections due to periodic vs. open "imaginary time"
boundary conditions  differ in interesting and important ways. 
Experimental implications for vortex lines are described briefly.    
\end{abstract}


\tableofcontents



\section{Introduction}
 \label{introd}

The theory of weakly interacting and/or dilute superfluid Bose systems
has been studied for several decades, and now constitutes a 
well--established branch of condensed--matter physics (see, e.g.,
Refs.~\cite{fetwal,nozpin,grifin}). More recently, upon utilizing the
path integral representation of many--particle quantum mechanics
\cite{npopov,negorl}, considerable progress has been reported for the
even more difficult task to quantitatively understand the physical
properties of {\it strongly} interacting boson superfluids such as
Helium 4 below the Lambda phase transition line \cite{ceperl}. In
addition to the detailed investigation of correlation functions in
boson systems with these and other many--particle methods
\cite{grisum,polcep,dalstr}, field--theoretic and
renormalization--group approaches have been put forward in order to
analyze these interesting systems
\cite{raswei,fishoh,marqes,kolstr,chafri,alebee,cascas}.

While the statistical mechanics of superfluid bosons as such is thus a
well--advanced field, historically there has been much less effort to
study the {\it influence of quenched disorder} on superfluid
properties; e.g., the effects of point disorder (even to lowest order
in the defect concentration) on correlation functions. Characteristic
quantities such as the condensate fraction and the superfluid density
have been systematically evaluated only in the past few years
\cite{khuang,giorgi,schakl}, in part motivated by the prospect of
boson localization and related quantum phase transitions driven by the
interplay of disorder and quantum fluctuations for superfluids in
porous media
\cite{mahale,fiswei,sinrok,lzhang,matrul,wallin,hatano,pazisc}.

Additional impetus for the investigation of disordered bosons stems
from studies of magnetic flux lines in type--II high--temperature
superconductors (for recent reviews, see Refs.~\cite{blater,alamos}),
motivated by the formal analogy of the quantum mechanics of
two--dimensional bosons with the statistical mechanics of
(2+1)--dimensional directed lines \cite{nelson,nelseu,kamnel,nelrev}. 
As is readily seen from the path integral formulation of the
many--particle quantum mechanics for superfluids \cite{npopov,negorl},
the corresponding particle world lines behave as directed elastic
strings with periodic boundary conditions in the direction of
propagation $z$ (imaginary time), with the boson mass $m$ representing
an effective line tension ${\tilde \epsilon_1}$, and quantum
fluctuations $\propto \hbar$ mapping on thermal fluctuations 
$\propto T$ (see Sec.~\ref{pathin}). This mathematical correspondence
can be further exploited for the case of magnetic flux lines, as
summarized in Table~\ref{analog}, with the two--dimensional boson
density $n$ and conjugate chemical potential $\mu$ translating to the
magnetic flux density $B$ and external magnetic field $H$,
respectively. Similarly, the (imaginary) velocity ${\bf v}$ and
superfluid density $\rho_s$ correspond to a transverse magnetic field
${\bf h}$ and the inverse tilt modulus $c_{44}^{-1}$ of the flux line
ensemble.

As is depicted in Fig.~\ref{phases}, in a pure system, the superfluid
phase of bosons is represented by a highly entangled line liquid 
\footnote{With the appropriate periodic boundary conditions, this
reflects the occurrence of macroscopic particle exchange processes;
compare the very suggestive pictures and discussions in
Ref.~\cite{ceperl}.}, while the normal--fluid phase corresponds to a
disentangled liquid of flux lines. Finally, the hexagonal Abrikosov
flux lattice is represented by a two--dimensional crystalline solid in
the boson language \cite{nelseu}. The prospects for finding an
intermediate ``supersolid'' phase with both macroscopic phase
coherence and crystalline order (not shown in the figure) may in fact
be more promising for flux lines \cite{frnefi,balnel}. If periodic
boundary conditions are applied for the superconductor along the
magnetic--field direction, the fact that in two dimensions a
second--order Berezinskii--Kosterlitz--Thouless transition separates
the normal and superfluid phases implies that there should be a sharp
distinction between the entangled and disentangled line liquid states
[the Kosterlitz--Thouless transition temperature as function of the
boson density $T_{\rm KT}(n)$ mapping on a critical sample length
$L(B)$ as function of the magnetic field]. However, if instead one
assumes {\it free} rather than periodic boundary conditions, which
should in fact be more realistic for most flux line problems, the
correspondence between inverse temperature and sample size along the
field direction is only approximate, and there need not be a sharp
phase transition between an entangled and disentangled flux liquid
\cite{fislee}. This observation demonstrates that the issue of
boundary conditions becomes important when the above correspondence is
utilized, and that open boundary conditions may lead to very different
physical behavior for $L < \infty$ from what would ``naively'' be
inferred from the boson picture.

For magnetic flux lines in high--temperature superconductors, disorder
changes the phase diagram in essential ways. E.g., the presence of
uncorrelated point defects destroys the crystalline order of the
Abrikosov lattice \cite{larkin}, and may lead to a new
disorder--dominated ground state, termed the vortex glass
\cite{fisher}. For the application of high--$T_c$ superconductors in a
magnetic field, an effective mechanism for flux pinning is required,
in order to minimize dissipative losses caused by the Lorentz--force
induced vortex motion and thus retain a truly superconducting state,
defined by a vanishing linear resistivity. For that purpose linear 
damage tracks have been produced by heavy--ion irradiation leading to
columnar defects which turned out to be very effective in localizing
the flux lines \cite{blater,alamos}. Theoretically, this corresponds
to a two--dimensional boson localization transition; the new ground
state of vortices pinned to columnar defects is called the Bose glass
\cite{lyutov,nelvin,larvin} (see also Ref.~\cite{nelnor}). We have
recently further exploited the quantum analogy, in this case with
disordered semiconductors (the Coulomb glass), in order to show that
long--range intervortex repulsions in the Bose glass produce a highly
correlated ground state and an additional reduction in vortex
transport \cite{taunel}. Such considerations may eventually allow
controlled ``tailoring'' of defects in such a way that magnetic flux
lines are bound most effectively. For example, the introduction of a
certain amount of ``splay'', i.e., variation in columnar defect tilt
angles was suggested to drastically improve flux pinning
\cite{hwaled,lednel,devsca}. It was also demonstrated that columnar
defects effectively increase the interlayer Josephson coupling in the
pancake vortex regime \cite{koshel}. In addition, the pronounced
effect of planar disorder (as is frequently present in the high--$T_c$ 
ceramics in the form of twin boundaries) was studied \cite{marvin},
and recently also related to the so--called peak effect \cite{lamavi}. 
The investigation of the effect of transverse magnetic fields trying
to tilt the flux lines away from, say, an array of parallel columnar
pins, leads to an interesting localization problem in 
{\it non--Hermitian} quantum mechanics \cite{hatnel}.

In this article, we compare and contrast the behavior of boson
superfluids and vortex liquids in the presence of various types of
disorder, emphasizing effects due to the different boundary
conditions. Coherent--state path integral techniques
\cite{npopov,negorl} provide a unified framework to discuss both
problems. The ``open'' boundary conditions appropriate to vortex
liquids are illustrated in Fig.~\ref{phases}(c), and differ from the
periodic boundary conditions relevant to real bosons by a set of
independent integrations over the entry and exit points of the
directed lines, see Sec.~\ref{flines}.

Some of the varieties of disorder treated here (to leading order in
the disorder strength) are compared for vortex lines and superfluids
in Table~\ref{disvar}. Point disorder in three--dimensional vortex
arrays corresponds to a rather unphysical random space-- and
time--dependent potential applied to, say, superfluid Helium~4 films.
However, columnar and planar pins in vortex liquids are directly
relevant to boson superfluid films on disordered substrates. Columnar
pins, of course, correspond to ``point'' disorder in two--dimensional
superfluids, as would occur on an amorphous substrate, while random
planar disorder for vortex liquids is represented by random lines
etched (say, via microlithography) onto an otherwise regular 
substrate. As part of this review, we rederive via the coherent--state
method results for point disorder in superfluids \cite{khuang,giorgi},
and produce new results for random line disorder perturbing two-- and
three--dimensional bosons, and random planes interacting with a
three--dimensional superfluid. Among other results, we find that the
superfluid density and condensate fraction in a zero--temperature
superfluid film can be significantly reduced by depositing a set of
lines with random positions and orientations in the substrate.

To first order in the defect concentration, the influence of weak
point disorder on correlation functions and thermodynamic properties
of the flux liquid was calculated in the limit $L \to \infty$ by
Nelson and Le~Doussal \cite{nelled,nelala}; in the present paper, we
shall extend these previous investigations, which are based on the
path integral representation of the partition sum and the Gaussian
approximation, to $d$--dimensional generalizations of the disorder
displayed in Table~\ref{disvar}, as well as two families of
symmetrically tilted columnar defects and randomly splayed extended
disorder. Our calculations of how disorder affects the superfluid
density generalize a method developed by Hwa {\it et al.} to determine
the effect of splayed columnar defects on the flux line tilt modulus
\cite{hwaled}. Our investigation of the role played by different
(namely periodic and free) boundary conditions (first discussed in
Ref.~\cite{nelseu}) elucidates the subtle differences in the
underlying physics of boson superfluids and magnetic flux line liquids
\footnote{In a different context, finite--size effects have also been
studied by Marchetti and Nelson, using a hydrodynamic approach
\cite{marnel}.}. We obtain explicit results for the finite--size
corrections for a number of thermodynamic quantities, such as the
boson condensate fraction $n_0$ and the inverse tilt modulus
$c_{44}^{-1}$ (superfluid density $\rho_s$); these predictions should
be subject to direct experimental tests. By invoking the boson analogy 
for a finite system ``naively'', i.e., using periodic boundary
conditions, one would expect $n_0$ and $c_{44}^{-1}$ to decrease as a
powers of $1/L$, because thermal excitations --- 
$k_{\rm B} T \sim 1/L$ --- reduce the Bose condensate fraction and the
superfluid density at nonzero temperatures (see Sec.~\ref{pspebc});
for open boundary conditions, however, it turns out that both these
quantities {\it increase} (see Sec.~\ref{psopbc})! This behavior
arises because at the surfaces the probability distribution for the
time--averaged position of a wiggling flux line is determined by the
bosonic ``wave function'' \cite{nelseu,nelnor}, and is therefore
larger than in the bulk, where this quantity is given by the ``wave
function'' squared (a Gaussian for a free random walker); see
Sec.~\ref{flines} and Fig.~\ref{wavfun}. Near the surfaces, the
enhanced thermal wandering facilitates entanglement, and renders the
system softer with respect to tilting, effectively {\it increasing}
$n_0$ and the overall $c_{44}^{-1}$.

Most of our final results on vortex liquids are summarized for weak,
delta--function--like interactions between lines in the direction
perpendicular to $z$. This restriction is easily relaxed by inserting
the full Fourier--transformed pair potential [see Eq.~(\ref{fouint})]
into the appropriate formulas. However, our results for flux liquids
are {\it also} limited to regimes where the interactions between lines
are strictly {\it local} in $z$, the magnetic--field direction. 
Although this restriction facilitates comparisons with nonrelativistic 
bosons, it does limit straightforward applications of our results to
high magnetic fields in superconductors. The most problematic quantity
at high fields is the vortex tilt modulus $c_{44}$, which has a large
compressive contribution due to the interactions nonlocal in $z$. As
discussed by Larkin and Vinokur \cite{larvin} for isotropic vortex
liquids using a nonlocal generalization of the boson mapping
\cite{feiges}, the renormalized tilt modulus may be written as
\begin{equation}
        c_{44} = {B^2 \over 4 \pi} 
                \left( 1 + {1 \over 4 \pi \lambda^2 n_s} \right) \ ,
 \label{rentil}
\end{equation}
where $\lambda$ is the London penetration depth and $n_s$ denotes the
superfluid number density. The first term is the compressive
contribution due to nonlocality in $z$. Nonlocality is expected to be
much less important in the second ``vortex part'' of the tilt modulus
$c^v_{44}$, which is inversely proportional to the superfluid density
of the equivalent boson system 
\footnote{The correlation function which gives the superfluid density
(see Secs.~\ref{pathin} and \ref{puresy} below) depends only on 
{\it fluctuations} in the tangents to the vortex lines about
configurations parallel to the $z$ axis. Locality in $z$ is an
excellent approximation for these fluctuations, see App.~A of the
second part of Ref.~\cite{nelvin}.}. We thus expect that our results
may be used as an approximation to this second term of
Eq.~(\ref{rentil}) at high fields. The generalization of the
Larkin--Vinokur formula to anisotropic superconductors with effective
mass ratio $m_\perp / m_z$ has recently been derived by Geshkenbein
\cite{geshpc}, whose result we quote here for completeness
\begin{equation}
        c_{44} = {B^2 \over 4 \pi} \left( 1 + 
                {m_\perp / m_z \over 4 \pi \lambda^2 n_s} \right) \ .
 \label{anisot}
\end{equation}

This article is intended to be self--contained and pedagogical, and we
therefore include sufficient introductory material. In the following
section, we review the description of both the quantum mechanics of
superfluid bosons and the statistical mechanics of magnetic flux lines
by coherent--state path integrals. The important role of Galilean
invariance (for bosons) and affine transformations (for flux lines)
will be discussed. In Sec.~\ref{puresy}, we study pure systems, first
at zero temperature (i.e., $L \to \infty$ for flux lines), and then
also the lowest--order  corrections for finite temperatures (finite
sample thickness for the case of magnetic vortices). In
Sec.~\ref{disbos}, the disorder contributions to density, momentum,
and vorticity correlations, condensate fraction (depletion), and
superfluid density are evaluated for bosonic systems (periodic
boundary conditions) to lowest order in the defect concentration, and
explicit formulas are presented for point, linear, and planar disorder
in two and three space dimensions. In Sec.~\ref{disfll}, we discuss
magnetic flux lines in high--temperature superconductors, for which
open boundary conditions provide a better description, and determine
the corresponding physical quantities, namely the magnitude of the
``boson'' order parameter, density and tilt correlations, and the
renormalization of the tilt modulus by the disorder. Due to the
generality of these computations the resulting formulas are rather
lengthy; therefore we present explicit expressions for the cases of
point disorder, splayed columnar defects, parallel extended and two
families of symmetrically tilted defects in Sec.~\ref{expdis}. Our
findings for new kinds of correlated disorder in boson superfluids,
and the effects of different boundary conditions, are summarized and
discussed in Sec.~\ref{sumdis}. In the Appendices, we derive the
fluctuation formula of Hwa {\it et al.} \cite{hwaled} for the tilt
modulus in very thick samples (in the thermodynamic limit) using
affine transformations, discuss the ``phase--only approximation'' and
its limitations, present a detailed derivation of the correlation
functions and their disorder contributions for the case of open
boundary conditions, and list some useful formulas required for
performing the Matsubara frequency sums and momentum integrals.


\section{Path integral representation for bosons and flux lines with
         disorder} 
 \label{pathin}

In this section, we set the stage for our investigations of weakly
interacting superfluid bosons and magnetic flux lines subject to
disorder. By writing down the corresponding path integral
representations for the partition sum, we elucidate the close formal
relationship between the two distinct physical problems, and point out
the differences. We also comment on the important role of Galilean
invariance and affine transformations, respectively.

\subsection{Superfluid bosons}
 \label{bosons}

Consider the Hamiltonian for $N$ bosonic particles in $d$ space
dimensions, with mass $m$, interacting via the scalar pair potential
$V(r)$, and subject to an ``external'' disorder potential 
$V_D({\bf r})$,
\begin{equation}
        {\cal H}_N^{\rm bos} = \sum_{i=1}^N \Biggl[ 
                        - \, {\hbar^2 \over 2 m} \bbox{\nabla}_i^2 
                                        + V_D({\bf r}_i) \Biggr] 
        + \frac{1}{2} \sum_{i\not=j} V(|{\bf r}_i-{\bf r}_j|) \ ,
 \label{bosham}
\end{equation}
with the understanding that the corresponding $N$--particle wave
function is symmetric with respect to all possible permutations. As a
consequence, the trace leading to the partition sum is to be
restricted to these totally symmetric bosonic states $\{ n \}$ with
energy eigenvalues $E_n^{\rm bos}$,
\begin{equation}
        Z_N^{\rm bos} = \!\!\!\!\!
        \sum_{{\rm bosonic \; states} \; \{ n \}} \!\!\!\!\! 
                                        e^{- \beta E_n^{\rm bos}} \ ,
 \label{bossum}
\end{equation}
where $\beta = 1 / k_{\rm B} T$.

By applying the standard methods of rewriting the Hamiltonian
(\ref{bosham}) in the language of second quantization and transforming
to coherent states, the following imaginary--time path integral
representation for the grand--canonical partition function is readily
derived \cite{npopov,negorl},
\begin{equation}
        Z_{\rm gr}^{\rm bos} = \int_{\begin{array}{l} 
                \scriptsize \psi^*(r, \beta \hbar) = \psi^*(r,0) \\
        \scriptsize \psi(r, \beta \hbar) = \psi(r,0) \end{array}} \!
        {\cal D}[\psi({\bf r},\tau)] {\cal D}[\psi^*({\bf r},\tau)]
                        \; e^{- S[\psi^*,\psi] / \hbar} \ ,
 \label{bosgrz}
\end{equation}
with the action
\begin{eqnarray}
        S[\psi^*,\psi] = \int_0^{\beta \hbar} \!\! d\tau \int \! d^dr
        \, \Biggl\{ &&\hbar \psi^*({\bf r},\tau) \,
                {\partial \psi({\bf r},\tau) \over \partial \tau}
        + {\hbar^2 \over 2 m} \, | \bbox{\nabla} \psi({\bf r},\tau)|^2
        - \Bigl[ \mu + V_D({\bf r},\tau) \Bigr] \,
                                |\psi({\bf r},\tau)|^2 \nonumber \\
        &&+ \frac{1}{2} \int \! d^dr' \, V(|{\bf r}-{\bf r}'|) \,
        |\psi({\bf r}',\tau)|^2 \, |\psi({\bf r},\tau)|^2 \Biggr\} \ ,
 \label{bosact}
\end{eqnarray}
where $\mu$ is the chemical potential. The constraints for the fields
$\psi$ and $\psi^*$ in Eq.~(\ref{bosgrz}) stem from the
permutation symmetry requirement and thus reflect the boson
statistics; they can be interpreted as periodic boundary conditions
for an imaginary--time slab which at finite temperature is of
``length'' $\beta \hbar$.

In the spirit of Landau--Ginzburg mean--field theory, we apply the
method of steepest descent and find a nonlinear Schr\"odinger equation
as the stationarity condition \cite{nozpin}
\begin{eqnarray}
        0 = {\delta S[\psi^*,\psi] \over \delta \psi^*({\bf r},\tau)}
        = &&\hbar \, {\partial \psi({\bf r},\tau) \over \partial \tau}
        - {\hbar^2 \over 2 m} \, \bbox{\nabla}^2 \psi({\bf r},\tau)
        - \Bigl[ \mu + V_D({\bf r},\tau) \Bigr] \psi({\bf r},\tau)
        \nonumber \\
        &&+ \int \! d^dr' \, V(|{\bf r}-{\bf r}'|) \,
        |\psi({\bf r}',\tau)|^2 \, \psi({\bf r},\tau) \ .
 \label{stdesc}
\end{eqnarray}  
Upon specializing to short--range interactions with
$V({\bf r}) = V_0 \delta({\bf r})$, where $V_0 = V({\bf q} = {\bf 0})$
and $V({\bf q}) = \int \! d^dr V({\bf r}) e^{-i{\bf q}{\bf r}}$ is the 
Fourier--transformed potential, taking the configurational average
(denoted by an overbar), and assuming that 
$|\psi({\bf r},\tau)|^2 \approx n_0$ independent of ${\bf r}$ and 
$\tau$, this yields $\Bigl[ - (\mu + \overline{V_D}) + n_0 V_0 \Bigr]
\sqrt{n_0} = 0$, with solutions
\begin{eqnarray}
        n_0 = 0 &&\quad {\rm for} \; \mu + \overline{V_D} \leq 0 
                                                        \nonumber \\
        n_0 = (\mu + \overline{V_D}) / V_0 &&\quad {\rm for} \;
                                        \mu + \overline{V_D} > 0 \ .
 \label{landau}
\end{eqnarray}
The presence of disorder shifts the transition point to a phase with
nonzero Bose condensate fraction $n_0$, which in a pure system would
occur when $\mu > 0$. In the above mean--field approximation, the free
energy in the superfluid phase becomes
\begin{equation}
        F_{\rm gr}(T,\Omega,\mu) = - k_{\rm B} T \ln Z_{\rm gr} =
        - \left[ \mu + \overline{V_D} \right]^2 \Omega / 2 V_0 \ ,
 \label{landfe}
\end{equation}
where $\Omega = \int \! d^dr$ denotes the volume of the
system. Therefore the mean particle density is
\begin{equation}
        n = - \Omega^{-1} \left( \partial F_{\rm gr} / \partial \mu
                                \right)_{T,\Omega} \approx n_0
 \label{densty}
\end{equation}
in this approximation, and the mass density 
$\rho \approx m n_0$. Similarly, the isothermal compressibility
becomes 
\begin{equation}
        \kappa_T = n^{-2} \left( \partial n / \partial \mu
                \right)_{T,\Omega} \approx (n_0^2 V_0)^{-1} \ ,
 \label{comprs}
\end{equation}
which implies for the macroscopic sound velocity (with 
$\kappa_S \approx \kappa_T$)
\begin{equation}
        c_1 = (n m \kappa_S)^{-1/2} \approx (n_0 V_0 / m)^{1/2} \ .
 \label{soundv}
\end{equation}
Notice that $c_1$ vanishes both for $n_0 \to 0$ and $V_0 \to 0$.

Eq.~(\ref{densty}) implies that within mean--field theory, which
neglects fluctuations, all particles enter the condensate in the
superfluid phase. When density and phase fluctuations are taken into
account, however, one finds that actually the condensate fraction is
$n_0 < n$ (see Sec.~\ref{puresy} and
Refs.~\cite{fetwal,nozpin,npopov}). The Bose condensate is formed by
those particles which take part in macroscopic ring exchange
processes on scales larger than the thermal de Broglie wavelength;
the required symmetrization of the many--particle wavefunction then
leads to phase coherence. In the path integral picture, superfluidity
and $n_0 \not= 0$ can be related to a macroscopic fraction of world
lines entangled and connected by virtue of the periodic boundary
conditions, as is illustrated in Ref.~\cite{ceperl}. Note that $n_0$,
the thermodynamic {\it order parameter}, is conceptually quite
distinct from the superfluid density $\rho_s$, which is a 
{\it transport coefficient} (see Sec.~\ref{puresy}), defined as an
appropriate limit of a dynamical response function
\cite{fetwal,nozpin,grifin,npopov}; in terms of the path integral
representation, $\rho_s$ corresponds to the mean--square world line
winding number \cite{ceperl}.
 
If we now allow for an inhomogeneous condensate [including an
inhomogeneous chemical potential $\mu({\bf r},\tau)$], and transform to
real time according to $\tau \to i t$, Eq.~(\ref{stdesc}) becomes
\begin{equation}
        i \hbar \, {\partial \psi({\bf r},t) \over \partial t} =
        - {\hbar^2 \over 2 m} \, \bbox{\nabla}^2 \psi({\bf r},t)
        - \left( \mu + \overline{V_D} \right) \psi({\bf r},t)
        + V_0 |\psi({\bf r},t)|^2 \, \psi({\bf r},t) \ .
 \label{nlschr}
\end{equation}  
It is then convenient to introduce the ``polar'' representation in
terms of the new fields $\pi({\bf r},t)$ and $\Theta({\bf r},t)$,
\begin{equation}
        \psi({\bf r},t) = \sqrt{n_0 + \pi({\bf r},t)} \, 
                                e^{i \Theta({\bf r},t)} \ ,
 \label{polrep}
\end{equation}
implying that the density fluctuations are described by the $\pi$
field,
\begin{equation}
        n({\bf r},t) = |\psi({\bf r},t)|^2 = n_0 + \pi({\bf r},t) \ ,
 \label{densfl}
\end{equation}
and furthermore assume small fluctuations from the mean condensate
density only: $\psi \approx \sqrt{n_0} \left( 1 + \pi / 2 n_0 \right)
e^{i \Theta}$ [the average of the phase field $\Theta({\bf r},t)$ is
taken to be zero]. We choose the constant $n_0$ in Eq.~(\ref{polrep})
to be {\it exactly} equal to the average boson order parameter
squared,
\begin{equation}
        n_0 = | \langle \psi({\bf r},t) \rangle |^2 \ ,
 \label{boseop}
\end{equation}
which is a direct measure of the macroscopic occupation number of the 
${\bf p} = {\bf 0}$ momentum state. Note that in general  
$n_0 \not= \langle n({\bf r},t) \rangle = \langle |\psi({\bf r},t)|^2 
\rangle$, although we will often set $n_0 \approx n$, because 
$\langle \pi({\bf r},t) \rangle \approx 0$ in the Gaussian
approximation [see Eqs.~(\ref{densfl}) and (\ref{densty})]. At 
$T = 0$, $n_0 = n$ only for the ideal (non--interacting) Bose gas.  

Upon defining the superfluid velocity according to
\begin{equation}
        {\bf v}_s({\bf r},t) = 
                \hbar \bbox{\nabla} \Theta({\bf r},t) / m \ ,
        \quad \bbox{\nabla} \times {\bf v}_s = {\bf 0} \ ,
 \label{sflvel}
\end{equation}
the imaginary part of Eq.~(\ref{nlschr}) then yields the continuity
equation for the irrotational superfluid flow \cite{nozpin},
\begin{equation}
        {\partial n({\bf r},t) \over \partial t} + \bbox{\nabla} \cdot
        \Bigl[ n({\bf r},t) {\bf v}_s({\bf r},t) \Bigr] = 0 \ .
 \label{conteq}
\end{equation}
On the other hand, from the real part of Eq.~(\ref{nlschr}) the
characteristic correlation length
\begin{equation}
        \xi^2 = \hbar^2 / 4 m n_0 V_0 = \hbar^2 / 4 m^2 c_1^2
 \label{corlen}
\end{equation}
is inferred, and if the variation of the density $\pi({\bf r},t)$ is
slow on a scale $\xi$, $\xi^2 \bbox{\nabla}^2 \pi \ll \pi$, the real
part of Eq.~(\ref{nlschr}) becomes
\begin{equation}
        m \, {\partial {\bf v}_s({\bf r},t) \over \partial t} +
        \bbox{\nabla} \left[ \frac{m}{2} {\bf v}_s({\bf r},t)^2
        \right] = - \bbox{\nabla} \mu({\bf r},t)
 \label{sfflow}
\end{equation}
(note that $\pi = \delta n = n^2 \kappa_T \delta \mu \approx
\delta \mu / V_0$), which is the fundamental equation of superfluid
flow \cite{nozpin}. Finally, this may be cast into an alternative
form, using the fact that superfluid flow is irrotational,
$\bbox{\nabla} \times {\bf v}_s = {\bf 0}$,
\begin{equation}
        m \, {d {\bf v}_s({\bf r},t) \over d t} \equiv
        m \, {\partial {\bf v}_s({\bf r},t) \over \partial t} +
        \left[ {\bf v}_s({\bf r},t) \cdot \bbox{\nabla} \right] 
        {\bf v}_s({\bf r},t) = - \bbox{\nabla} \mu({\bf r},t) \ ,
 \label{eulerf}
\end{equation}
which can be interpreted as Euler's equation for nonviscous flow,
driven by a chemical potential gradient.

In Secs.~\ref{puresy} and \ref{disbos}, we extend this analysis to
incorporate both fluctuations in the fields $\pi$ and $\Theta$ (in
the Gaussian approximation) and in the disorder potential $V_D({\bf
r})$ (to lowest nontrivial order). First, however, we discuss the
boson analogy for the statistical mechanics of magnetic flux lines in
some detail.

\subsection{Magnetic flux lines}
 \label{flines}

In order to describe the large--scale properties of magnetic vortices
in high--$T_c$ materials, which are strongly type--II superconductors,
we work in the London limit, and define each flux line by its
(two--dimensional) trajectory ${\bf r}_i(z)$, as it traverses the
superconducting sample of length $L$ parallel to the external magnetic
field ${\bf H}$, which is taken to be aligned along the $z$
direction. The ensuing Gibbs free energy for $N$ such vortices subject
to a disorder potential $V_D({\bf r})$, is \cite{nelseu,nelled}
\begin{eqnarray}
        G_N({\bf H}) = &&\int_0^L \! dz \left\{ \sum_{i=1}^N
        \left( \epsilon_1 \left[ 1 + {m_\perp \over m_z} \biggl\vert 
                {d {\bf r}_i(z) \over d z} \biggr\vert^2 \right]^{1/2} 
                + V_D \Bigl[ {\bf r}_i(z) \Bigr] \right)
        + \frac{1}{2} \sum_{i\not=j} 
        V \Bigl(|{\bf r}_i(z)-{\bf r}_j(z)|\Bigr) \right\} \nonumber\\
        &&- {{\bf H} \over 4 \pi} \int \! d^3r \, {\bf b}({\bf r}) \ ;
 \label{gibbfe}
\end{eqnarray}
here, $\epsilon_1 = \epsilon_0 \ln \kappa$ is the effective line
tension, $\epsilon_0 = \left( \phi_0 / 4 \pi \lambda \right)^2$ sets
the energy scale, $\phi_0 = h c / 2 e$ is the magnetic flux quantum,
$\kappa = \lambda / \xi$ is the ratio of the London penetration depth
$\lambda$ and the correlation length $\xi$, and the material
anisotropy is embodied in the effective mass ratio $m_\perp / m_z$ 
($ \ll 1$ for high--temperature superconductors). Finally, the
screened logarithmic intervortex repulsion is given by
\begin{equation}
        V(r) = 2 \epsilon_0 K_0(r/\lambda) \ ,
 \label{vorrep}
\end{equation}
where $K_0(x)$ is a modified Bessel function, with the asymptotic
behavior $K_0(x) \propto - \ln x$ for $x \to 0$, and 
$K_0(x) \propto x^{-1/2} e^{-x}$ for $x \to \infty$. In Fourier space
this potential reads
\begin{equation}
        V(q) = V_0 / (1 + \lambda^2 q^2) \ ,
 \label{fouint}
\end{equation}
where $V_0 = \phi_0^2 / 4 \pi$. Note that we have assumed here that
the vortex interaction is local in $z$, which is a good approximation
when the flux lines are essentially straight and parallel to $z$, so
that $\langle | d {\bf r} / d z|^2 \rangle \ll 1$. In the same spirit,
we can then expand the square root in Eq.~(\ref{gibbfe}), and
reexpress the integral over the magnetic flux density 
${\bf b}({\bf r})$ in terms of $\phi_0$ and $N$, which leads to
\begin{equation}
        G_N({\bf H}) = \mu N L + F_N[\{ {\bf r}_i(z) \}] \ ,
 \label{legtra}
\end{equation}
where we have introduced the ``chemical potential'' [see
Eq.~(\ref{gcansm}) below]
\begin{equation}
        \mu = H \, {\phi_0 \over 4 \pi} - \epsilon_1 
                = {\phi_0 \over 4 \pi} \left( H - H_{c_1} \right) \ ,
 \label{chmpot}
\end{equation}
which vanishes at the upper critical field 
$H_{c_1} = 4 \pi \epsilon_1 / \phi_0$, and the vortex free--energy
functional 
\begin{equation}
        F_N[\{ {\bf r}_i(z) \}] = \int_0^L \! dz 
        \left\{ \sum_{i=1}^N \left( {{\tilde \epsilon}_1 \over 2}
        \biggl\vert {d {\bf r}_i(z) \over d z} \biggr\vert^2 
                        + V_D \Bigl[ {\bf r}_i(z) \Bigr] \right)
        + \frac{1}{2} \sum_{i\not=j}
        V \Bigl( |{\bf r}_i(z)-{\bf r}_j(z)| \Bigr) \right\} \ ,
 \label{vortfe}
\end{equation}
with the modified line tension 
${\tilde \epsilon}_1 = (m_\perp / m_z) \epsilon_1$.

A full statistical treatment requires a summation of 
$e^{- G_N / k_{\rm B} T}$ over all possible vortex trajectories 
$\{ {\bf r}_i(z) \}$; thus the partition sum becomes
\begin{equation}
        Z_{\rm gr}^{\rm fl} = \sum_{N=0}^\infty {1 \over N!}
                        e^{\mu N L / k_{\rm B} T} Z_N^{\rm fl} \ ,
 \label{gcansm}
\end{equation}
where the canonical partition function for a system of $N$ lines reads
\begin{equation}
        Z_N^{\rm fl} = \prod_{i=1}^N \int \! {\cal D}[{\bf r}_i(z)]
                e^{- F_N[\{ {\bf r}_i(z) \}] / k_{\rm B} T} \ .
 \label{linsum}
\end{equation}
With Eq.~(\ref{vortfe}) this already has the form of a
quantum--mechanical partition function in the path integral
representation, i.e., for the world lines of $N$ particles of ``mass''
${\tilde \epsilon}_1$ moving through ``imaginary time'' $z$,
interacting with potential $V(r)$, and subject to a disorder potential
$V_D({\bf r})$. According to Eq.~(\ref{chmpot}), the external magnetic
field plays the role of a chemical potential, and at $T=0$ vortices
will start to penetrate the sample when  $H > H_{c_1}$; furthermore,
the areal particle density is related to the magnetic flux via 
$n = B / \phi_0$. 

To render this quantum analogy of the classical statistical mechanics
problem of directed lines more precise \cite{nelseu,nelala}, we note
that the partition sum (\ref{linsum}) can be rewritten in terms of the
transfer matrix $e^{- {\cal H}_N^{\rm fl} L / k_{\rm B} T}$ connecting
neighboring constant--$z$ slices, i.e., as an integral over a
quantum--mechanical matrix element (in ``imaginary time''),
\begin{equation}
        Z_N^{\rm fl} = \prod_{i=1}^N \int \! d{\bf r}_i'
                \prod_{i=1}^N \int \! d{\bf r}_i \; 
                \langle {\bf r}_1' \cdots {\bf r}_N' | 
                        e^{- {\cal H}_N^{\rm fl} L / k_{\rm B} T}
                                | {\bf r}_1 \cdots {\bf r}_N \rangle \ ,
 \label{tramat}
\end{equation}
where the states $| {\bf r}_1 \cdots {\bf r}_N \rangle$ and 
$\langle {\bf r}_1' \cdots {\bf r}_N' |$ describe the entry and exit
points of the vortices, respectively, and the Hamiltonian is [compare
Eq.~(\ref{bosham})]
\begin{equation}
        {\cal H}_N^{\rm fl} = \sum_{i=1}^N \Biggl[
                - \, {(k_{\rm B} T)^2 \over 2 {\tilde \epsilon}_1}
                        \bbox{\nabla}_i^2 + V_D({\bf r}_i) \Biggr] 
        + \frac{1}{2} \sum_{i\not=j} V(|{\bf r}_i-{\bf r}_j|) \ .
 \label{fllham}
\end{equation}
By inserting a complete set of many--particle energy eigenstates 
$| n \rangle$ in Eq.~(\ref{tramat}), and introducing the
zero--momentum state
\begin{equation}
        | {\bf p}_1 = {\bf 0} \cdots {\bf p}_N = {\bf 0} \rangle =
                \prod_{i=1}^N \int \! d{\bf r}_i \;
                        | {\bf r}_1 \cdots {\bf r}_N \rangle \ ,
 \label{zermom}
\end{equation}
we find
\begin{eqnarray}
        Z_N^{\rm fl} &&= 
        \langle {\bf p}_1 = {\bf 0} \cdots {\bf p}_N = {\bf 0} | 
                        e^{- {\cal H}_N^{\rm fl} L / k_{\rm B} T}
        | {\bf p}_1 = {\bf 0} \cdots {\bf p}_N = {\bf 0} \rangle 
                                                        \nonumber \\
        &&= \sum_n \left\vert \langle n 
        | {\bf p}_1 = {\bf 0} \cdots {\bf p}_N = {\bf 0} \rangle 
                \right\vert^2 e^{- E_n^{\rm fl} L / k_{\rm B} T} \ .
 \label{flsum1}
\end{eqnarray}
The ``quantum--mechanical'' interpretation of this matrix element is
that of a system prepared in the ground state 
$| {\bf p}_1 = {\bf 0} \cdots {\bf p}_N = {\bf 0} \rangle$ of the 
{\it ideal} Bose gas at $z = 0$. It is then allowed to evolve in
imaginary time under the influence of interactions and disorder. After
``time'' $L$, this interacting superfluid is projected back onto the
ideal Bose gas ground state.

To see that only boson states are involved in the statistics of these
fictitious quantum particles with ``free'' boundary conditions, note
that the zero--momentum state (\ref{zermom}) is symmetric under
permutations, and therefore only those eigenstates of the
permutation--symmetric Hamiltonian (\ref{fllham}) contribute to the 
partition sum (\ref{flsum1}) that are permutation--symmetric
themselves. Hence the summation effectively includes bosonic
many--particle states only,
\begin{equation}
        Z_N^{\rm fl} = \!\!\!\!\! 
        \sum_{{\rm bosonic \; states} \; \{ n \}} \!\!\!\!\! 
        \left\vert \langle n 
        | {\bf p}_1 = {\bf 0} \cdots {\bf p}_N = {\bf 0} \rangle 
                \right\vert^2 e^{- E_n^{\rm bos} L / k_{\rm B} T} \ .
 \label{flsum2}
\end{equation}

Upon comparing with the ``pure'' bosonic partition function
(\ref{bossum}), we see that there is a difference arising from the
weights involved in the projection onto the zero--momentum ground
state. Yet in the limit $L \to \infty$, only the lowest--energy
bosonic state contributes to the sum (\ref{flsum2}), and upon
identification of the respective quantities in Table~\ref{analog}, the
quantum mechanics of superfluid bosons at $T = 0$ becomes completely
equivalent to the statistical mechanics of directed lines in the
thermodynamic limit. If we impose {\it periodic} boundary conditions
on the vortices, i.e. use a toroidal geometry in the field direction,
the exact correspondence of Eqs.~(\ref{flsum2}) and (\ref{bossum})
even survives for finite sample lengths (flux lines) / nonzero
temperatures (bosons), and a coherent--state path integral
representation for the grand--canonical partition function
(\ref{gcansm}) may be derived, which is precisely of the form of
Eqs.~(\ref{bosgrz}),(\ref{bosact}). Note that thermal fluctuations
$\propto k_{\rm B} T$ correspond to quantum fluctuations $\propto
\hbar$, while the torus length $L$ maps onto the inverse boson
temperature $\beta \hbar$. In this case, a sharp
Berezinskii--Kosterlitz--Thouless transition is to be expected from an
entangled (superfluid) phase at large $L$ to a disentangled (normal
fluid) phase at small $L$ \cite{nelson,nelseu}.

For magnetic flux lines in most superconducting samples, however,
periodic boundary conditions are rather artificial, and a more
adequate description of finite systems may be obtained using 
{\it open} boundary conditions as embodied in Eq.~(\ref{flsum2}),
where the line exit and entry points are integrated over freely,
corresponding to an ideal Bose gas with
\begin{equation}
        \psi({\bf r},0) = \psi({\bf r},L) = \sqrt{n} \ .
 \label{freebc}
\end{equation}
Note that this rigid constraint at the boundaries means that the
``depletion'' $n - n_0 = \langle | \psi |^2 \rangle - 
| \langle \psi \rangle |^2$ (see Sec.~\ref{pspebc}) vanishes
identically at the top and bottom of the sample. As is shown in
Ref.~\cite{nelseu}, within the mean--field approximation the sole
effect of the projection amplitude in (\ref{flsum2}) is a shift of the
chemical potential according to
\begin{equation}
        {\tilde \mu} = \mu - {k_{\rm B} T \over L} \, 
                                        \ln {N \over Z_1} \ , 
 \label{mureno}
\end{equation}
where $Z_1$ represents the partition function for a single flux
line. Physically, this amounts to a downward entropic renormalization
of the upper critical field $H_{c_1}$ [see Eq.~(\ref{chmpot})], due to 
thermal fluctuation of vortices. With the replacement 
$\mu \to {\tilde \mu}$, and the correspondences of Table~\ref{analog},
the results of the previous Sec.~\ref{bosons} can then be taken over. 
Fluctuation effects may furthermore be described by a coherent--state
path integral with an action of the form (\ref{bosact}), provided the
additional constraints on the fields $\psi$ and $\psi^*$, or $\pi$ and
$\Theta$, respectively, are explicitly taken into account, see
Ref.~\cite{nelseu} and Sec.~\ref{psopbc} below. In this geometry, the
exact correspondence between $L$ and $\beta \hbar$ in a real boson
system is lost, and the sharp phase transition from an entangled to a
disentangled flux liquid may in fact be replaced by a smooth crossover
\cite{fislee}. We also remark that the bosonic field operators ${\hat
\psi}^\dagger$ and ${\hat \psi}$ correspond to magnetic monopole and
anti--monopole creation operators in the flux line picture.

The vortex free energy (\ref{vortfe}) was derived for a constant
magnetic field ${\bf H}$ in the $z$ direction. The response to
``tilt'', i.e., to a small slowly varying magnetic field 
${\bf H}_\perp({\bf r},z)$ perpendicular to $z$ is also of
interest. Such perturbations couple directly to the ``tangent field''
of the flux lines,
\begin{equation}
        {\bf t}({\bf r},z) = \sum_{i=1}^N {d {\bf r}_i(z) \over dz} \,
        \delta^{(2)} \biggl( {\bf r}-{\bf r}_i(z) \biggr) \ , 
 \label{tanfld}
\end{equation}
and the free energy (\ref{vortfe}) is modified according to
\footnote{To derive this result, note that the perpendicular component
of the magnetic field ${\bf b}_\perp({\bf r},z)$ in Eq.~(\ref{gibbfe})
satisfies the London equation, ${\bf b}_\perp({\bf r},z) =
(m_z / m_\perp) \lambda^2 \bbox{\nabla}_\perp^2 
{\bf b}_\perp({\bf r},z) + \lambda^2 \partial_z^2 
{\bf b}_\perp({\bf r},z) + \phi_0 \, {\bf t}({\bf r},z)$. Upon
substituting for ${\bf b}_\perp({\bf r},z)$ in Eq.~(\ref{gibbfe}), the
derivative terms on the right--hand side can be eliminated via an
integration by parts with negligible error, provided the variations in
the external field ${\bf H}_\perp({\bf r},z)$ are sufficiently
slow. For a treatment of the response to more general field
perturbations, see Appendix A of Ref.~\cite{nelled}.}
\begin{eqnarray}
        F_N &&\longrightarrow F_N - {\phi_0 \over 4 \pi} \int \! d^3r
        \, {\bf H}({\bf r},z) \cdot {\bf t}({\bf r},z) \nonumber \\
        &&\longrightarrow F_N - {\tilde \epsilon}_1 \sum_{i=1}^N
                \int_0^L \! dz \, {\bf v}({\bf r}_i,z) \cdot 
                {d {\bf r}_i(z) \over dz} \ ,
 \label{tilfre}
\end{eqnarray}
where
\begin{equation}
        {\bf v}({\bf r},z) \equiv {\phi_0 \over 
        4 \pi {\tilde \epsilon}_1} \, {\bf H}_\perp({\bf r},z) \ .
 \label{velcty}
\end{equation}
The field ${\bf v}({\bf r},z)$ thus couples to the flux lines like a
vector potential in this ``Lagrangian'' formulation of the physics. In
the transfer matrix representation of the vortex partition function,
the Hamiltonian (\ref{fllham}) becomes \cite{nelled}
\begin{equation}
        {\cal H}_N^{\rm fl} = {1 \over 2 {\tilde \epsilon}_1}
        \sum_{j=1}^N \Biggl[ {k_{\rm B} T \over i} \bbox{\nabla}_j
        + i {\tilde \epsilon}_1 {\bf v}({\bf r}_j,z) \Biggr]^2
                + \sum_{i=1}^N V_D({\bf r}_i)
        + \frac{1}{2} \sum_{i\not=j} V(|{\bf r}_i-{\bf r}_j|) \ .
 \label{tilham}
\end{equation}
Note that the tilt field generates an {\it imaginary} vector potential
in this Hamiltonian formulation. If ${\bf H}_\perp$ is constant, then
tilting the external field acts like an imaginary Galilean boost with
velocity $i{\bf v} = i\phi_0 {\bf H}_\perp / 4\pi {\tilde \epsilon}_1$ 
on the fictitious bosons.

As a simple, but useful application of the mapping of vortex physics
onto quantum mechanics, we briefly consider the thermal wandering of a
single flux line ($N = 1$) in a sample of length $L$, starting at the
origin ${\bf r} = {\bf 0}$, and ending at the opposite surface at
${\bf r} = {\bf r}_\perp$ \cite{nelvin,nelnor}. With
Eqs.~(\ref{tramat}) and (\ref{fllham}), the constrained partition
function for this situation reads
\begin{equation}
        Z({\bf r}_\perp,{\bf 0};L) = \langle {\bf r}_\perp | 
        e^{- {\cal H}_1^{\rm fl} L / k_{\rm B}T} | {\bf 0} \rangle \ ,
 \label{conpar}
\end{equation}
and the probability distribution for the vortex tip position at the
upper surface becomes
\begin{equation}
        {\cal P}({\bf r}_\perp;L) = Z({\bf r}_\perp,{\bf 0};L) \Bigg /
        \int \! d^dr_\perp \, Z({\bf r}_\perp,{\bf 0};L) = 
        {\sum_n \langle n | {\bf 0} \rangle \, \langle 
        {\bf r}_\perp | n \rangle \, e^{- E_n^{\rm fl} L / k_{\rm B}T} 
        \over \sum_n \langle n | {\bf 0} \rangle \int \! d^dr_\perp \,
                                \langle {\bf r}_\perp | n \rangle \, 
                        e^{- E_n^{\rm fl} L / k_{\rm B}T}} \ ,
 \label{surwan}
\end{equation}
where in the second equation we have inserted a complete set of energy
eigenstates $| n \rangle$ with eigenvalues $E_n^{\rm fl}$. For thick
samples, $L \to \infty$, the ground state dominates the above sums,
and therefore 
\begin{equation}
        {\cal P}({\bf r}_\perp;L) = {\psi_0({\bf r}_\perp) \over
        \int \! d^dr_\perp \, \psi_0({\bf r}_\perp)} \left[ 1 +
        {\cal O} \left( e^{-(E_1^{\rm fl}-E_0^{\rm fl})L / k_{\rm B}T}
        \right) \right] \ ;
 \label{wvfunc}
\end{equation}
here, $\psi_0({\bf r}_\perp) = \langle {\bf r}_\perp | 0 \rangle$ is
the ground state ``wave function'' with eigenvalue $E_0^{\rm fl}$,
and $E_1^{\rm fl}$ is the energy of the first excited state. The
probability distribution for the flux line to have wandered a distance
$r_\perp$ at the surface is thus proportional to the ground state
wave function [note that $\psi_0({\bf r}_\perp)$ is positive and
nodeless]. 

On the other hand, if we are interested in the more general problem of
one flux line entering the sample at ${\bf r} = {\bf r}_i$ and leaving
it at ${\bf r}_f$, and ask what is it its average thermal wandering
probability distribution ${\tilde {\cal P}}({\bf r};L)$ at height $z$
far away from the two boundaries, the answer is
\begin{equation}
        {\tilde {\cal P}}({\bf r};L) = {\tilde Z}({\bf r};L) \Bigg /
                        \int \! d^dr \, {\tilde Z}({\bf r};L) \ ,
 \label{bulwan}
\end{equation}
where, using the definition (\ref{conpar}),
\begin{equation}
        {\tilde Z}({\bf r};L) = \int \! d^dr_i \int \! d^dr_f \,
        Z({\bf r}_f,{\bf r};L-z) \, Z({\bf r},{\bf r}_i;z) \ .
 \label{parfol}
\end{equation}
Upon inserting a complete set of energy eigenstates, for thick samples
the final result is
\begin{equation}
        {\tilde {\cal P}}({\bf r};L) = {\psi_0({\bf r})^2 \over
        \int \! d^dr \, \psi_0({\bf r})^2} \left[ 1 + {\cal O} 
        \left( e^{- (E_1^{\rm fl}-E_0^{\rm fl}) L / k_{\rm B} T}
                                                \right) \right] \ ,
 \label{sqwvfn}
\end{equation}
i.e., in the bulk, this probability distribution is proportional to
the square of the ground state wave function. Note that for periodic
boundary conditions along the $z$ direction, which rule out any
special surface effects, Eq.~(\ref{sqwvfn}) describes the thermal
wandering in the entire sample. On the other hand, for open boundary
conditions the surface behavior is different from that in the bulk,
and Eq.~(\ref{wvfunc}) applies. We therefore conclude that near the
surfaces, thermal flux line wandering is enhanced for open boundary
conditions, as depicted in Fig.~\ref{wavfun}.

We comment finally on the meaning of the ``boson'' order parameter for
directed lines. On a formal level, superfluidity is usually
accompanied by long--range order in the correlation function
\begin{equation}
        {\tilde G}({\bf r},z;{\bf r}',z') = 
        \langle \psi({\bf r},z) \psi^*({\bf r}',z') \rangle \ ,
 \label{correl}
\end{equation}
specifically
\begin{equation}
        \lim_{|{\bf r}-{\bf r}'| \to \infty} 
        {\tilde G}({\bf r},z;{\bf r}',z') = \langle \psi({\bf r},z) 
        \rangle \langle \psi^*({\bf r}',z') \rangle = n_0 \not= 0 \ ,
 \label{lgrord}
\end{equation}
where the average is evaluated via coherent--state path integrals. To
understand what this long--range order means for vortices
\cite{nelala}, it is helpful to consider first a situation where the
order is manifestly {\it short}--ranged, in the Abrikosov flux
lattice. Figure~\ref{monpol} shows that the operators 
${\hat \psi}$ and ${\hat \psi}^\dagger$ create vacancy and
interstitial strings in the vortex crystal, terminating in ``magnetic
monopoles''. The composite operator in Eq.~(\ref{correl}) creates an
extra line at $({\bf r}',z')$ (i.e., a column of interstitials in the
solid), and destroys an existing line at $({\bf r},z)$, creating a
column of vacancies. The lowest--energy configuration is then a line
of vacancies (for $z' > z$) or interstitials (for $z' < z$) connecting
the two points with an energy $\sigma s$ proportional to the length
$s$ of this ``string''. It follows that the correlation function
(\ref{correl}) decays exponentially to zero (i.e., 
$\propto e^{- \sigma |{\bf r}-{\bf r}'| / k_{\rm B} T}$) for large
separations in the crystalline phase. In an entangled line 
{\it liquid}, on the other hand, vacancies and interstitials ``melt''
away and the asymptotic string tension $\sigma$ vanishes, implying
long--range order (\ref{lgrord}) in 
${\tilde G}({\bf r},z;{\bf r}',z')$. In this limit, 
$\langle \psi({\bf r},z) \rangle = \langle \psi^*({\bf r}',z') \rangle
= e^{- E_{\rm mon} / k_{\rm B} T}$, where $E_{\rm mon}$ is the energy
of an isolated ``magnetic monopole''.

\subsection{Galilean invariance of the pure system and affine
            transformations}
 \label{galinv}

Before we proceed with the investigation of fluctuation effects in
boson superfluids and flux line liquids, we discuss the fundamental
symmetries of the pure systems, namely invariance with respect to
Galilean transformations and uniform tilts, respectively. We begin
with the disorder--free boson Hamiltonian (\ref{bosham}), and study
the effect of a Galilean boost with constant velocity 
${\bf v} = {\rm const.}$,
\begin{eqnarray}
        {\bf r} &&\to {\bf r}' = {\bf r} + {\bf v} t \ , \quad
        t \to t' = t \ , \nonumber \\
        {\bf P} &&\to {\bf P}' = {\bf P} + N m {\bf v} \ , \nonumber \\
        E &&\to E' = E + {\bf v} \cdot {\bf P} + \frac{1}{2} Nm v^2 \ ,
 \label{galtra}
\end{eqnarray}
where ${\bf P}$ is the total momentum, and $E$ the total energy of the
$N$--particle system.

In the coherent--state path integral representation (with imaginary
time $\tau = i t$), the particle density $n({\bf r},\tau)$, momentum
density ${\bf g}({\bf r},\tau)$, particle current 
${\bf j}({\bf r},\tau)$, and energy density 
${\tilde {\cal H}}({\bf r},\tau) = {\cal H}({\bf r},\tau) - 
\mu n({\bf r},\tau)$ read
\begin{eqnarray}
        n({\bf r},\tau) &&= | \psi({\bf r},\tau) |^2 \ ,
 \label{parden} \\
        {\bf g}({\bf r},\tau) &&= m {\bf j}({\bf r},\tau) =
        {\hbar \over 2 i} \Bigl[ \psi^*({\bf r},\tau) 
        \bbox{\nabla} \psi({\bf r},\tau) - \psi({\bf r},\tau) 
        \bbox{\nabla} \psi^*({\bf r},\tau) \Bigr] \ ,
 \label{momden} \\
        {\tilde {\cal H}}({\bf r},\tau) &&= 
        {\hbar^2 \over 2 m} | \bbox{\nabla} \psi({\bf r},\tau) |^2
                                - \mu | \psi({\bf r},\tau) |^2 
        + \frac{1}{2} \int \! d^dr' V({\bf r}-{\bf r}') 
        | \psi({\bf r}',\tau) |^2 \; | \psi({\bf r},\tau) |^2 \ .
 \label{hamden}
\end{eqnarray}
A Galilean transformation (\ref{galtra}) is then represented by a
unitary transformation acting on the fields,
\begin{equation}
        \psi({\bf r},\tau) \to \psi'({\bf r}',\tau') =
        \psi({\bf r},\tau) \, e^{i m {\bf v} \cdot {\bf r} /\hbar} \ ;
 \label{unitra}
\end{equation}
as is easily seen, this leaves the particle density invariant,
\begin{equation}
        n'({\bf r}',\tau') = n({\bf r},\tau) \ ,
 \label{dentra}
\end{equation}
while the transformed momentum and energy densities become
\begin{eqnarray}
        {\bf g}'({\bf r}',\tau') &&= {\bf g}({\bf r},\tau)
                                + m {\bf v} \, n({\bf r},\tau) \ ,
 \label{momtra} \\
        {\tilde {\cal H}}'({\bf r}',\tau') &&= 
                                {\tilde {\cal H}}({\bf r},\tau) 
                                + {\bf v} \cdot {\bf g}({\bf r},\tau) 
                        + \frac{1}{2} m v^2 \, n({\bf r},\tau) \ ,
 \label{hamtra}
\end{eqnarray}
in accord with Eqs.~(\ref{galtra}). Note that the third term in
Eq.~(\ref{hamtra}) may be absorbed into a shift of the chemical
potential,
\begin{eqnarray}
        {\cal H}'({\bf r}',\tau') &&= {\cal H}({\bf r},\tau)
        + {\bf v} \cdot {\bf g}({\bf r},\tau) \ , 
 \label{enetra} \\
        \mu' &&= \mu - m v^2 / 2 \ .
 \label{chmtra} 
\end{eqnarray}

In the polar representation
\begin{equation}
        \psi({\bf r},\tau) = \sqrt{n_0 + \pi({\bf r},\tau)} \, 
                                e^{i \Theta({\bf r},\tau)} \ ,
 \label{pithet}
\end{equation}
the effect of the Galilean transformation is
\begin{eqnarray}
        \pi'({\bf r}',\tau') &&= \pi({\bf r},\tau) \ , 
 \label{pitran} \\
        \Theta'({\bf r}',\tau') &&= \Theta({\bf r},\tau) 
                                + m {\bf v} \cdot {\bf r} / \hbar \ .
 \label{thtran}
\end{eqnarray}
Upon introducing the superfluid velocity according to
Eq.~(\ref{sflvel}), 
\begin{equation}
        {\bf v}_s({\bf r},\tau) = 
                \hbar \bbox{\nabla} \Theta({\bf r},\tau) / m \ , \quad
        {\bf g}({\bf r},\tau) = m {\bf v}_s({\bf r},\tau)
                                                 n({\bf r},\tau)  \ ,
 \label{sfvmom}
\end{equation}
one finds
\begin{equation}
        {\bf v}_s'({\bf r}',\tau')={\bf v}_s({\bf r},\tau)+{\bf v} \ .
 \label{sfvtra}
\end{equation}

For the case of directed line liquids, the Galilean boost
(\ref{galtra}) translates into an affine transformation, namely a
uniform tilt away from the $z$ axis,
\begin{equation}
        {\bf r} \to 
        {\bf r}' = {\bf r} + {\bf h} z / {\tilde \epsilon}_1 \ , \quad
        z \to z' = z \ , \quad
        {d {\bf r} \over dz} \to {d {\bf r}' \over dz'} =
        {d {\bf r} \over dz} + {{\bf h} \over {\tilde \epsilon}_1} \ ,
 \label{afftra}
\end{equation}
with ${\bf h} = {\rm const.}$, which may be interpreted as a
transverse external magnetic field. As discussed above, the uniform
tilt ${\bf h} / {\tilde \epsilon}_1 \equiv \phi_0 {\bf H}_\perp / 
4 \pi {\tilde \epsilon}_1$ corresponds to a Galilean transformation
with {\it imaginary} velocity $i {\bf v}$ (see Table~\ref{analog}). 
This imaginary boost velocity means that the phase does {\it not}
change in the simple fashion described by Eq.~(\ref{thtran}). In the
corresponding quantum--mechanical problem one has to consider the
effect of imaginary vector potentials leading to the appearance of
non--Hermitian operators \cite{hatnel}. The transformation formulas
for the energy density may be read off from the previous expressions;
the shift in the chemical potential becomes
\begin{equation}
        \mu' = \mu + h^2 / 2 {\tilde \epsilon}_1 \ .
 \label{afchtr}
\end{equation}

We finally remark that under a Galilean boost (\ref{galtra}) or affine
transformation (\ref{afftra}), the disorder potential is modified
according to
\begin{equation}
        V_D({\bf r}(\tau),\tau) \to 
                        V_D({\bf r}'+ i {\bf v} \tau',\tau') \ , 
 \label{deftra}
\end{equation}  
and
\begin{equation}
        V_D({\bf r}(z),z) \to 
                V_D({\bf r}'-{\bf h} z' / {\tilde \epsilon_1},z') \ ,
 \label{afdotr}
\end{equation}  
respectively. Therefore, disorder breaks ``Galilean'' invariance;
however, for uncorrelated defects for which the disorder correlator is
a constant, the symmetry is recovered by taking the quenched disorder
average. In the Appendix~\ref{galtil}, we review how these affine
transformations can be exploited for the calculation of the disorder
contribution to the line tilt modulus in the thermodynamic limit 
$L \to \infty$ \cite{hwaled}. A method which is more convenient for
finite $L$ is discussed in the next section.


\section{Fluctuations and correlation functions in pure systems}
 \label{puresy}

We are now ready to evaluate correlation functions for our model
action (\ref{bosact}). We shall henceforth regularly use the boson
notation and language, but nevertheless keep the application to flux
liquids in mind as well. We start by rederiving and summarizing the
known results for the pure system with periodic (bosonic) boundary
conditions \cite{npopov}, before we extend our analysis to open
boundary conditions (Sec.~\ref{psopbc}), which are more appropriate to
flux line systems, in the presence of uncorrelated and correlated
disorder (Secs.~\ref{disbos},\ref{disfll}).

\subsection{Correlation functions in the Gaussian ensemble with
            periodic boundary conditions (Bogoliubov approximation)}
 \label{harcor}

To separate density and phase fluctuations, it is convenient to use
the polar representation (\ref{pithet}) for the fields $\psi$ and
$\psi^*$. Note that the path integral in the partition sum
(\ref{bosgrz}) is originally defined by integration over
${\rm Re} \, \psi = \sqrt{n_0 + \pi} \cos \Theta$ and
${\rm Im} \, \psi = \sqrt{n_0 + \pi} \sin \Theta$; the Jacobian
corresponding to the above transformation is $1/2$, and being constant
may be absorbed into the path integral measure. As we shall only be
dealing with small fluctuations, we extend the integration range for
the phase field $\Theta$ from $[-\pi;\pi]$ to $[-\infty;+\infty]$. 
Thus the grand--canonical partition function becomes a functional
integral over the independent variables $\pi$ and $\Theta$,
\begin{equation}
        Z_{\rm gr}^{\rm bos} = \int_{\begin{array}{l} 
                \scriptsize \pi(r,\beta\hbar) = \pi(r,0) \\
        \scriptsize \Theta(r,\beta\hbar) = \Theta(r,0) \end{array}}
        \! {\cal D}[\pi({\bf r},\tau)] {\cal D}[\Theta({\bf r},\tau)]
        \; e^{- \left( S_0[\pi,\Theta] + S_{\rm int}[\pi,\Theta] 
                                                \right) / \hbar} \ ,
 \label{pithps}
\end{equation}
where, upon splitting the disorder potential into its mean value and
the fluctuations around it, $V_D({\bf r},\tau) = \overline{V_D} +
\delta V_D({\bf r},\tau)$ and using Eq.~(\ref{landau}) in the
superfluid phase, $\mu + \overline{V_D} = n_0 V_0$, the harmonic and
``interacting'' parts of the action read
\begin{eqnarray}
        S_0[\pi,\Theta] = \int_0^{\beta \hbar} \!\! d\tau \int \! d^dr
        \Biggl\{ &&i \hbar \pi({\bf r},\tau) \,
                {\partial \Theta({\bf r},\tau) \over \partial \tau}
        + {\hbar^2 \over 8 m n_0} \,
                \left[ \bbox{\nabla} \pi({\bf r},\tau) \right]^2
        + {\hbar^2 n_0 \over 2 m} \,
                \left[ \bbox{\nabla} \Theta({\bf r},\tau) \right]^2
                                                        \nonumber \\
        &&- \delta V_D({\bf r},\tau) 
                        \Bigl[ n_0 + \pi({\bf r},\tau) \Bigr] \,
        + \frac{1}{2} \int \! d^dr' \, V(|{\bf r}-{\bf r}'|) \,
        \pi({\bf r}',\tau) \, \pi({\bf r},\tau) \Biggr\} \nonumber \\
        &&- \frac{1}{2} n_0^2 V_0 \beta \hbar \Omega \ ,
 \label{haract}
\end{eqnarray}
and
\begin{equation}
        S_{\rm int}[\pi,\Theta] = \int_0^{\beta \hbar} \!\! d\tau 
        \int \! d^dr \Biggl\{ {- \hbar^2 \over 8 m n_0} \, 
                {\pi({\bf r},\tau) \over n_0 +\pi({\bf r},\tau)} 
                \left[ \bbox{\nabla} \pi({\bf r},\tau) \right]^2
        + {\hbar^2 \over 2 m} \, \pi({\bf r},\tau) \left[ 
        \bbox{\nabla} \Theta({\bf r},\tau) \right]^2 \Biggr\} \ ,
 \label{intact}
\end{equation}
respectively. This separation of the action amounts essentially to a
splitting into single-- and multi--quasiparticle excitations. In this
paper, we shall entirely neglect the anharmonic terms (\ref{intact}),
and restrict ourselves to the Gaussian approximation. Note that
taking into account the nonlinear fluctuation terms (\ref{intact})
perturbatively would amount to an expansion in powers of 
$\hbar^2$ ($\sim T^2$ for flux lines); this would then include further
contributions from the disorder potential which enter via their linear
coupling to the density fluctuations $\pi$.

The harmonic action (\ref{haract}) is of course immediately
diagonalized via (discrete) Fourier transformation,
\begin{equation}
        \psi({\bf r},\tau) = {1 \over \beta \hbar \Omega} 
                \sum_{{\bf q},\omega_m} \psi({\bf q},\omega_m) 
        e^{i \left( {\bf q}\cdot{\bf r} - \omega_m \tau \right)} \ , 
 \label{foutra} 
\end{equation}
with the inverse
\begin{equation}
        \psi({\bf q},\omega_m) = \int \! d^dr 
                \int_0^{\beta \hbar} \!\! d\tau \, \psi({\bf r},\tau)
        e^{- i \left( {\bf q}\cdot{\bf r} - \omega_m \tau \right)} \ .
 \label{fouinv}
\end{equation}
In the thermodynamic limit, the discrete wavevector sum is as usual
replaced by an integral,
$\sum_{\bf q} \ldots \to \Omega (2 \pi)^{-d} \int d^dq \ldots$; the
periodic boundary conditions in imaginary time on the other hand
enforce the discrete bosonic Matsubara frequencies to be
\begin{equation}
        \omega_m = m \, {2 \pi \over \beta \hbar} \; , \quad 
                                m = 0, \pm 1, \pm 2, \ldots \ ,
 \label{matsub}
\end{equation}
which, similarly to the wavenumbers, approach a continuum as 
$\beta \to \infty$, $\sum_m \ldots \to \beta \hbar / (2 \pi)^{-1} 
\int_{-\infty}^\infty d \omega \ldots$.

For pure systems, i.e., if $\delta V_D({\bf r},\tau) = 0$, the
effective action in the Gaussian approximation can be written in the 
matrix form
\begin{equation}
        S_0[\pi,\Theta] = {1 \over 2 \beta \hbar \Omega}
        \sum_{{\bf q},\omega_m} \Bigl( \Theta(-{\bf q},-\omega_m) \, ,
        \, \pi(-{\bf q},-\omega_m) \Bigr) \, {\bf A}({\bf q},\omega_m)
        \, \left( \begin{array}{c} \Theta({\bf q},\omega_m) \\
                        \pi({\bf q},\omega_m) \end{array} \right) 
        - \frac{1}{2} n_0^2 V_0 \beta \hbar \Omega \ ,
 \label{puhars}
\end{equation}
with
\begin{equation}
        {\bf A}({\bf q},\omega_m) = \left( \begin{array}{cc}
                        n_0 \hbar^2 q^2 / m & - \hbar \omega_m \\ 
        \hbar \omega_m & V(q) + \hbar^2 q^2 / 4 m n_0 \end{array} 
                                                        \right) \ .
 \label{matrix}
\end{equation}  
The density and phase correlation functions can therefore be obtained
by simple inversion of the matrix (\ref{matrix}),
\begin{equation}
        \hbar {\bf A}^{-1}({\bf q},\omega_m) = 
        {1 \over \omega_m^2 + \epsilon_{\rm B}(q)^2 / \hbar^2}
        \left( \begin{array}{cc} 
        m \epsilon_{\rm B}(q)^2 / n_0 \hbar^3 q^2 & \omega_m \\ 
                - \omega_m & n_0 \hbar q^2 / m \end{array} \right) \ ;
 \label{matinv}
\end{equation}
here, $\epsilon_{\rm B}(q)$ denotes the Bogoliubov quasi--particle
spectrum
\begin{equation}
        \epsilon_{\rm B}(q) = \sqrt{n_0 \hbar^2 q^2 V(q) / m +
                                (\hbar^2 q^2 / 2 m)^2} \ ,
 \label{bogspe}
\end{equation}
which at long wavelengths displays a linear dispersion,
\begin{equation}
        \epsilon_{\rm B}(q) \to \hbar c_1 q \quad {\rm for} \quad 
                                                        q \to 0 \ ,
 \label{lindis}
\end{equation}
where $c_1$ is the speed of sound from Eq.~(\ref{soundv}).

We can assign a physical meaning to the Bogoliubov spectrum by
investigating the density--density correlation function,
\begin{equation}
        S({\bf r},\tau;{\bf r}',\tau') = 
        \langle n({\bf r},\tau) n({\bf r}',\tau') \rangle -n^2 \approx
        \langle \pi({\bf r},\tau) \pi({\bf r}',\tau') \rangle \ .
 \label{dencor}
\end{equation}
Note that by adding a source term to the action, i.e., by performing a
Legendre transformation according to $S[\delta \mu] = S[\psi^*,\psi] +
\int_0^{\beta \hbar} \! d\tau \int \! d^dr \, \delta \mu({\bf r},\tau) 
|\psi({\bf r},\tau)|^2$, we can also write this in the form of a
functional derivative,
\begin{equation}
        S({\bf r},\tau;{\bf r}',\tau') = 
        \hbar^2 \, {\delta^2 \ln Z_{\rm gr}[\delta \mu] \over
                                \delta [\delta \mu({\bf r},\tau)] 
                                \delta [\delta \mu({\bf r}',\tau')]}
                                        \Bigg \vert_{\delta \mu = 0} =
        \hbar \,{\delta \langle \pi({\bf r},\tau) \rangle_{\delta \mu}
                        \over \delta [\delta \mu({\bf r}',\tau')]}
                                \Bigg \vert_{\delta \mu = 0} \ ,
 \label{denres}
\end{equation}
which shows that the imaginary--time correlation function is actually
a quantum--mechanical response function. From Eq.~(\ref{matinv}), in
Fourier space we immediately find as a consequence of translation
invariance in space and time
\begin{equation}
        S({\bf q},\omega_m;{\bf q}',\omega_{m'}) = 
        S({\bf q},\omega_m) \, (2 \pi)^d \delta({\bf q}+{\bf q}') \,
                        \beta \hbar \, \delta_{m,-m'} \ , 
 \label{redcor}
\end{equation}
with the reduced density correlation function
\begin{equation}
        S({\bf q},\omega_m) = {n_0 \hbar q^2 / m \over
                \omega_m^2 + \epsilon_{\rm B}(q)^2 / \hbar^2} \ ;
 \label{fouden} 
\end{equation}
by performing the simple Matsubara frequency sum [Eq.~(\ref{bfmat2})
in App.~\ref{bosmat}] we can readily transform back to imaginary time,
\begin{equation}
        S({\bf q},\tau) = {1 \over \beta \hbar} \sum_m 
                        S({\bf q},\omega_m) e^{-i \omega_m \tau} =
        {n_0 \hbar^2 q^2 \over 2 m \epsilon_{\rm B}(q)} \,
                {e^{(\beta - \tau/\hbar) \epsilon_{\rm B}(q)} +
                        e^{\epsilon_{\rm B}(q) \tau \hbar} \over 
                        e^{\beta \epsilon_{\rm B}(q)} - 1} \ ,
 \label{timstr}
\end{equation}
with the static limit
\begin{equation}
        S({\bf q},0) = {n_0 \hbar^2 q^2 \over 2 m \epsilon_{\rm B}(q)}
                \, \coth {\beta \epsilon_{\rm B}(q) \over 2} \ ,
 \label{ststrf}
\end{equation}
which is precisely the static structure factor in the Bogoliubov
approximation for weakly interacting Bose gases \cite{fetwal,nozpin}. 
Note that $S({\bf q},0)$ is inversely proportional to the excitation
spectrum when $\beta \to \infty$. If we now analytically continue to
$t = -i \tau$, and carry out the usual Fourier integral, we arrive at
\begin{equation}
        S({\bf q},\omega) = \int \! dt \, S({\bf q},t) e^{i \omega t}
                = {n_0 \hbar^2 q^2 \over 2 m \epsilon_{\rm B}(q)} \, 
                {2 \pi \over e^{\beta \epsilon_{\rm B}(q)} - 1}
                \left[ e^{\beta \epsilon_{\rm B}(q)}
        \delta \Bigl( \omega - \epsilon_{\rm B}(q) / \hbar \Bigr) +
        \delta \Bigl( \omega + \epsilon_{\rm B}(q) / \hbar \Bigr)
                                                        \right] \ ,
 \label{dynstr}
\end{equation}
i.e., the dynamic structure factor directly measured in scattering
experiments. One therefore expects quasiparticle peaks with the
dispersion relation (\ref{bogspe}), thermally excited with a
Bose--Einstein distribution (note the detailed--balance factor 
$e^{\beta \epsilon_{\rm B}(q)}$). The Gaussian approximation for the
density and phase fluctuations produces exactly the same results as
the familiar Bogoliubov approach in many--particle quantum mechanics,
where a canonical transformation for the creation and annihilation
operators is employed in order to diagonalize the second--quantized
quasi--particle Hamiltonian \cite{fetwal,nozpin}. In the
long--wavelength limit, the elementary excitations of the Bose
condensate are therefore no longer of the free--particle form
$\epsilon_0(q) = \hbar^2 q^2 / 2 m$, but display the linear dispersion
(\ref{lindis}) and may thus be interpreted as phonon modes. Depending
on the form of the interaction potential $V(q)$, the dispersion
relation may display the roton minimum characteristic of strongly 
interacting Bose liquids like superfluid Helium 4. However, our
harmonic approximation is only valid either for weakly interacting
systems where $V(q)$ is sufficiently small, or for dilute Bose gases
for which the pair potential may be strong, but where the scattering
length $a = m V_0 / 4 \pi$ is small compared to the average
interparticle spacing. In that case, one can exactly sum all the
ladder diagrams  describing multiple scattering events, with the 
result that eventually $m V(q)$ is to be replaced by the effective
s--wave scattering amplitude $f_0 = m V_0$ \cite{fetwal,nozpin}; the
dimensionless coupling in the Bogoliubov theory, serving as an
expansion parameter in the weak--coupling limit, is in fact 
$g^2 \propto n^{(d-2)/d} V_0$ \cite{fishoh}. In both situations the
quasiparticle dispersion is a monotonically increasing function, and
shows no roton minimum. 

We also remark that a second route to arrive at the result
(\ref{dynstr}), is to remember that 
$S({\bf q},\omega_m) = \hbar \chi_n({\bf q},\omega_m)$ is the density
{\it response} function; by analytic continuation in frequency space,
$\omega_m \to i(\omega + i \eta)$, $\eta \to 0^+$, and splitting
$\chi_n({\bf q},\omega) = \chi_n'({\bf q},\omega) + 
i \chi_n''({\bf q},\omega)$, where $\chi_n'({\bf q},\omega) =
[\chi_n({\bf q},\omega+i\eta) + \chi_n({\bf q},\omega-i\eta)]/2$
and $\chi_n''({\bf q},\omega) = [\chi_n({\bf q},\omega+i\eta) -
\chi_n({\bf q},\omega-i\eta)]/2i$ denote its components which are
continuous and discontinuous across the real axis, respectively, one
may then infer the dynamic structure factor (in real time), i.e., the
corresponding {\it correlation} function, from the quantum--mechanical
fluctuation--dissipation theorem
\begin{equation}
        2 \hbar \chi_n''({\bf q},\omega) = 
        \left( 1 - e^{-\beta\hbar\omega} \right) S({\bf q},\omega) \ .
 \label{fldsth}
\end{equation}

As a second important correlation function, we consider the momentum
(mass current) correlator in the Gaussian approximation, defined
as
\begin{equation}
        C_{ij}({\bf r},\tau;{\bf r}',\tau') = 
        \langle g_i({\bf r},\tau) g_j({\bf r}',\tau') \rangle -
                        \langle g_i({\bf r},\tau) \rangle 
                        \langle g_j({\bf r}',\tau') \rangle \ .
 \label{momcor}
\end{equation}
According to Eq.~(\ref{sfvmom}), 
${\bf g} = m {\bf j} = \hbar (n_0 + \pi) \bbox{\nabla} \Theta$;
inserting this into Eq.~(\ref{momcor}), we can factorize the
four--point correlators into all possible two--point functions (i.e.,
apply Wick's theorem), while the three--point correlators vanish in
the Gaussian ensemble. In Fourier space, this yields
\begin{eqnarray}
        &&C_{ij}({\bf q},\omega_m;{\bf q}',\omega_m') = 
        - \hbar^2 n_0^2 q_i q_j' \langle \Theta({\bf q},\omega_m)
                \Theta({\bf q}',\omega_m') \rangle \nonumber \\
        &&- {1 \over (\beta \Omega)^2}
        \sum_{{\bf p},\omega_n} \sum_{{\bf p}',\omega_n'} p_i p_j'
        \biggl[ \langle \pi({\bf q}-{\bf p},\omega_{m-n}) 
                \pi({\bf q}'-{\bf p}',\omega_{m'-n'}) \rangle
        \langle \Theta({\bf p},\omega_n) 
                \Theta({\bf p}',\omega_{n'}) \rangle \nonumber \\
        &&\qquad \qquad \qquad \qquad
                + \langle \pi({\bf q}-{\bf p},\omega_{m-n}) 
                                \Theta({\bf p}',\omega_{n'}) \rangle
        \langle \Theta({\bf p},\omega_n) 
        \pi({\bf q}'-{\bf p}',\omega_{m'-n'}) \rangle \biggr] \ .
 \label{gaumcr}
\end{eqnarray}
Note that this quantity is {\it not} simply proportional to the phase
correlation function as obtained from Eq.~(\ref{matinv}),
\begin{equation}
        \langle \Theta({\bf q},\omega_m) 
                                \Theta({\bf q}',\omega_m') \rangle =
        {m \epsilon_{\rm B}(q)^2 / n_0 \hbar^3 q^2 \over
        \omega_m^2 + \epsilon_{\rm B}(q)^2 / \hbar^2} \, (2 \pi)^d
        \delta({\bf q}+{\bf q}') \, \beta \hbar \, \delta_{m,-m'} \ .
 \label{phacor}
\end{equation}
[In the Appendix~\ref{poappr}, we discuss the ``phase--only
approximation'' in some detail, which neglects the additional
correlators in Eq.~(\ref{gaumcr})]. Using the Gaussian correlation
functions to be read off from Eq.~(\ref{matinv}), one finds for the
transverse current response function
\begin{eqnarray}
        &&\chi_\perp({\bf q},\omega_m) \equiv {1 \over (d-1) \hbar}
                \sum_{ij} P_{ij}^T({\bf q}) C_{ij}({\bf q},\omega_m)
                                                        \nonumber \\ 
        &&\qquad = {1 \over (d-1)\beta \Omega} \sum_{{\bf p},\omega_n}
        {q^2 p^2 - ({\bf q}{\bf p})^2 \over q^2 p^2}
        {({\bf q}-{\bf p})^2 \epsilon_{\rm B}(p)^2 / \hbar^2 
                                + p^2 \omega_{m-n} \omega_n \over
        [\omega_{m-n}^2+\epsilon_{\rm B}(|{\bf q}-{\bf p}|)^2/\hbar^2] 
                [\omega_n^2 + \epsilon_{\rm B}(p)^2 / \hbar^2]} \ .
 \label{trmres}
\end{eqnarray}
Here, the transverse projector is $P_{ij}^T({\bf q}) = \delta_{ij} -
P_{ij}^L({\bf q})$, with $P_{ij}^L({\bf q}) = q_i q_j / q^2$. Note
that $\chi_\perp({\bf q},\omega_m)$ would have vanished entirely, had
we not taken the nonlinear terms in Eq.~(\ref{gaumcr}) into account,
because the contribution from the phase fluctuations in
Eq.~(\ref{gaumcr}) is purely longitudinal 
\footnote{The above factorization of the four--point correlation
function actually goes beyond the Gaussian approximation (tree
level). In field--theory language, it corresponds to a computation of
the one--loop contributions to the composite--operator cumulant
$\langle [\pi \nabla_i \Theta] [\pi \nabla_j \Theta] \rangle$. 
Therefore, in order to consistently evaluate the longitudinal current
response function to this order, one would have to include the
one--loop corrections to the phase correlations as well, which would
involve the nonlinear vertices of Eq.~(\ref{intact}). In the limits
$\omega_m \to 0$ and ${\bf q} \to {\bf 0}$, the loop terms actually
cancel, and the ensuing result for the total density is 
$\rho = \lim_{q \to 0} \chi_\parallel({\bf q},0) \approx n_0 m$.}.

In the limit $\omega_m \to 0$ (first), followed by 
${\bf q} \to {\bf 0}$, the transverse current response function yields
the norma--fluid density, for this quantity is defined precisely as
the transport coefficient relating to the system's response to
transverse motion \cite{fetwal,nozpin,grifin,npopov,ceperl}. This
yields
\begin{equation}
        \rho_n = \lim_{q \to 0} \chi_\perp({\bf q},0) = 
        {1 \over d \beta \Omega} \sum_{{\bf p},\omega_n} p^2 \,
        {\epsilon_{\rm B}(p)^2 / \hbar^2 - \omega_n^2 \over \left[
        \epsilon_{\rm B}(p)^2 / \hbar^2 + \omega_n^2 \right]^2} \ ;
 \label{norfld}
\end{equation}
upon performing the Matsubara frequency sum [Eq.~(\ref{dfmsm4}) in
App.~\ref{bosmat}], we find
\begin{equation}
        \rho_n = {\beta \hbar^2 \over 4 d} 
                \int \! {d^dq \over (2 \pi)^d} \left( {q \over 
                \sinh [\beta \epsilon_{\rm B}(q) / 2] } \right)^2 \ ,
 \label{nflden}
\end{equation}
and therefore $\rho_n(T=0) = 0$; at finite ``temperatures'',
Eq.~(\ref{nflden}) will be evaluated explicitly in two and three
dimensions in Sec.~\ref{pspebc}.

It is interesting to note that the transverse momentum response
function is actually intimately related to the vorticity correlation
function, which (in $d=2$ and $d=3$ dimensions) is defined as
\begin{equation}
        V_{ij}({\bf r},\tau;{\bf r}',\tau') = m^{-2} \Big\langle 
        \Bigl[ \bbox{\nabla} \times {\bf g}({\bf r},\tau) \Bigr]_i 
        \Bigl[ \bbox{\nabla}' \times {\bf g}({\bf r}',\tau)' \Bigr]_j 
                                                \Big\rangle \ .
 \label{vorcor}
\end{equation}
In Fourier space, and after factorizing according to Wick's theorem,
this becomes in terms of the density and phase fluctuations
\begin{eqnarray}
        V_{ij}({\bf q},\omega_m;{\bf q}',\omega_{m'}) 
        &&= \sum_{k,l,k',l'} \epsilon_{ikl} \epsilon_{jk'l'} 
                                        {1 \over (m \beta \Omega)^2} 
                \sum_{{\bf p},\omega_n} \sum_{{\bf p}',\omega_{n'}} 
        (q_k-p_k) p_l (q_{k'}'-p_{k'}') p_{l'}' \times \nonumber \\
        &&\times \biggl[ \langle \pi({\bf q}-{\bf p},\omega_{m-n}) 
                \pi({\bf q}'-{\bf p}',\omega_{m'-n'}) \rangle
        \langle \Theta({\bf p},\omega_n) 
                \Theta({\bf p}',\omega_{n'}) \rangle \nonumber \\
        &&\qquad + \langle \pi({\bf q}-{\bf p},\omega_{m-n}) 
                                \Theta({\bf p}',\omega_{n'}) \rangle
        \langle \Theta({\bf p},\omega_n) 
        \pi({\bf q}'-{\bf p}',\omega_{m'-n'}) \rangle \biggr] \ .
 \label{fouvor}
\end{eqnarray}
Following the same procedure as above, one may demonstrate explicitly
that the vorticity correlations are purely transverse, 
$V_\parallel({\bf q},\omega_m) = \sum_{ij} P_{ij}^L({\bf q})
V_{ij}({\bf q},\omega_m) = 0$, and after some algebra finds
\begin{equation}
        m^2 V_\perp({\bf q},\omega_m) = 
                        \hbar q^2 \chi_\perp({\bf q},\omega_m) \ .
 \label{momvor}
\end{equation}
It is therefore the intrinsic vorticity fluctuations that render the
transverse current response function and the normal--fluid density
(\ref{nflden}) nonzero at positive temperatures. These fluctuations
would be confined to the vortex cores in a ``phase--only''
Kosterlitz--Thouless picture of finite--temperature superfluid films.

Finally, it is natural for flux lines to introduce the ``tilt''
correlation function by considering an action perturbed by an
imaginary ``velocity'' field ${\bf v}({\bf r},\tau)$ given by
Eq.~(\ref{velcty}),
\begin{equation}
        S[{\bf v}] = S[\psi^*,\psi] + \int_0^{\beta \hbar} \! d\tau \!
        \int \! d^dr \, [i{\bf v}({\bf r},\tau) \cdot 
        {\bf g}({\bf r},\tau) - mv({\bf r},\tau)^2 n({\bf r},\tau) /2] 
 \label{shfact}
\end{equation}
[compare Eq.~(\ref{hamtra})], and then defining \cite{nelled}
\begin{eqnarray}
        T_{ij}({\bf r},\tau;{\bf r}',\tau') &&= {\hbar^2 \over m^2}
                \, {\delta^2 \ln Z_{\rm gr}[{\bf v}] \over 
                \delta v_i({\bf r},\tau) \delta v_j({\bf r}',\tau')}
                        \Bigg \vert_{{\bf v} = {\bf 0}} = \langle 
        t_i({\bf r},\tau) t_j({\bf r}',\tau') \rangle \nonumber \\
        &&= (n \hbar / m) \, 
        \delta_{ij} \delta({\bf r}-{\bf r}') \delta(\tau-\tau') 
                - C_{ij}({\bf r},\tau;{\bf r}',\tau') / m^2 \ ,
 \label{tilcor} 
\end{eqnarray}
which in Fourier space becomes
\begin{equation}
        T_{ij}({\bf q},\omega_m) = (n \hbar / m) \, \delta_{ij} - 
                                C_{ij}({\bf q},\omega_m) / m^2 \ .
 \label{foutil}
\end{equation}
Thus, tangent--tangent correlations for flux lines are intimately
related to momentum correlations in the equivalent superfluid. To
extract the tilt modulus from Eq.~(\ref{foutil}), one has to take 
${\bf q} \to {\bf 0}$ first, followed by $\omega_m \to 0$. These
limits mimic the actual physical situation for open boundary
conditions, reflecting the response to an external tilt field which
acts on the top and bottom faces (perpendicular to $z$) of a
superconducting slab \cite{hatnel,nelled}. We can again isolate a
matrix of ``transport'' coefficients, namely
\begin{equation}
        {c^v_{ij}}^{-1} = (n^2 \hbar)^{-1} \lim_{\omega_m \to 0}
        T_{ij}({\bf 0},\omega_m) = {1 \over n m} \left( \delta_{ij} - 
        {1 \over n m \beta \Omega} \sum_{{\bf p},\omega_n} p_i p_j \, 
        {\epsilon_{\rm B}(p)^2 / \hbar^2 - \omega_n^2 \over \left[ 
        \epsilon_{\rm B}(p)^2/\hbar^2+\omega_n^2 \right]^2} \right)\ .
 \label{tilmat}
\end{equation}
For an isotropic system this matrix reduces to the tilt modulus,
${c^v_{ij}}^{-1} = {c^v_{44}}^{-1} \delta_{ij}$, with 
\begin{equation}
        c^v_{44} = (n m)^2 (n m - \rho_n)^{-1} = \rho^2 / \rho_s \ ,
 \label{tilmod}
\end{equation}
where $\rho_s = \rho - \rho_n$ is the superfluid density, and we have
set $\rho = n m$. In the flux liquid notation, $\rho_n = 0$
for (disorder--free) thick samples ($L \to \infty$), and thus 
$\rho_s = \rho = n m$ and $c^v_{44} = n m = n {\tilde \epsilon}_1$ in
the vortex language.

At this point we mention that in an Abrikosov flux lattive the
transverse tilt correlations assume the form \cite{nelled}
\begin{equation}
        T_\perp({\bf q},\omega_m) = {n^2 \hbar \omega_m^2 \over 
                                c_{66} q^2 + n m \omega_m^2} \ ,
 \label{abrlat}
\end{equation}
which reduces to the transverse part of the flux liquid result
[cf. Eq.~(\ref{foutil})] when the shear modulus $c_{66}$ vanishes. 
Thus, in the {\it crystalline} phase, $c_{44}^v = n m$ as in the
liquid, but $\rho_s = 0$ as the result of the appearance of the shear 
modulus {\it and} the opposite order of limits ($\omega_m \to 0$ to be
taken first, then ${\bf q} \to {\bf 0}$).

As we shall discuss in detail in Secs.~\ref{disbos} and \ref{disfll},
the effect of disorder will be twofold. First, for disorder correlated
in the $xy$ plane, the tilt modulus tensor (\ref{tilmat}) will in
general be anisotropic (i.e., the normal and superfluid densities,
being transport coefficients, will depend on the direction of motion
with respect to the correlated defects); second, disorder will lead to
a renormalization of $\rho_n$ and hence $\rho_s$ and $c^v_{ij}$. A
phase transition into a localized phase of bosons / flux lines will
occur when eventually the superfluid density and therefore the inverse tilt
modulus vanish (in a certain direction at least). This ``generalized
Bose glass'' phase then constitutes a true superconductor with zero
linear resistivity \cite{nelvin}, while the flux liquid with 
$\rho_s \not= 0$ behaves like a normal metal \cite{blater}.

\subsection{Green's functions, condensate depletion, and normal fluid
            density for a superfluid Bose gas}
 \label{pspebc}

It is useful to define the following Green's functions for the Bose
system (with periodic boundary conditions in imaginary time),
\begin{eqnarray}
        G({\bf r},\tau;{\bf r}',\tau') &&= 
        \langle \psi({\bf r},\tau) \psi^*({\bf r}',\tau') \rangle 
                                                        - n_0 \ ,
 \label{grefun} \\
        G_{12}({\bf r},\tau;{\bf r}',\tau') &&= 
        \langle \psi({\bf r},\tau) \psi({\bf r}',\tau') \rangle - n_0 
 \label{angref}
\end{eqnarray}
(compare Refs.~\cite{fetwal,npopov}). For a translationally invariant
system one deduces from Dyson's equations the general structure
\begin{equation}
        G({\bf q},\omega_m) = \left[ i \omega_m + \hbar q^2 / 2 m
        - \mu / \hbar + \Sigma_{11}(-{\bf q},-\omega_m) \right] 
                                        / D({\bf q},\omega_m) \ ,
 \label{selfen}
\end{equation}
which defines the self--energy $\Sigma_{11}({\bf q},\omega_m)$, while
the anomalous Green's function $G_{12}({\bf q},\omega_m)$ takes the
form
\begin{equation}
        G_{12}({\bf q},\omega_m) = - \Sigma_{12}({\bf q},\omega_m) 
                                        / D({\bf q},\omega_m) \ ,
 \label{anself}
\end{equation}
and the denominator reads
\begin{eqnarray}
        &&D({\bf q},\omega_m) = - \left[ i \omega_m 
                - \frac{1}{2} \Bigl( \Sigma_{11}({\bf q},\omega_m) 
                - \Sigma_{11}(-{\bf q},-\omega_m) \Bigr) \right]^2
                                                        \nonumber \\
        &&\qquad + \left[ \hbar q^2 / 2 m - \mu / \hbar 
                + \frac{1}{2} \Bigl( \Sigma_{11}({\bf q},\omega_m) 
                + \Sigma_{11}(-{\bf q},-\omega_m) \Bigr) \right]^2 
        - \Big \vert \Sigma_{12}({\bf q},\omega) \Big \vert^2 \ .
 \label{grdeno}
\end{eqnarray}
For noninteracting bosons, $\Sigma_{11}({\bf q},\omega_m) = 
\Sigma_{12}({\bf q},\omega_m) = 0$, and hence the anomalous Green's
function vanishes, while the free propagator becomes
\begin{equation}
        G_0({\bf q},\omega_m)^{-1} = 
                - i \omega_m + \hbar q^2 / 2 m - \mu / \hbar \ .
 \label{frepro}
\end{equation}
Furthermore, it can be shown that $\Sigma_{11}({\bf 0},0) -
\Sigma_{12}({\bf 0},0) = \mu / \hbar$ to all orders in perturbation
theory \cite{fetwal}.

In order to calculate these Green's functions in the harmonic
approximation, we expand Eq.~(\ref{pithet}) to first order in the
density and phase fluctuations,
\begin{equation}
        \psi({\bf r},\tau) \approx \sqrt{n_0} \left[ 1 + 
        {\pi({\bf r},\tau) \over 2 n_0} + i \Theta({\bf r},\tau) -
                {\pi({\bf r},\tau)^2 \over 8 n_0^2} + 
        {i \over 2 n_0} \pi({\bf r},\tau) \Theta({\bf r},\tau) - 
                \frac{1}{2} \Theta({\bf r},\tau)^2 \right] \ .
 \label{psiexp}
\end{equation}
To leading order, this yields for the Green's functions in Fourier
space 
\begin{eqnarray}
        G({\bf q},\omega_m;{\bf q}',\omega_{m'}) &&\approx
        {1 \over 4 n_0} \langle \pi({\bf q},\omega_m) 
                                \pi({\bf q}',\omega_{m'}) \rangle -
        {i \over 2} \langle \pi({\bf q},\omega_m) 
                \Theta({\bf q}',\omega_{m'}) \rangle \nonumber \\
        &&\quad + {i \over 2} \langle \Theta({\bf q},\omega_m) 
                                \pi({\bf q}',\omega_{m'}) \rangle +
        n_0 \langle \Theta({\bf q},\omega_m)
                        \Theta({\bf q}',\omega_{m'}) \rangle \ ,
 \label{expgre} \\
        G_{12}({\bf q},\omega_m;{\bf q}',\omega_{m'}) &&\approx
        {1 \over 4 n_0} \langle \pi({\bf q},\omega_m) 
                                \pi({\bf q}',\omega_{m'}) \rangle +
        {i \over 2} \langle \pi({\bf q},\omega_m) 
                \Theta({\bf q}',\omega_{m'}) \rangle \nonumber \\
        &&\quad + {i \over 2} \langle \Theta({\bf q},\omega_m) 
                                \pi({\bf q}',\omega_{m'}) \rangle -
        n_0 \langle \Theta({\bf q},\omega_m)
                        \Theta({\bf q}',\omega_{m'}) \rangle \ .
 \label{expang} 
\end{eqnarray}
Upon inserting the results (\ref{matinv}) from the previous section,
one readily finds for the (reduced) Green's functions in Fourier space
[see Eq.~(\ref{redcor})]
\begin{eqnarray}
        G({\bf q},\omega_m) &&= 
        {i \omega_m + \hbar q^2 / 2 m + n_0 V(q) / \hbar \over 
                \omega_m^2 + \epsilon_{\rm B}(q)^2 / \hbar^2} \ ,
 \label{normgr} \\
        G_{12}({\bf q},\omega_m) &&= {- n_0 V(q) / \hbar \over
                \omega_m^2 + \epsilon_{\rm B}(q)^2 / \hbar^2} \ ,
 \label{anomgr}
\end{eqnarray}
and therefore with $\mu = n_0 V_0$
\begin{eqnarray}
        \Sigma_{11}({\bf q},\omega_m) &&= 
                        n_0 [ V_0 + V(q) ] / \hbar \; , \quad
        \Sigma_{12}({\bf q},\omega_m) = n_0 V(q) / \hbar \ ,
 \label{selbog} \\
        \quad D({\bf q},\omega_m) &&=
                \omega_m^2 + \epsilon_{\rm B}(q)^2 / \hbar^2 \ .
 \label{denbog}
\end{eqnarray}
Here $\epsilon_{\rm B}(q)$ denotes the Bogoliubov spectrum, already
introduced in Eq.~(\ref{bogspe}). The results
(\ref{normgr})--(\ref{denbog}), obtained in the Gaussian
approximation, are again identical to those from the Bogoliubov theory
applicable to either weakly interacting or dilute Bose systems
\cite{fetwal,nozpin}. Another useful representation for the Green's
functions consists in their separation into single poles and their
residues,
\begin{eqnarray}
        G({\bf q},\omega_m) &&= 
        {u(q)^2 \over - i \omega_m + \epsilon_{\rm B}(q) / \hbar} +
        {v(q)^2 \over i \omega_m + \epsilon_{\rm B}(q) / \hbar} \ ,
 \label{boggre} \\
        G_{12}({\bf q},\omega_m) &&= - u(q) v(q) \left(
        {1 \over - i \omega_m + \epsilon_{\rm B}(q) / \hbar} +
        {1 \over i \omega_m + \epsilon_{\rm B}(q) / \hbar} \right) \ ,
 \label{bogang}
\end{eqnarray}
with the weight functions
\begin{equation}
        \begin{array}{l} u(q)^2 \\ v(q)^2 \end{array} \Biggr\} =
        \frac{1}{2} \left( {n_0 V(q) + \hbar^2 q^2 / 2 m \over 
                        \epsilon_{\rm B}(q)} \pm 1 \right) \ ,
 \label{weight}
\end{equation}
obeying $u(q)^2 - v(q)^2 = 1$ and 
$u(q) v(q) = n_0 V(q) / 2 \epsilon_{\rm B}(q)$; these are precisely
the coefficients of the Bogoliubov transformation from the original
boson to quasiparticle creation and annihilation operators. 

We can now use the Green's function to calculate the average particle
number in the grand--canonical ensemble; carefully taking the normal
ordering of the original Bose field operators into account
\cite{npopov,negorl}, we have to evaluate (with $\eta \downarrow 0$)
\begin{eqnarray}
        N - N_0 
        &&= \int \! d^dr \, \Bigl( \langle \psi({\bf r},\tau-\eta) 
        \psi^*({\bf r},\tau+\eta) \rangle - n_0 \Bigr) \nonumber \\
        &&= \int \! {d^dq \over (2\pi)^d} \, {1 \over (\beta \hbar)^2} 
        \sum_{m,m'} e^{-i \left( \omega_m + \omega_{m'} \right) \tau}
                e^{i \left( \omega_m - \omega_{m'} \right) \eta}
                        G({\bf q},\omega_m;-{\bf q},\omega_{m'}) \ .
 \label{parnum}
\end{eqnarray}
If in addition time translation invariance holds, we find
\begin{equation}
        n - n_0 = \int \! {d^dq \over (2 \pi)^d} {1 \over \beta \hbar}
                        \sum_m e^{i \omega_m \eta} G({\bf q},\omega_m)
 \label{condep}
\end{equation}
(note the crucial exponential factor stemming from the time ordering).
Assuming that the particle density $n$ is known, Eq.~(\ref{condep})
can be regarded as a formula for the squared magnitude of the boson
order parameter, $n_0 = |\langle \psi({\bf r},\tau) \rangle|^2$.
Upon using Eq.~(\ref{boggre}) and Eq.~(\ref{bfmat1}) from
App.~\ref{bosmat}, we find the condensate depletion in terms of the
quasiparticle weights (\ref{weight}),
\begin{equation}
        n - n_0 =  \int \! {d^dq \over (2 \pi)^d} 
        \left[ v(q)^2 + {u(q)^2 + v(q)^2 \over 
                        e^{\beta \epsilon_{\rm B}(q)} - 1} \right] \ .
 \label{deplet}
\end{equation}
Because of the relation $\mu = n_0 V_0$, this actually constitutes an
implicit equation for the chemical potential. The first term in
Eq.~(\ref{deplet}) represents the zero--temperature depletion, i.e.,
the fraction of particles that are pushed out of the condensate as a 
consequence of interactions and quantum fluctuations; the second
contribution yields the additional depletion caused by thermal
excitations \footnote{The depletion can also be derived somewhat more
directly by using the definition (\ref{boseop}): for,
Eq.~(\ref{psiexp}) implies that 
$n_0 \equiv |\langle \psi \rangle|^2 \approx n_0 + \langle \pi \rangle
- \langle \pi^2 \rangle / 4 n_0 + i (\langle \pi \Theta \rangle - 
\langle \Theta \pi \rangle) / 2 - n_0 \langle \Theta^2 \rangle$, and
hence because of 
$n \equiv \langle |\psi|^2 \rangle = n_0 + \langle \pi \rangle$ one
finds $n - n_0 = \langle \pi \rangle = \langle \pi^2 \rangle / 4 n_0 -
i (\langle \pi \Theta \rangle - \langle \Theta \pi \rangle) / 2 + 
n_0 \langle \Theta^2 \rangle$ [compare Eq.~(\ref{expgre})], which
immediately leads to Eq.~(\ref{deplet}). Note again that formally the
contributions to these composite--operator averages amount to loop
integrals.}.

For short--range interactions, $V(q) \approx V_0 = 4 \pi a / m$, where $a$
denotes the (s--wave) scattering length, and Eq.~(\ref{deplet}) can be
evaluated explicitly; at $T = 0$, we have (with 
$x = \hbar q / \sqrt{8 \pi n_0 a}$)  
\begin{eqnarray}
        n(T=0) &&- n_0 = \int \! {d^dq \over (2 \pi)^d} \, v(q)^2
                                                        \nonumber \\
        &&= (2 a n_0 /\hbar^2)^{d/2} \Gamma(d/2)^{-1} \int_0^\infty \!
        x^{d-2} \left[ (2+x^2)^{1/2} -x -(2+x^2)^{-1/2} \right] dx \ .
 \label{zerdep}
\end{eqnarray}
In $d=1$, this integral diverges logarithmically at the lower limit,
which means there is no Bose condensation in one dimension, in
accordance with the Hohenberg--Mermin--Wagner theorem. For $d=2$, the
integral is equal to $1/2$, and hence
\begin{equation}
        d = 2 \, : \; n(T=0) - n_0 = n_0 a / \hbar^2 \ ,
 \label{20depl}
\end{equation}
while in three dimensions \cite{fetwal,nozpin,npopov}
\begin{equation}
        d = 3 \, : \; 
        n(T=0) - n_0 = (8 / 3 \sqrt{\pi}) (n_0 a / \hbar^2)^{3/2} \ .
 \label{30depl}
\end{equation}

Note that Eq.~(\ref{20depl}) with $\hbar \to k_{\rm B} T$, 
$m \to {\tilde \epsilon}_1$ and $V_0 = \phi_0^2 / 4 \pi$ yields the
``boson order parameter'' squared for vortex lines as function of
temperature in the limit of infinitely thick samples ($L \to \infty$),
\begin{equation}
        n_0(T) = n \left[ 1 + {\tilde \epsilon_1} 
        \left( \phi_0 / 4 \pi k_{\rm B} T \right)^2 \right]^{-1} \ .
 \label{entlin}
\end{equation}
The reduction in this quantity from unity is a measure of the
confining effects and diminished entanglement induced by the repulsive
vortex interactions.

Returning to the boson representation, we now calculate the
lowest--order finite--temperature corrections explicitly,
approximating the Bogoliubov spectrum by its phonon branch, 
$\epsilon_{\rm B}(q) \approx \hbar c_1 q$. Then 
$1 + 2 v(q)^2 \approx m c_1 / \hbar q$, and 
$n(T) - n_0 = \Delta n(T)$, with
\begin{equation}
        \Delta n(T) = \int \! {d^dq \over (2 \pi)^d} 
        {1 + 2 v(q)^2 \over e^{\beta \epsilon_{\rm B}(q)} - 1} \approx
        {\Gamma(d-1) \zeta(d-1) m \over 2^{d-1} \pi^{d/2} \Gamma(d/2)
                        \hbar^d c_1^{d-2}} \, (k_{\rm B} T)^{d-1} \ ,
 \label{findep}
\end{equation}
where Eq.~(\ref{integ1}) from App.~\ref{momint} was used. In two
dimensions, the $\zeta$ function diverges, and hence a nonzero boson
order parameter is only possible at zero temperature; at finite
temperatures, there is still a superfluid phase characterized by a
finite superfluid density $\rho_s$ (see below), and the phase
transition will be of the Berezinskii--Kosterlitz--Thouless type. For
$d=3$, Eq.~(\ref{findep}) yields \cite{fetwal,nozpin,npopov}
\begin{equation}
        d = 3 \, : \; 
        \Delta n(T) = m (k_{\rm B} T)^2 / 12 \hbar^3 c_1 \ .
 \label{3Tdepl}
\end{equation}

We can also utilize the Green's function (\ref{grefun}) to rederive
the normal--fluid density (\ref{nflden}) following a famous argument
due to Landau \cite{fetwal,nozpin,npopov}. Consider a superfluid
confined in a cylindrical pipe, whose walls are moved along its axis
at constant velocity ${\bf v}$ with respect to the superfluid. In this
situation, the superfluid velocity is zero, and the normal fluid
velocity is equal to ${\bf v}$, because according to the two--fluid
model only the normal fluid component is dragged along by the moving
walls. Thus
\begin{equation}
        \langle {\bf g}({\bf r},\tau) \rangle = 
        m \langle {\bf j}({\bf r},\tau) \rangle = \rho_n {\bf v} \ ;
 \label{curexp}
\end{equation}
the task now is to calculate the average current to first order in
${\bf v}$, which then, by comparison with Eq.~(\ref{curexp}), yields
$\rho_n$. The definition (\ref{momden}) leads, for spatially
translation--invariant systems, to (with $\eta \downarrow 0$) 
\begin{equation}
        \langle {\bf g}({\bf r},\tau) \rangle = 
                \int \! {d^dq \over (2 \pi)^d} \hbar {\bf q} 
                        {1 \over (\beta \hbar)^2} \sum_{m,m'}
        e^{-i (\omega_m+\omega_{m'}) \tau} e^{-i \omega_{m'} \eta} 
                G({\bf q},\omega_m;-{\bf q},\omega_{m'})_{-{\bf v}} \ ,
 \label{mdengr}
\end{equation}
which reduces further to
\begin{equation}
        \langle {\bf g}({\bf r},\tau) \rangle = 
        \int \! {d^dq \over (2 \pi)^d)} {{\bf q} \over \beta} \sum_m
                e^{i \omega_m \eta} G({\bf q},\omega_m)_{-{\bf v}} \ ,
 \label{mdtrgr}
\end{equation}
if translation invariance in imaginary time holds as well, which is of
course the case for a pure system with periodic boundary
conditions. The index ``$-{\bf v}$'' serves as a reminder that we need
the Green's function for a system with nonzero relative velocity
between the walls and the superfluid.

We can readily obtain the Green's function for moving walls from the
known quantity $G({\bf q},\omega_m)_{\bf 0}$ for fixed walls,
Eqs.~(\ref{normgr}),(\ref{boggre}), by performing a Galilean
transformation with constant velocity $-{\bf v}$. The Matsubara
frequencies in the transformed system are [see
Eqs.~(\ref{galtra}),(\ref{foutra})],
\begin{eqnarray}
        \psi'({\bf r}',\tau') &&= \psi({\bf r}+i{\bf v}\tau,\tau) =
                {1 \over \beta \hbar \Omega} \sum_{{\bf q},\omega_m} 
                                        \psi({\bf q},\omega_m)
        e^{i{\bf q}\cdot{\bf r}-i(\omega_m-i{\bf q}\cdot{\bf v})\tau}
                = \sum_{{\bf q}',\omega_m'} \psi'({\bf q}',\omega_m')
                e^{i{\bf q}'\cdot{\bf r}'-i\omega_m'\tau'} \nonumber\\ 
        &&\Longrightarrow \qquad {\bf q} \to {\bf q}' = {\bf q} 
                                                \quad , \qquad 
        \omega_m \to \omega_m' = \omega_m - i {\bf q}\cdot{\bf v} \ ; 
 \label{mattra}
\end{eqnarray}
notice that the periodic boundary conditions are also obeyed by the
transformed fields, $\psi'({\bf r}'-i{\bf v} \beta \hbar,\beta \hbar)
= \psi'({\bf r}',0)$.

Upon inserting the shifted Matsubara frequencies in
Eq.~(\ref{boggre}), we find
\begin{equation}
        G({\bf q},\omega_m)_{-{\bf v}} = 
        {1 +v(q)^2 \over - i \omega_m + \epsilon_{\rm B}(q) / \hbar 
                                        - {\bf q} \cdot {\bf v}} +
        {v(q)^2 \over i \omega_m + \epsilon_{\rm B}(q) / \hbar 
                                        + {\bf q} \cdot {\bf v}} \ ,
 \label{movgre}
\end{equation}
with the unchanged Bogoliubov coefficient $v(q)$, Eq.~(\ref{weight}).
Performing the Matsubara frequency sum [Eq.~(\ref{bfmat1})] finally 
yields Landau's formula
\begin{equation}
        \langle {\bf g}({\bf r},\tau) \rangle = 
        \int \! {d^dq \over (2 \pi)^d} {\hbar {\bf q} \over 
        e^{\beta [\epsilon_{\rm B}(q)-\hbar{\bf q}\cdot{\bf v}]} - 1}
 \label{lanfor}
\end{equation}
[the terms $\propto v(q)^2$ vanish, because their combination is
antisymmetric under ${\bf q} \to -{\bf q}$]. Eq.~(\ref{lanfor}) can be
interpreted as a sum over the momenta times a Bose--Einstein
quasiparticle occupation number in a system with moving walls (where
the energy eigenvalues are reduced by $\hbar {\bf q} \cdot {\bf v}$). 
An expansion to linear order in ${\bf v}$ then again leads to the
result (\ref{nflden}).

We now evaluate this expression in the phonon approximation,
$\epsilon_{\rm B}(q) \approx \hbar c_1 q$. Via integration by parts,
the integral in Eq.~(\ref{nflden}) becomes of the form (\ref{integ1}),
and hence
\begin{equation}
        \rho_n(T) = 
        {\Gamma(d+2) \zeta(d+1) \over 2^{d-1} \pi^{d/2} d \Gamma(d/2)}
                \, {(k_{\rm B} T)^{d+1} \over \hbar^d c_1^{d+2}} \ .
 \label{exnfld}
\end{equation}
Thus, in two dimensions we find
\begin{equation}
        d = 2 \, : \; 
        \rho_n = 3 \zeta(3) (k_{\rm B} T)^3 / 2 \pi \hbar^2 c_1^4 \ ,
 \label{2Tnfld}
\end{equation}
with $\zeta(3) \approx 1.202$, while for $d=3$
\cite{fetwal,nozpin,npopov}
\begin{equation}
        d = 3 \, : \; 
        \rho_n = 2 \pi^2 (k_{\rm B} T)^4 / 45 \hbar^3 c_1^5 \ .
 \label{3Tnfld}
\end{equation}

For a flux liquid with {\it periodic boundary conditions} in the $z$
direction along the magnetic field, the leading finite--size
corrections to the tilt modulus thus are ($d = 2$)
\begin{equation}
        \Delta c_{44}^{-1}(L) = - 3 \zeta(3) k_{\rm B} T / 
                2 \pi (n {\tilde \epsilon}_1)^2 c_1^4 L^3 \ .
 \label{pbfnc4}
\end{equation}
Thus, $c_{44}$ {\it increases} and it is {\it harder} to tilt vortices
in finite--size systems with periodic boundary conditions, as one
might expect. In the following section, we determine the corresponding
finite--size corrections for the condensate fraction $n_0$ and the
tilt modulus (normal--fluid density) for the more realistic case of
open boundary conditions.

We finally remark that a crucial point in the calculation of the
normal--fluid density is that both phase {\it and} density (amplitude)
fluctuations are taken into account properly. It is tempting, given
that the phase fluctuations are the dominant low--energy modes, to 
integrate out the $\pi$ fields in the harmonic action (\ref{haract}),
and then derive thermodynamic quantities like the depletion and the
normal--fluid density from the ensuing effective action. As is already
apparent from Eq.~(\ref{trmres}), however, essential parts of the 
transverse current correlations will then be missing, e.g., there will
be no vorticity in the system, unless vortices are explicitly
introduced by hand. Hence one finds $\rho_n = 0$ even for $T > 0$ in
this approximation. This unsatisfactory result arises because it is
actually not the leading terms (in a long--wavelength expansion), but
the {\it next--to--leading} contributions containing the amplitude
fluctuations, which produce a non--zero vorticity and normal fluid
density at finite temperatures. The phase--only approximation (for the
disordered boson system) and its limitations are more extensively
discussed in App.~\ref{poappr}.

\subsection{Finite--size corrections for open boundary conditions}
 \label{psopbc}

We now proceed to study finite--size corrections for the flux line
problem with free boundary conditions. To facilitate comparison with
the periodic boundary conditions in the previous section, we retain
the boson notation. According to Eq.~(\ref{freebc}), there can be no
density and phase fluctuations at the bounding surfaces for all 
${\bf r}$, 
\begin{equation}
        \pi({\bf r},0) = \pi({\bf r},\beta\hbar) = 0 =
        \Theta({\bf r},0) = \Theta({\bf r},\beta\hbar) \ ,
 \label{constr}
\end{equation}
reflecting ``ideal Bose gas'' boundary conditions at the top and
bottom of the sample. Eq.~\ref{constr} amounts to a constraint for the
path integrals as in Eq.~(\ref{bosgrz}) that has to be taken into
account explicitly. Upon introducing the vector notation
\begin{equation}
        \Upsilon({\bf r},\tau) = \left( \begin{array}{c}
                \Theta({\bf r},\tau) \\ \pi({\bf r},\tau)
                                        \end{array} \right) \ ,
 \label{vector}
\end{equation}
the two--point correlation functions with the constraints
(\ref{constr}) requires evaluating the path integral 
\begin{eqnarray}
        \langle \Upsilon({\bf r},\tau) 
                \Upsilon^T({\bf r}',\tau') \rangle &&= {\int \! 
        {\cal D}[\Upsilon({\bf r},\tau)] \, \Upsilon({\bf r},\tau) 
                \Upsilon^T({\bf r}',\tau') \, e^{-S[\Upsilon]/\hbar}
        \, \prod_{\bf r} \delta \Bigl( \Upsilon({\bf r},0) \Bigr)
        \over \int \! {\cal D}[\Upsilon({\bf r},\tau)] \,
                e^{-S[\Upsilon]/\hbar} \, 
        \prod_{\bf r} \delta \Bigl( \Upsilon({\bf r},0) \Bigr)} 
                                                        \nonumber \\
        &&= {\int \! {\cal D}[\lambda({\bf r})] 
        {\cal D}[\Upsilon({\bf r},\tau)] \, \Upsilon({\bf r},\tau) 
                \Upsilon^T({\bf r}',\tau') \, e^{-S[\Upsilon]/\hbar 
        + i \int \! d^dr \, \lambda({\bf r}) \Upsilon({\bf r})}
        \over \int \! {\cal D}[\lambda({\bf r})] 
        {\cal D}[\Upsilon({\bf r},\tau)] \, e^{-S[\Upsilon]/\hbar 
        + i \int \! d^dr \, \lambda({\bf r}) \Upsilon({\bf r})}} \ ,
 \label{cnpath}
\end{eqnarray}
where in the second line a {\it static} auxiliary field 
$\lambda({\bf r})$ has been introduced to enforce the constraints. In
the Gaussian approximation, the correlation functions obeying the
boundary conditions (\ref{constr}) can then be reduced in a
straightforward manner to those with periodic boundary conditions that
are encoded in the matrix ${\bf A}^{-1}({\bf q},\omega_m)$ of
Eq.~(\ref{matinv}). For the pure case, this has been worked out in
detail in Ref.~\cite{nelseu}, with the result
\begin{equation}
        \langle \Upsilon({\bf q},\omega_m) 
                \Upsilon^T({\bf q}',\omega_{m'}) \rangle = (2 \pi)^d 
        \delta({\bf q}+{\bf q}') \hbar {\bf A}^{-1}({\bf q},\omega_m)
        \left[ \beta \hbar \delta_{m,-m'} - {\bf A}({\bf q})
                        {\bf A}^{-1}({\bf q},-\omega_{m'}) \right] \ , 
 \label{pbccrf}
\end{equation}
where ${\bf A}({\bf q})$ denotes the inverse of 
${\bf A}^{-1}({\bf q},\tau=0) = (\beta \hbar)^{-1} \sum_m 
{\bf A}^{-1}({\bf q},\omega_m)$,
\begin{equation}
        {\bf A}({\bf q}) = 
        \left( \begin{array}{cc} 
                n_0 \hbar^2 q^2 / m \epsilon_{\rm B}(q) & 0 \\ 
                0 & m \epsilon_{\rm B}(q) / n_0 \hbar^2 q^2
        \end{array} \right)     
        2 \hbar \tanh {\beta \epsilon_{\rm B}(q) \over 2} \ .
 \label{stamat}
\end{equation}
In App.~\ref{obcder}, the analogous formula is derived for a general
disordered system, and the pure result follows from Eq.~(\ref{opdisc})
by simply setting the disorder correlator to zero. Note that
translation invariance only holds in the $d$ perpendicular directions,
while the second term in Eq.~(\ref{pbccrf}), representing the
corrections from the open--boundary--conditions constraint, explicitly
breaks translational invariance along the direction of the lines.

Upon multiplying the matrices (\ref{stamat}) and (\ref{matinv}), we
find for the density correlation function with open boundary
conditions
\begin{eqnarray}
        &&S({\bf q},\omega_m;{\bf q}',\omega_{m'}) = 
        (2 \pi)^d \delta({\bf q}+{\bf q}') \, {n_0 \hbar q^2 / m \over
                \omega_m^2 + \epsilon_{\rm B}(q)^2 / \hbar^2} 
        \times \nonumber \\ &&\qquad \times \left( \beta \hbar 
                \delta_{m,-m'} - {2\hbar \over \epsilon_{\rm B}(q)}
                \, \tanh {\beta \epsilon_{\rm B}(q) \over 2} \,
        {\omega_m \omega_{m'} + \epsilon_{\rm B}(q)^2 / \hbar^2 \over 
        \omega_{m'}^2 + \epsilon_{\rm B}(q)^2 / \hbar^2} \right) \ .
 \label{opbcde}
\end{eqnarray}
The frequency sums required for the equal--time correlation function
follow from Eqs.~(\ref{bsmat3}) and (\ref{dmsum2}), with the result
\begin{eqnarray}
        &&S({\bf q},\tau;{\bf q}',\tau) = {1 \over (\beta \hbar)^2} 
                \sum_{m,m'} e^{-i (\omega_m + \omega_{m'}) \tau} \,
                S({\bf q},\omega_m;{\bf q}',\omega_{m'}) =
                S({\bf q},0) \, (2 \pi)^d \delta({\bf q}+{\bf q}') \ ,
 \label{eqtcor} \\
        &&S({\bf q},0) = 
                {n_0 \hbar^2 q^2 \over 2 m \epsilon_{\rm B}(q)} \, 
        \coth {\beta \epsilon_{\rm B}(q) \over 2} \, \left( 1 - 
        {1 \over \cosh^2 [\beta \epsilon_{\rm B}(q) / 2]} \right) =
        {n_0 \hbar^2 q^2 \over 2 m \epsilon_{\rm B}(q)} \,
                        \tanh {\beta \epsilon_{\rm B}(q) \over 2} \ .
 \label{opstrf}
\end{eqnarray}
Compared to the corresponding result for periodic boundary conditions
(\ref{ststrf}), the hyperbolic cotangent function has been replaced by
a hyperbolic tangent, as if the excitations on the superfluid ground
state were governed by a fermionic rather than a bosonic distribution
function [cf. Eq.~(\ref{bfmat2})]; accordingly, the limit 
$S({\bf q}) = n_0 \hbar^2 q^2 / 2m \epsilon_{\rm B}(q)$ for 
$\beta \hbar \to \infty$ is approached from {\it below} for open
boundary conditions, as opposed to from above for periodic boundary
conditions. 

Similarly, the Green's function (\ref{expgre}) reads
\begin{eqnarray}
        &&G({\bf q},\omega_m;{\bf q}',\omega_{m'}) =
        (2 \pi)^d \delta({\bf q}+{\bf q}') \times \nonumber \\
        &&\quad \times 
        \Biggl[ {\hbar q^2 / 2 m + n_0 V(q) / \hbar \over 
        \omega_m^2 + \epsilon_{\rm B}(q)^2 / \hbar^2} \left( \beta 
        \hbar \delta_{m,-m'} - {2\hbar \over \epsilon_{\rm B}(q)}
                \, \tanh {\beta \epsilon_{\rm B}(q) \over 2} \,
        {\omega_m \omega_{m'} + \epsilon_{\rm B}(q)^2 / \hbar^2 \over 
        \omega_{m'}^2 + \epsilon_{\rm B}(q)^2 / \hbar^2} \right)
        \nonumber \\ &&\qquad + 
        {1 \over \omega_m^2 + \epsilon_{\rm B}(q)^2 / \hbar^2} \left( 
        i \omega_m \beta \hbar \delta_{m,-m'} - {2 \epsilon_{\rm B}(q)
        \over \hbar} \, \tanh {\beta \epsilon_{\rm B}(q) \over 2} \,
                {i (\omega_m - \omega_{m'}) \over \omega_{m'}^2 + 
                \epsilon_{\rm B}(q)^2 / \hbar^2} \right) \Biggr] \ ,
 \label{opgref}
\end{eqnarray}
from which we can infer the average particle number with
Eq.~(\ref{parnum}) using the Matsubara frequency sums in
App.~\ref{bosmat}; instead of Eq.~(\ref{deplet}) for periodic boundary
conditions, the result for the depletion now is
\begin{eqnarray}
        n - n_0 &&= \int \! {d^dq \over (2 \pi)^d} 
        \left[ v(q)^2 + \Bigl[ u(q)^2 + v(q)^2 \Bigr] \left( 
        {1 \over e^{\beta \epsilon_{\rm B}(q)} - 1} - {1 \over 
        \sinh [\beta \epsilon_{\rm B}(q)]} \right) \right] \nonumber\\
        &&= \int \! {d^dq \over (2 \pi)^d} 
                        \left[ v(q)^2 - {u(q)^2 + v(q)^2 \over 
                        e^{\beta \epsilon_{\rm B}(q)} + 1} \right] \ ;
 \label{opdepl}
\end{eqnarray}
i.e., the finite--``temperature'' (finite--size) corrections change
sign, and the corresponding ``excitations'' are governed by a
Fermi--Dirac distribution (with zero ``chemical potential'') rather
than a Bose--Einstein function as in the case of periodic boundary
conditions.

With the aid of Eq.~(\ref{integ2}), we can evaluate the ``depletion'',
which is now actually a condensate {\it enhancement}, in the phonon
approximation. Remarkably, in two dimensions the open boundary
conditions lead to a {\it finite} correction,
\begin{equation}
        d = 2 \, : \; 
        \Delta n(T) = - (\ln 2) m (k_{\rm B} T) / 2 \pi \hbar^2 \ ,
 \label{2opdep}
\end{equation}
in contrast to the logarithmic divergence one obtains for the case of
periodic boundary conditions. The possibility of long--range order at
finite temperatures in two dimensions arises, due to the static
surface ``field'' $\lambda({\bf r})$ which enforces the 
open--boundary--conditions constraints at $\tau = 0$ and 
$\tau = \beta \hbar$. This field strengthens the order parameter at
the free surfaces, which promotes condensation in the bulk, when
the thickness in the imaginary--time direction is finite. For higher
dimensions we find
\begin{equation}
        d \geq 3 \, : \; 
        \Delta n(T) = - \left( 1 - {1 \over 2^{d-2}} \right)
        {\Gamma(d-1) \zeta(d-1) \over 2^{d-1} \pi^{d/2} \Gamma(d/2)} \, 
        {m (k_{\rm B} T)^{d-1} \over \hbar^d c_1^{d-2}} \ ;
 \label{dopdep}
\end{equation}
thus, in three dimensions,
\begin{equation}
        d = 3 \, : \; 
        \Delta n(T) = - m (k_{\rm B} T)^2 / 24 \hbar^3 c_1 \ ,
 \label{3opdep}
\end{equation}
which is ($-1/2$) times the depletion for periodic boundary
conditions, Eq.~(\ref{3Tdepl}). As we have already discussed in the
Introduction, the physical reason for the signs in
Eqs.~(\ref{2opdep})--(\ref{3opdep}) is in the flux line language that 
thermal wandering is stronger at the surfaces, and thus line
entanglement is facilitated, see Fig.~\ref{wavfun}.

For the same reason, with open boundary conditions one expects the
superfluid density $\rho_s$ (and the inverse tilt modulus 
$c_{44}^{-1}$) to be {\it larger} at finite $\beta \hbar$ than in the
``zero--temperature'' limit, and thus the finite--``temperature'' /
finite--size corrections should have opposite signs compared to the
bosonic case with periodic boundary conditions [Eqs.(\ref{nflden}) and
(\ref{exnfld})]. By repeating Landau's argument for tilted lines, we
obtain the Green's function from Eq.~(\ref{opgref}) upon replacing 
$\omega_m \to \omega_m - i {\bf q} \cdot {\bf v}$ and  
$\omega_{m'} \to \omega_{m'} + i {\bf q} \cdot {\bf v}$
[cf. Eq.~(\ref{mattra})]. Then expanding to first order in 
${\bf q} \cdot {\bf v}$ and performing the Matsubara frequency sums
with the aid of App.~\ref{bosmat}, yields the normal--fluid density,
which has now become a function of imaginary time,
\begin{equation}
        \rho_n(\tau) = {\hbar \over d} 
        \int \! {d^dq \over (2 \pi)^d} \, q^2 \left( 2 \tau 
        {\sinh [(\beta \hbar - 2 \tau) \epsilon_{\rm B}(q) / \hbar]
                \over \sinh [\beta \epsilon_{\rm B}(q)] } -
        {\beta \hbar \over 4 \sinh^2 [\beta \epsilon_{\rm B}(q)/2] }
                                                        \right) \ .
 \label{itrhon}
\end{equation}
In the center of the bulk ($\tau = \beta \hbar / 2$) this reduces to
\begin{equation}
        \rho_n(\tau = \beta \hbar / 2) = - {\beta \hbar^2 \over 4 d}
                \int \! {d^dq \over (2 \pi)^d} \left( {q \over 
                \sinh [\beta \epsilon_{\rm B}(q) / 2] } \right)^2 \ ,
 \label{opnfld}
\end{equation}
which is just Eq.~(\ref{nflden}) with a {\it negative} sign. This
negative result does not of course have the usual interpretation of
two--fluid superfluid hydrodynamics, i.e., of a physical ``particle
density'', once flux line boundary conditions are applied. In the
phonon approximation, we find explicitly
\begin{equation}
        \rho_n(T) = - \, 
        {\Gamma(d+2) \zeta(d+1) \over 2^{d-1} \pi^{d/2} d \Gamma(d/2)}
                \, {(k_{\rm B} T)^{d+1} \over \hbar^d c_1^{d+2}} \ ,
 \label{opexfl}
\end{equation}
and in two and three dimensions one has
\begin{eqnarray}
        d = 2 \, : \; &&\rho_n = 
        - 3 \zeta(3) (k_{\rm B} T)^3 / 2 \pi \hbar^2 c_1^4 \ ,
 \label{2opfld} \\
        d = 3 \, : \; &&\rho_n = 
        - 2 \pi^2 (k_{\rm B} T)^4 / 45 \hbar^3 c_1^5 \ .
 \label{3opfld}
\end{eqnarray}

For the finite--size correction of the tilt modulus with open boundary
conditions this implies (in the flux line notation, $d = 2$)
\begin{equation}
        \Delta c_{44}^{-1}(L) = 3 \zeta(3) k_{\rm B} T / 
                2 \pi (n {\tilde \epsilon}_1)^2 c_1^4 L^3 \ .
 \label{opfnc4}
\end{equation}
As anticipated, the system with open boundary conditions can be more
{\it easily} tilted than one with periodic boundary conditions
[Eq.~(\ref{pbfnc4})].


\section{Disorder contributions for superfluid bosons
         (periodic boundary bonditions)} 
 \label{disbos}

In the preceding section, we studied defect--free systems of
superfluid bosons and flux liquids, and thereby established notations,  
introduced the relevant quantities, explained our approximations, and
discussed the relevance of boundary conditions. We now determine the
influence of weak disorder on density and phase correlations, the
depletion, and the normal--fluid density (tilt modulus). We start by
investigating weakly interacting or dilute superfluid Bose gases, or
flux liquids with periodic boundary conditions along the
magnetic--field direction. The case of open boundary conditions more
appropriate to vortices will be treated in Sec.~\ref{disfll}. We shall
derive our results for general disorder first, specializing at the end
to specific impurity correlators (Sec.~\ref{pbexdo}).

\subsection{Density, current, and vorticity correlations}
 \label{pbcorr}

We retain our approximation to work in the Gaussian ensemble described
by the action (\ref{haract}) only, neglecting the nonlinear terms
(\ref{intact}). In Fourier space, the harmonic action, including the
disorder terms, reads
\begin{eqnarray}
        S_0[\pi,\Theta] &&= 
        {1 \over 2 \beta \hbar \Omega} \sum_{{\bf q},\omega_m} 
        \Biggl[ \Bigl( \Theta(-{\bf q},-\omega_m)  \, , \, 
        \pi(-{\bf q},-\omega_m) \Bigr) \, {\bf A}({\bf q},\omega_m)
        \, \left( \begin{array}{c} \Theta({\bf q},\omega_m) \\
                \pi({\bf q},\omega_m) \end{array} \right) \nonumber \\
        &&\qquad \qquad - 2 \Bigl( \Theta(-{\bf q},-\omega_m) \, , \, 
        \pi(-{\bf q},-\omega_m) \Bigr) \, \left( \begin{array}{c} 0 \\ 
        \delta V_D({\bf q},\omega_m) \end{array} \right) \Biggr]
        - \frac{1}{2} n_0^2 V_0 \beta \hbar \Omega \ ,
 \label{dshars}
\end{eqnarray}
where the harmonic coupling matrix ${\bf A}({\bf q},\omega_m)$ is
given by Eq.~(\ref{matrix}). In order to obtain the two--point
correlation functions to first order in the disorder correlator
\begin{equation}
        \Delta({\bf q},\omega_m;{\bf q}',\omega_{m'}) =
        \overline{\delta V_D({\bf q},\omega_m) 
                        \delta V_D({\bf q}',\omega_{m'})} \ ,
 \label{discor}
\end{equation}
we introduce the shifted fields (note that the Jacobian of this
transformation is 1)
\begin{eqnarray}
        \left( \begin{array}{c} {\tilde \Theta}({\bf q},\omega_m) \\
                {\tilde \pi}({\bf q},\omega_m) \end{array} \right) &&=
        \left( \begin{array}{c} \Theta({\bf q},\omega_m) \\
                        \pi({\bf q},\omega_m) \end{array} \right) 
        - {\bf A}^{-1}({\bf q},\omega_m) \, \left( \begin{array}{c} 0 
        \\ \delta V_D({\bf q},\omega_m) \end{array}\right) \nonumber\\
        &&= \left( \begin{array}{c} \Theta({\bf q},\omega_m) \\
                        \pi({\bf q},\omega_m) \end{array} \right) - \,
                {\delta V_D({\bf q},\omega_m) \over \omega_m^2 + 
        \epsilon_{\rm B}(q)^2 / \hbar^2} \left( \begin{array}{c} 
        \omega_m / \hbar \\ n_0 q^2 / m \end{array} \right) \ ; 
 \label{shflds} 
\end{eqnarray}
upon performing the quenched disorder average we find 
\footnote{For a derivation of these results using the replica trick,
see Ref.~\cite{nelled}, App.~B.}
\begin{eqnarray}
        \overline{\langle \Theta({\bf q},\omega_m) 
                        \Theta({\bf q}',\omega_{m'}) \rangle} &&=
        {m \epsilon_{\rm B}(q)^2 / n_0 \hbar^3 q^2 \over
        \omega_m^2 + \epsilon_{\rm B}(q)^2 / \hbar^2} \, (2 \pi)^d 
        \delta({\bf q}+{\bf q}') \, \beta \hbar \delta_{m,-m'}
        \nonumber \\ &&\quad + {\omega_m \omega_{m'} / \hbar^2 \over
                        [\omega_m^2 + \epsilon_{\rm B}(q)^2 / \hbar^2]
        [\omega_{m'}^2 + \epsilon_{\rm B}(q')^2 / \hbar^2]} \,
                \Delta({\bf q},\omega_m;{\bf q}',\omega_{m'}) \ ,
 \label{dsthth} \\
        \overline{\langle \Theta({\bf q},\omega_m) 
                \pi({\bf q}',\omega_{m'}) \rangle} &&= {\omega_m \over
        \omega_m^2 + \epsilon_{\rm B}(q)^2 / \hbar^2} \, (2 \pi)^d  
        \delta({\bf q}+{\bf q}') \, \beta \hbar \delta_{m,-m'}
        \nonumber \\ &&\quad + {n_0 {q'}^2 \omega_m / \hbar m \over 
                        [\omega_m^2 + \epsilon_{\rm B}(q)^2 / \hbar^2]
        [\omega_{m'}^2 + \epsilon_{\rm B}(q')^2 / \hbar^2]} \,
                \Delta({\bf q},\omega_m;{\bf q}',\omega_{m'}) \ ,
 \label{dsthpi} \\
        \overline{\langle \pi({\bf q},\omega_m) 
        \pi({\bf q}',\omega_{m'}) \rangle} &&= {n_0 \hbar q^2 /m \over 
        \omega_m^2 + \epsilon_{\rm B}(q)^2 / \hbar^2} \, (2 \pi)^d 
        \delta({\bf q}+{\bf q}') \, \beta \hbar \delta_{m,-m'} 
        \nonumber \\ &&\quad + {n_0^2 q^2 {q'}^2 / m^2 \over
                        [\omega_m^2 + \epsilon_{\rm B}(q)^2 / \hbar^2]
        [\omega_{m'}^2 + \epsilon_{\rm B}(q')^2 / \hbar^2]} \,
                \Delta({\bf q},\omega_m;{\bf q}',\omega_{m'}) \ .
 \label{dspipi}
\end{eqnarray}

Eq.~(\ref{dspipi}), of course, gives the dynamic structure factor in
harmonic approximation to first order in the defect correlator. In
most situations, translational invariance in space and time will be
restored statistically, i.e.,
\begin{equation}
        \Delta({\bf q},\omega_m;{\bf q}',\omega_{m'}) = 
        \Delta({\bf q},\omega_m) \, (2 \pi)^d \delta({\bf q}+{\bf q}')
        \, \beta \hbar \delta_{m,-m'} \ ;
 \label{trinvc}
\end{equation}
in that case, one finds that the disorder contributions to the
structure factor have the typical Lorentzian--squared form,
\begin{equation}
        \overline{S({\bf q},\omega_m)} = {n_0 \hbar q^2 / m
        \over \omega_m^2 + \epsilon_{\rm B}(q)^2 / \hbar^2} +
        \left( {n_0 q^2 / m \over \omega_m^2 + \epsilon_{\rm B}(q)^2
                / \hbar^2} \right)^2 \Delta({\bf q},\omega_m)  \ .
 \label{dodecr}
\end{equation}
The corresponding static structure factor becomes, using
Eqs.~(\ref{bsmat3}) and (\ref{dfmsm1}) of App.~\ref{bosmat},
\begin{equation}
        \overline{S({\bf q},\tau=0)} = 
        {n_0 \hbar^2 q^2 \over 2 m \epsilon_{\rm B}(q)} \, 
                        \coth {\beta \epsilon_{\rm B}(q) \over 2} 
        + \left( {n_0 q^2 \over m} \right)^2 {1 \over \beta \hbar} 
        \sum_n {\Delta({\bf q},\omega_n) \over
                [\omega_n^2 + \epsilon_{\rm B}(q)^2 / \hbar^2]^2} \ .
 \label{dostst}
\end{equation}

With the help of Eqs.~(\ref{gaumcr}),(\ref{fouvor}) and some tedious
algebra, we can also compute the current and vorticity correlations to
first order in the defect correlator. Upon assuming that the system is
statistically translation--invariant and isotropic, we can define
longitudinal and transverse components of the current correlation
function, and as in the pure case the transverse current response
function is closely related with the ({\it purely} transverse)
vorticity correlations, 
\begin{eqnarray}
        m^2 \, \overline{V_\perp({\bf q},\omega_m)} &&= \hbar q^2 \,
        \overline{\chi_\perp({\bf q},\omega_m)} \nonumber \\ 
        &&= {\hbar \over (d-1) \beta \Omega} \sum_{{\bf p},\omega_n} 
        {[q^2 p^2 - ({\bf q}{\bf p})^2] 
        [({\bf q}-{\bf p})^2 \epsilon_{\rm B}(p)^2 / \hbar^2 + 
                                p^2 \omega_{m-n} \omega_n] \over p^2
        [\omega_{m-n}^2+\epsilon_{\rm B}(|{\bf q}-{\bf p}|)^2/\hbar^2] 
        [\omega_n^2 + \epsilon_{\rm B}(p)^2 / \hbar^2]} \nonumber \\
        &&\quad + {n_0 \over (d-1) m \beta \Omega} 
        \sum_{{\bf p},\omega_n} {q^2 p^2 - ({\bf q}{\bf p})^2 \over
        [\omega_{m-n}^2+\epsilon_{\rm B}(|{\bf q}-{\bf p}|)^2/\hbar^2] 
        [\omega_n^2 + \epsilon_{\rm B}(p)^2 / \hbar^2]} \times 
        \nonumber \\ &&\qquad \times \Biggl[ {({\bf q}-{\bf p})^2 
        \over p^2} \, {({\bf q}-{\bf p})^2 \epsilon_{\rm B}(p)^2 / 
                        \hbar^2 + p^2 \omega_{m-n}\omega_n \over 
        \omega_{m-n}^2+\epsilon_{\rm B}(|{\bf q}-{\bf p}|)^2/\hbar^2}
                \, \Delta({\bf q}-{\bf p},\omega_{m-n}) \nonumber \\ 
        &&\qquad \qquad \qquad \qquad \qquad - 
        {({\bf q}-{\bf p})^2 \omega_n^2 - p^2 \omega_{m-n} \omega_n
                \over \omega_n^2 + \epsilon_{\rm B}(q)^2 / \hbar^2} \, 
                                \Delta({\bf p},\omega_n) \Biggr] \ .
 \label{trvrcr}
\end{eqnarray}
For general, ``time--dependent'' configurations of disorder, the
average mass current does not necessarily vanish; e.g., uniformly
moving defects ``drag'' a certain fraction of the particles along,
even at $T = 0$. Indeed, upon using Eq.~(\ref{dsthpi}), one finds
\begin{eqnarray}
        \langle {\bf g}({\bf q},\omega_m) \rangle &&=
        {i \hbar \over \beta \Omega} \sum_{{\bf p},\omega_n} {\bf p}
        \langle \pi({\bf q}-{\bf p},\omega_{m-n}) 
                        \Theta({\bf p},\omega_n) \rangle \nonumber \\
        &&= i {n_0 \over m} {1 \over \beta \hbar \Omega}
        \sum_{{\bf p},\omega_n} {{\bf p} ({\bf q}-{\bf p})^2 \omega_n 
        \, \Delta({\bf q}-{\bf p},\omega_{m-n};{\bf p},\omega_n) \over
        [\omega_{m-n}^2+\epsilon_{\rm B}(|{\bf q}-{\bf p}|)^2/\hbar^2] 
        [\omega_n^2 + \epsilon_{\rm B}(p)^2 / \hbar^2]} \ .
 \label{traped}
\end{eqnarray}

Upon returning to Eq.~(\ref{trvrcr}), we find for the normal fluid
density in an isotropic system
\begin{eqnarray}
        \overline{\rho_n} = 
        \lim_{q \to 0} \overline{\chi_\perp({\bf q},0)} &&= 
        {1 \over d \beta \Omega} \sum_{{\bf p},\omega_n} p^2 \,
        {\epsilon_{\rm B}(p)^2 / \hbar^2 - \omega_n^2 \over \left[
        \epsilon_{\rm B}(p)^2 / \hbar^2 + \omega_n^2 \right]^2}
        \nonumber \\ &&\quad + 
        {n_0 \over d m \beta \hbar \Omega} \sum_{{\bf p},\omega_n} p^4
        \, {\epsilon_{\rm B}(p)^2/\hbar^2 - 3 \omega_n^2 \over \left[
        \epsilon_{\rm B}(p)^2 / \hbar^2 + \omega_n^2 \right]^3} \,
        \Delta({\bf p},\omega_n) \ ;
 \label{donfld}
\end{eqnarray}
we shall evaluate this for interesting special cases in
Sec.~\ref{pbexdo}. For static disorder, i.e.,
\begin{equation}
        \Delta({\bf q},\omega_m) =
                \Delta({\bf q}) \, \beta \hbar \delta_{m,0} \ ,
 \label{statdf}
\end{equation}
Eq.~(\ref{donfld}) further simplifies to
\begin{equation}
        \overline{\rho_n} = {\beta \hbar^2 \over 4 d} 
                \int \! {d^dq \over (2 \pi)^d} \left( {q \over 
                \sinh [\beta \epsilon_{\rm B}(q) / 2] } \right)^2
        + {n_0 \over d m} \int {d^dq \over (2 \pi)^d} \, 
                \left( {\hbar q \over \epsilon_{\rm B}(q)} \right)^4
                \, \Delta({\bf q}) \ .
 \label{stdnfd}
\end{equation}
An identical formula for the disorder--induced renormalization of the
normal--fluid density was derived by Giorgini, Pitaevskii, and
Stringari \cite{giorgi}. Note that the defect contribution to
(\ref{stdnfd}) is {\it temperature--independent}, see
Sec.~\ref{obtilt}.

For flux liquids, Eqs.~(\ref{foutil}) and (\ref{trvrcr}) lead to the
vortex contribution to the tilt modulus 
\begin{eqnarray}
        \overline{{c^v_{ij}}^{-1}} = (n^2 \hbar)^{-1} 
        \lim_{\omega_m \to 0} \overline{T_{ij}({\bf 0},\omega_m)} &&= 
        {1 \over n m} \Biggl( \delta_{ij} - {1 \over n m \beta \Omega}
        \sum_{{\bf p},\omega_n} p_i p_j \, {\epsilon_{\rm B}(p)^2 /
        \hbar^2 - \omega_n^2 \over \left[ \epsilon_{\rm B}(p)^2 / 
                \hbar^2 + \omega_n^2 \right]^2} \nonumber \\ &&\quad 
        - {1 \over m^2 \beta \hbar \Omega} \sum_{{\bf p},\omega_n} 
        p_i p_j p^2 \, {\epsilon_{\rm B}(p)^2 /\hbar^2 - 3 \omega_n^2 
        \over \left[ \epsilon_{\rm B}(p)^2 / \hbar^2 + \omega_n^2 
                \right]^3} \, \Delta({\bf p},\omega_n) \Biggr) \ ,
 \label{dotilt}
\end{eqnarray}
which for an isotropic system agrees with Eq.~(\ref{tilmod}). We also
remark that the last term in Eq.~(\ref{dotilt}) can formally be
written as
\begin{eqnarray}
        &&- {1 \over m^2} \int \! {d^dq \over (2\pi)^d} \, q_i q_j q^2
        {1 \over \beta \hbar} \sum_m {\epsilon_{\rm B}(q)^2 / \hbar^2 
        - 3 \omega_m^2 \over \left[ \epsilon_{\rm B}(q)^2 / \hbar^2 
        + \omega_m^2 \right]^3} \, \Delta({\bf q},\omega_m)\nonumber\\ 
        &&\qquad = {1 \over 2 n_0 m \hbar} \int \! 
                                {d^dq \over (2 \pi)^d} \, q_i q_j 
        {1 \over \beta \hbar} \sum_m \Delta({\bf q},\omega_m) \, 
        {\partial^2 \over \partial\omega_m^2} S({\bf q},\omega_m) \ , 
 \label{sqderf}
\end{eqnarray}
where $S({\bf q},\omega_m)$ is the density correlation function of the
pure system (\ref{fouden}). Eq.~(\ref{dotilt}) is thus the
generalization to finite systems of the tilt modulus formula of Hwa 
{\it et al.} \cite{hwaled}. In App.~\ref{galtil}, we review how the
defect contribution to the tilt modulus can be obtained using an
affine transformation; it is, however, not straightforward to use the
same idea and method to obtain finite--size corrections. In the
following subsection, a more general expression for the normal fluid
density and the tilt modulus will be derived, valid even when
translational invariance along the $\tau$ direction is not restored
statistically.

\subsection{Condensate fraction and normal--fluid density}
 \label{pbcofr}

Eqs.~(\ref{dsthth})--(\ref{dspipi}) yield the disorder--averaged
Green's function in harmonic approximation (\ref{expgre}), which
reads, to first order in the disorder correlator,
\begin{eqnarray}
        \overline{G({\bf q},\omega_m;{\bf q}',\omega_{m'})} &&=
        {i \omega_m + \hbar q^2 / 2 m + n_0 V(q) / \hbar \over 
                \omega_m^2 + \epsilon_{\rm B}(q)^2 / \hbar^2} \, 
        (2 \pi)^d \delta({\bf q}+{\bf q}') \, \beta \hbar 
        \delta_{m,-m'} \nonumber \\ &&\quad + {n_0 \over \hbar^2} \, 
        {(i\omega_m + \hbar q^2/2m) (-i\omega_{m'} + \hbar {q'}^2/2m) 
                \over [\omega_m^2 + \epsilon_{\rm B}(q)^2 / \hbar^2] 
                [\omega_{m'}^2 + \epsilon_{\rm B}(q')^2 / \hbar^2]} \, 
                \Delta({\bf q},\omega_m;{\bf q}',\omega_{m'}) \ .
 \label{dogref}
\end{eqnarray}

With Eq.~(\ref{parnum}), we find the following general formula for the
depletion in a disordered Bose superfluid, i.e., the reduction in the
order parameter magnitude due to disorder, quantum and thermal
fluctuations, 
\begin{eqnarray}
        &&\overline{n(\tau)}-n_0 = \! \int \!\! {d^dq \over (2 \pi)^d} 
        \Biggl\{ \left[ v(q)^2 + {1 + 2 v(q)^2 \over 
        e^{\beta \epsilon_{\rm B}(q)} - 1} \right] \nonumber \\ 
        &&+ {n_0 \over \hbar^2 \Omega} {1 \over (\beta \hbar)^2} \!
                        \sum_{m,m'}\! e^{-i(\omega_m+\omega_{m'})\tau}
        {(i \omega_m + \hbar q^2/2m) (-i \omega_{m'} + \hbar q^2/2m)
                \over [\omega_m^2 + \epsilon_{\rm B}(q)^2 / \hbar^2] 
                [\omega_{m'}^2 + \epsilon_{\rm B}(q)^2 / \hbar^2]}
        \Delta({\bf q},\omega_m;-{\bf q},\omega_{m'}) \Biggr\}
 \label{dodepl}
\end{eqnarray}
This expression depends on imaginary time, when the disorder breaks
translation invariance. For a statistically translation--invariant
system (\ref{trinvc}), the disorder contribution in Eq.~(\ref{dodepl})
becomes
\begin{equation}
        \overline{n_\Delta} = {n_0 \over \hbar^2} 
        \int \! {d^dq \over (2 \pi)^d} \, {1 \over \beta \hbar} \sum_m
        {(\hbar q^2/2m)^2 - \omega_m^2 \over [\epsilon_{\rm B}(q)^2 / 
        \hbar^2 + \omega_m^2]^2} \, \Delta({\bf q},\omega_m) \ ,
 \label{trdepl}
\end{equation}
which for static defects (\ref{statdf}) simplifies further to yield
the {\it temperature--independent} result
\begin{equation}
        \overline{n_\Delta} = 
        {n_0 \over 4 m^2} \int \! {d^dq \over (2 \pi)^d} \, 
        \left( {\hbar q \over \epsilon_{\rm B}(q)} \right)^4 \, 
                                                \Delta({\bf q}) \ .
 \label{stdode}
\end{equation}
These expressions will be evaluated explicitly for interesting cases
in Sec.~\ref{pbexdo}. But we already see that for isotropic static
disorder there is an intimate relation between the defect
contributions to the depletion and the normal--fluid density, which are
both temperature--independent, namely \cite{giorgi,schakl}
\begin{equation}
        m \, \overline{n_\Delta} = 
                        {d \over 4} \, \overline{\rho_{n\,\Delta}} \ .
 \label{rnnrel}
\end{equation}

We conclude this section by deriving the effect of disorder on the
normal--fluid density, using Landau's arguments. This analysis yields
the most general expressions for the tilt modulus tensor; note that
Eq.~(\ref{dotilt}) requires a statistically translation--invariant
system (which is the most common situation), while the elegant
derivation using affine transformations in App.~\ref{galtil}
\cite{hwaled} is directly applicable only in the zero--temperature
limit ($\beta \hbar \to \infty$).

For a superfluid in motion with respect to the surrounding walls {\it
and} the intrinsic disorder, we have seen that the corresponding
Green's function is simply obtained from the one for the system at
rest by shifting the Matsubara frequencies according to 
$\omega_m \to \omega_m - i {\bf q}\cdot{\bf v}$ and 
$\omega_{m'} \to \omega_{m'} - i {\bf q}'\cdot{\bf v}$
[cf. Eq.~(\ref{mattra})]. Because the disorder correlator in the
comoving reference frame reads
\begin{equation}
        \Delta({\bf q},\omega_m;{\bf q}',\omega_{m'})_{\bf v} =
        \Delta({\bf q},\omega_m+i{\bf q}\cdot{\bf v};
                {\bf q}',\omega_{m'}+i{\bf q}'\cdot{\bf v}) \ ,
 \label{mvdcor}
\end{equation}
these transformations imply that the disorder correlator to be
inserted is effectively the one at rest. Thus, 
\begin{eqnarray}
        &&\overline{G({\bf q},\omega_m;
                        {\bf q}',\omega_{m'})_{-{\bf v}}} = 
                {i \omega_m + \hbar q^2 / 2 m + n_0 V(q) / \hbar + 
        {\bf q}\cdot{\bf v} \over (\omega_m - i{\bf q}\cdot{\bf v})^2
                + \epsilon_{\rm B}(q)^2 / \hbar^2} \, (2 \pi)^d 
                \delta({\bf q}+{\bf q}') \, \beta \hbar \delta_{m,-m'} 
        \nonumber \\ &&\quad + {n_0 \over \hbar^2} \, 
                {(i \omega_m + \hbar q^2 / 2 m + {\bf q}\cdot{\bf v}) 
        (-i \omega_{m'} + \hbar {q'}^2 / 2 m - {\bf q}'\cdot{\bf v})
        \over [(\omega_m-i{\bf q}\cdot{\bf v})^2 + 
                        \epsilon_{\rm B}(q)^2/\hbar^2] 
        [(\omega_{m'}-i{\bf q}'\cdot{\bf v})^2 + 
                        \epsilon_{\rm B}(q')^2 / \hbar^2]} \, 
                \Delta({\bf q},\omega_m;{\bf q}',\omega_{m'}) \ ;
 \label{domvgf}
\end{eqnarray}
Upon expanding Eq.~(\ref{domvgf}) to second order in 
${\bf q}\cdot{\bf v}$ and ${\bf q}'\cdot{\bf v}$, and making use of
Eqs.~(\ref{mdengr}) and (\ref{curexp}), we find for the normal--fluid
density, which due to the disorder has become a tensor and in general
a function of $\tau$ as well, 
\begin{eqnarray}
        \overline{\rho_{n \, ij}(\tau)} &&= {\beta \hbar^2 \over 4d}
        \, \delta_{ij} \int \! {d^dq \over (2\pi)^d} \left( {q \over 
        \sinh [\beta \epsilon_{\rm B}(q)/2]} \right)^2 \nonumber \\
        &&\quad + {n_0 \over \hbar\Omega} 
                        \int \! {d^dq \over (2\pi)^d}\, q_i q_j \, 
        {1 \over (\beta \hbar)^2} \sum_{m,m'}
        {e^{-i(\omega_m+\omega_{m'})\tau} 
                \Delta({\bf q},\omega_m;-{\bf q},\omega_{m'}) \over
                        [\omega_m^2 + \epsilon_{\rm B}(q)^2 / \hbar^2] 
                [\omega_{m'}^2 + \epsilon_{\rm B}(q)^2 / \hbar^2]}
         \times \nonumber \\ &&\qquad \qquad \qquad \qquad \times
        \Biggl( {(-i \omega_{m'} + \hbar q^2/2m) 
        [i \omega_m \hbar q^2/m - \omega_m^2 + 
                                \epsilon_{\rm B}(q)^2/\hbar^2] \over
        \omega_m^2 + \epsilon_{\rm B}(q)^2 / \hbar^2} \nonumber \\
        &&\qquad \qquad \qquad \qquad \quad +
        {(i \omega_m + \hbar q^2/2m) [-i \omega_{m'} \hbar q^2/m 
                - \omega_{m'}^2 + \epsilon_{\rm B}(q)^2/\hbar^2] \over
        \omega_{m'}^2 + \epsilon_{\rm B}(q)^2 / \hbar^2} \Biggr) \ .
 \label{dorhon}
\end{eqnarray}
If translational invariance is statistically restored by averaging
over the disorder, the defect contribution to Eq.~(\ref{dorhon})
reduces to the equivalent of Eq.~(\ref{dotilt}), because all the
terms in the numerator of (\ref{dorhon}) which are linear in
$\omega_m$ then vanish. For an isotropic system, the normal--fluid
density becomes a scalar, see Eq.~(\ref{donfld}). For possibly
anisotropic {\it static} disorder, the renormalization of the normal
fluid density becomes
\begin{equation}
        \overline{\rho_{n \, ij}} = {\beta \hbar^2 \over 4 d} \, 
        \delta_{ij} \int \! {d^dq \over (2 \pi)^d} \left( {q \over 
                \sinh [\beta \epsilon_{\rm B}(q) / 2] } \right)^2
        + {n_0 \over m} \int {d^dq \over (2 \pi)^d} \, q_i q_j
                \left( {\hbar^2 q \over \epsilon_{\rm B}(q)^2}
                                \right)^2 \, \Delta({\bf q}) \ ,
 \label{stdorn}
\end{equation}
which generalizes Eq.~(\ref{stdnfd}).

\subsection{Results for point--like, linear, and planar defects}
 \label{pbexdo}

We now evaluate the density correlation function and the effective
contributions to the depletion and the normal--fluid density caused by
pinning to static disorder explicitly for the cases of point and
(spatially) extended defects. We work in the boson picture, and use
the words ``point'',``line'', and ``plane'' to indicate the dimension 
of the disorder in the $d$--dimensional spacelike direction ${\bf
r}$. According to Eqs.~(\ref{stdode}) and (\ref{stdorn}), the defect
contributions for static disorder are independent of boson
temperature; i.e., in the case of periodic boundary conditions there
will be no finite--size corrections in the imaginary--time direction
for the corresponding quantities in type--II superconductors. For a
flux liquid, such ``static'' disorder translates to defects 
{\it correlated} along the magnetic--field direction, and ``point''
and ``linear'' defects in $d=2$ map on columnar pins and planar
disorder, respectively (see Table~\ref{disvar} for an explicit
comparison in $d = 2$). We shall return to these cases, as well as to
``true'' (uncorrelated in ${\bf r}$ {\it and} $\tau$) point disorder
and tilted extended defects in Sec.~\ref{expdis}, where we shall also
compare the results obtained for periodic and open boundary
conditions.

For boson point defects, the disorder correlator reads
\begin{equation}
        \Delta({\bf r},\tau;{\bf r}',\tau') = 
                \Delta \, \delta({\bf r}-{\bf r}') \quad , \qquad 
        \Delta({\bf q},\omega_m) = 
                                \Delta \, \beta \hbar \delta_{m,0} \ .
 \label{bopdcr}
\end{equation}
This immediately yields the density correlation function
\begin{equation}
        \overline{S({\bf q},\omega_m)} = {n_0 \hbar q^2 / m \over 
        \omega_m^2 + \epsilon_{\rm B}(q)^2 / \hbar^2} + \Delta \left( 
        {n_0 \hbar^2 q^2 \over m \epsilon_{\rm B}(q)^2} \right)^2
        \beta \hbar \delta_{m,0} \ ,
 \label{dopdsq}
\end{equation}
and hence for the static structure factor
\begin{equation}
        \overline{S({\bf q},\tau=0)} = 
        {n_0 \hbar^2 q^2 \over 2 m \epsilon_{\rm B}(q)} \, 
        \coth {\beta \epsilon_{\rm B}(q) \over 2} + \Delta \left( 
        {n_0 \hbar^2 q^2 \over m \epsilon_{\rm B}(q)^2} \right)^2 \ ;
 \label{dopdst}
\end{equation}
see also Eq.~(\ref{dostst}).

Furthermore, 
\begin{equation}
        m \, \overline{n_\Delta} = 
        \frac{d}{4} \, \overline{\rho_{n\,\Delta}} = 
        \Delta \, {n_0 \over 4 m} \int \! {d^dq \over (2 \pi)^d} \, 
        \left( {\hbar q \over \epsilon_{\rm B}(q)} \right)^4 \approx
        \Delta \, {m^3 n_0^{(d/2)-1} a^{(d/2)-2} \over 
                                2^{5-d} \pi^2 \Gamma(d/2) \hbar^d} 
        \int_0^\infty \! {x^{d-1} \over (1 + x^2)^2} \, dx \ ,
 \label{dopdfd}
\end{equation}
where in the final equation $V(q) \approx V_0 = 4 \pi a / m$ was
assumed ($x = \hbar q / 4 \sqrt{\pi a n_0}$). In two dimensions, one
finds 
\begin{equation}
        d = 2 \, : \; m \, \overline{n_\Delta} = 
                \overline{\rho_{n\,\Delta}} / 2 = 
                                \Delta m^3 / 16 \pi^2 \hbar^2 a \ , 
 \label{dopdf2}
\end{equation}
while in $d=3$ \cite{khuang,giorgi}
\begin{equation}
        d = 3 \, : \; m \, \overline{n_\Delta} = 
                3 \overline{\rho_{n\,\Delta}} / 4 = 
                \Delta m^3 n_0^{1/2} / 8 \pi^{3/2} \hbar^3 a^{1/2} \ .
 \label{dopdf3}
\end{equation}

We can readily generalize these results to the case of static parallel
defects, extended along a $d_\parallel$--dimensional subspace 
${\bf r}_\parallel$. The disorder is therefore uncorrelated only in
the $d_\perp$ transverse spatial directions ${\bf r}_\perp$, i.e.,
\begin{equation}
        \Delta({\bf r},\tau;{\bf r}',\tau') = 
        \Delta \, \delta({\bf r}_\perp-{\bf r}'_\perp) \quad , \qquad 
        \Delta({\bf q},\omega_m) = \Delta \, (2 \pi)^{d_\parallel} 
        \delta({\bf q}_\parallel) \, \beta \hbar \delta_{m,0} \ .
 \label{boedcr}
\end{equation}
The dimensions $d_\parallel = 1$ and thus $d_\perp = d-1$ correspond
to line defects, while $d_\parallel = 2$, $d_\perp = d-2$ describe
planar disorder, etc. The corresponding dynamic structure factor then
reads
\begin{equation}
        \overline{S({\bf q},\omega_m)} = {n_0 \hbar q^2 / m \over 
        \omega_m^2 + \epsilon_{\rm B}(q)^2 / \hbar^2} + \Delta \left( 
        {n_0 \hbar^2 q_\perp^2 \over m \epsilon_{\rm B}(q_\perp)^2} 
        \right)^2 (2 \pi)^{d_\parallel} \delta({\bf q}_\parallel) \,
                                        \beta \hbar \delta_{m,0} \ ,
 \label{doedsq}
\end{equation}
and the static structure factor becomes
\begin{equation}
        \overline{S({\bf q},\tau=0)} = 
        {n_0 \hbar^2 q^2 \over 2 m \epsilon_{\rm B}(q)} \, 
        \coth {\beta \epsilon_{\rm B}(q) \over 2} + \Delta \left( 
        {n_0 \hbar^2 q_\perp^2 \over m \epsilon_{\rm B}(q_\perp)^2} 
        \right)^2 (2 \pi)^{d_\parallel} \delta({\bf q}_\parallel) \ .
 \label{doedst}
\end{equation}

The disorder--induced depletion follows from Eq.~(\ref{dopdfd}) by
replacing $d$ with $d_\perp$,
\begin{equation}
        \overline{n_\Delta} = \Delta \, {n_0 \over 4 m^2} 
        \int \! {d^{d_\perp}q_\perp \over (2 \pi)^{d_\perp}} \, \left(
        {\hbar q_\perp \over \epsilon_{\rm B}(q_\perp)} \right)^4 = 
        \Delta \, {m^2 n_0^{(d_\perp/2)-1} a^{(d_\perp/2)-2} \over 
                2^{5-d_\perp} \pi^2 \Gamma(d_\perp/2) \hbar^{d_\perp}} 
        \int_0^\infty \! {x^{d_\perp-1} \over (1 + x^2)^2} \, dx \ .
 \label{doedfd}
\end{equation}
The normal--fluid density, defined as the transport coefficient
characterizing the linear response of the system to motion with
respect to its boundaries, will now depend on the direction relative
to the disorder. If this motion is enforced along the disorder
direction ${\bf r_\parallel}$, the defects will have no effect [see
Eq.~(\ref{stdorn}) with $q_i = q_{\parallel \, i}$],
\begin{equation}
        \overline{\rho_{n \, \parallel \, \Delta}} = 0 \ ;
 \label{pdoenf}
\end{equation}
on the other hand, if the probe acts transverse to the defects, we
find a similar result to Eq.~(\ref{dopdfd}) (for point disorder, all
directions are transverse to the defects), namely
\begin{equation}
        m \, \overline{n_\Delta} = 
        {d_\perp \over 4} \, \overline{\rho_{n \, \perp \, \Delta}} \ .
 \label{ednrnr}
\end{equation}
For linear defects in two dimensions or planar disorder in $d=3$
($d_\perp = 1$) this yields 
\begin{equation}
        d_\perp = 1 \, : \; m \, \overline{n_\Delta} = 
                \overline{\rho_{n \, \perp \, \Delta}} / 4 = 
                \Delta m^3 / 64 \pi^{3/2} \hbar n_0^{1/2} a^{3/2} \ ,
 \label{doedf1}
\end{equation}
whereas for line defects in three dimension
\begin{equation}
        d_\perp = 2 \, : \; m \, \overline{n_\Delta} = 
                \overline{\rho_{n \, \perp \, \Delta}} / 2 = 
                                \Delta m^3 / 16 \pi^2 \hbar^2 a \ ,
 \label{doedf2}
\end{equation}
which is, of course, identical to the result (\ref{dopdf2}) for point
defects in $d=2$. 

In the framework to the Bogoliubov approximation, which amounts to an
expansion with respect to both thermal and quantum fluctuations, the
disorder--induced depletion for $d_\perp < 4$ is smaller than the
corresponding enhancement of the normal--fluid density,
Eq.~(\ref{ednrnr}). The bosons will become localized when  
$\rho_{s \, \perp} \to 0$ (i.e., the corresponding ``tilt'' modulus 
$c^v_\perp$ diverges !). The emerging new low--temperature phase of
localized  bosons is the Bose glass
\cite{mahale,fiswei,sinrok,lzhang,matrul,wallin,hatano,pazisc} (we use
this term in a generalized form which also includes localization by
extended defects), and is characterized by a vanishing superfluid
density $\rho_s = 0$. In the corresponding ``Bose glass'' phase of
localized magnetic flux lines, the tilt modulus transverse to the
defect directions diverges, which implies the ``transverse Meissner
effect'', namely the absence of any response of the superconductor
towards transverse magnetic fields (up to some critical tilt angle),
see Sec.~\ref{expcon}.

As a final investigation of how disorder affects superfluids, consider
a Helium film on a substrate with disorder in the form of lines
(deposited, say, via microlithography) with random positions and
orientations. Imagine that the lines are arrayed in ``families''
indexed by a two--dimensional unit vector ${\bf {\hat n}}$
perpendicular to every number of a family. Within a single family, the
Fourier--transformed disorder correlator must have the form
\begin{equation}
        \Delta_{\hat n}({\bf q},\omega_m) = 
                2 \Delta \delta^{(1)}({\bf {\hat n}}\cdot{\bf q}) \,
                \beta \hbar \delta_{m,0} \ ,
 \label{ranlin}
\end{equation}
where the one--dimensional delta function restricts only the
wavevector component parallel to ${\bf {\hat n}}$. Upon using the
integral representation of the delta function, 
$\delta(x) = \int_{-\infty}^\infty e^{ixs} ds / 2 \pi$, we can compute
the quenched average over families in different directions by
integrating over ${\hat n} = (\cos \phi, \sin \phi)$ to find
\begin{equation}
        \Delta({\bf q},\omega_m) \equiv {1 \over 2 \pi} 
        \int_0^{2 \pi} \! d \phi \, \Delta_{\hat n}({\bf q},\omega_m)
                = {\Delta \over q} \, \beta \hbar \delta_{m,0} \ .
 \label{rnlncr}
\end{equation}
Results for the density correlations, depletion and normal--fluid
density can be obtained by inserting this correlator into 
Eqs.~(\ref{dodecr}), (\ref{stdode}), and (\ref{stdnfd}); thus we find, 
by deviding the corresponding results (\ref{doedsq}), (\ref{doedst})
for point defects by $q$, the density correlation function
\begin{equation}
        \overline{S({\bf q},\omega_m)} = {n_0 \hbar q^2 / m \over 
                \omega_m^2 + \epsilon_{\rm B}(q)^2 / \hbar^2} + 
                {\Delta \over q} \left( {n_0 \hbar^2 q_\perp^2 \over 
                        m \epsilon_{\rm B}(q_\perp)^2} \right)^2 
                (2 \pi)^{d_\parallel} \delta({\bf q}_\parallel) \,
                                        \beta \hbar \delta_{m,0} \ ,
 \label{sqrlin}
\end{equation}
and the static structure factor
\begin{equation}
        \overline{S({\bf q},\tau=0)} = 
        {n_0 \hbar^2 q^2 \over 2 m \epsilon_{\rm B}(q)} \, \coth 
        {\beta \epsilon_{\rm B}(q) \over 2} + {\Delta \over q} \left( 
        {n_0 \hbar^2 q_\perp^2 \over m \epsilon_{\rm B}(q_\perp)^2} 
        \right)^2 (2 \pi)^{d_\parallel} \delta({\bf q}_\parallel) \ .
 \label{strlin}
\end{equation}
 Similarly, we find in analogy with Eq.~(\ref{dopdfd})
\begin{equation}
        m \, \overline{n_\Delta} = 
        \frac{d}{4} \, \overline{\rho_{n\,\Delta}} = \Delta \, 
        {n_0 \over 4 m} \int \! {d^dq \over (2 \pi)^d} \, {1 \over q} 
        \left( {\hbar q \over \epsilon_{\rm B}(q)} \right)^4 \approx
        \Delta \, {m^3 n_0^{(d-3)/2} a^{(d-5)/2} \over 
                        2^{7-d} \pi^{5/2} \Gamma(d/2) \hbar^{d-1}} 
        \int_0^\infty \! {x^{d-2} \over (1 + x^2)^2} \, dx \ ,
 \label{dprlin}
\end{equation}
which is logarithmically divergent in $d = 1$, but yields a finite
result in two dimensions,
\begin{equation}
        d = 2 \, : \; m \, \overline{n_\Delta} = 
                \overline{\rho_{n\,\Delta}} / 2 = 
                \Delta m^3 / 128 \pi^{3/2} n_0^{1/2} \hbar a^{3/2} \ ,
 \label{dprln2}
\end{equation}
whereas in $d = 3$
\begin{equation}
        d = 3 \, : \; m \, \overline{n_\Delta} = 
                3 \overline{\rho_{n\,\Delta}} / 4 = 
                                \Delta m^3 / 16 \pi^3 \hbar^2 a \ .
 \label{dprln3}
\end{equation}
We see that, although the correlator (\ref{ranlin}) is singular as 
$q \to 0$, the disorder contributions to Eqs.~(\ref{stdnfd}) and
(\ref{stdode}) remain infrared--convergent (for $d > 1$). The
depletion and normal--fluid density are, however, {\it increased} due
to the correlations, relative to the same amount of disorder scattered
about in a ``point--like'' fashion.


\section{Disorder contributions for flux liquids
         (open boundary conditions)}  
 \label{disfll}

We now investigate the influence of disorder on the correlation
functions, ``boson order parameter'', and the tilt modulus for flux
liquids with open boundary conditions along the magnetic--field
direction, using the results obtained in App.~\ref{obcder}. In this
and the following section, we employ the flux line language, but
retain the boson superfluid notation, in order to facilitate
comparison with the results for the case of periodic boundary
conditions derived in Sec.~\ref{disbos}. We shall discuss the
equations for {\it general} disorder correlators, which lead to rather
lengthy and not very transparent formulas. Specific results for a
number of interesting types of defects are tabulated in
Sec.~\ref{expdis}.

\subsection{Density and tilt correlations}
 \label{obcorr}

Upon carrying out the matrix products in Eq.~(\ref{opdisc}), we arrive
at the following explicit results for the correlation functions with
open boundary conditions in harmonic approximation, to first order in
the disorder correlator,
\begin{eqnarray}
        &&\overline{\langle \Theta({\bf q},\omega_m)
                \Theta({\bf q}',\omega_{m'}) \rangle} = \nonumber \\
        &&\quad = {m \epsilon_{\rm B}(q)^2 / n_0 \hbar^3 q^2 \over
        \omega_m^2 + \epsilon_{\rm B}(q)^2 /\hbar^2} \left( 
        \beta\hbar \delta_{m,-m'} - {2\hbar \over \epsilon_{\rm B}(q)}
        \, \tanh {\beta \epsilon_{\rm B}(q) \over 2} \, 
        {\omega_m \omega_{m'} + \epsilon_{\rm B}(q)^2 / \hbar^2 \over 
        \omega_{m'}^2 + \epsilon_{\rm B}(q)^2 / \hbar^2} \right) 
                (2 \pi)^d \delta({\bf q}+{\bf q}') \nonumber \\ 
        &&+ {1 / \hbar^2 \over [\omega_m^2 + \epsilon_{\rm B}(q)^2 
        / \hbar^2] [\omega_{m'}^2 + \epsilon_{\rm B}(q')^2 /\hbar^2]} 
        \sum_{n,n'} \left( \omega_m \delta_{m,n} - 
                {2 \epsilon_{\rm B}(q) \over \beta \hbar^2} \, \tanh 
        {\beta \epsilon_{\rm B}(q) \over 2} \, {\omega_m + \omega_n 
        \over \omega_n^2 + \epsilon_{\rm B}(q)^2 / \hbar^2} \right) 
                                                \times \nonumber \\ 
        &&\qquad \qquad \times \left( \omega_{m'} \delta_{m',n'} - 
        {2 \epsilon_{\rm B}(q') \over \beta \hbar^2} \, \tanh {\beta 
        \epsilon_{\rm B}(q') \over 2} \, {\omega_{m'}+\omega_{n'}
        \over \omega_{n'}^2 + \epsilon_{\rm B}(q')^2 / \hbar^2} 
        \right) \Delta({\bf q},\omega_n;{\bf q}',\omega_{n'}) \ ,
 \label{dothth} \\
        &&\overline{\langle \Theta({\bf q},\omega_m) 
        \pi({\bf q}',\omega_{m'}) \rangle} = \nonumber \\ &&\quad =
        {\beta \hbar \over \omega_m^2 + \epsilon_{\rm B}(q)^2/\hbar^2}
        \left( \omega_m \delta_{m,-m'} - {2 \epsilon_{\rm B}(q) \over 
        \beta \hbar^2} \, \tanh {\beta \epsilon_{\rm B}(q) \over 2} \,
        {\omega_m - \omega_{m'} \over 
        \omega_{m'}^2 + \epsilon_{\rm B}(q)^2 / \hbar^2} \right) 
                (2 \pi)^d \delta({\bf q}+{\bf q}') \nonumber \\
        &&+ {n_0 {q'}^2 / \hbar m \over [\omega_m^2 + 
        \epsilon_{\rm B}(q)^2 / \hbar^2] [\omega_{m'}^2 + 
        \epsilon_{\rm B}(q')^2 / \hbar^2]} \sum_{n,n'} \left( 
        \omega_m \delta_{m,n} - {2 \epsilon_{\rm B}(q) \over \beta 
        \hbar^2} \, \tanh {\beta \epsilon_{\rm B}(q) \over 2} \, 
        {\omega_m + \omega_n \over \omega_n^2 + \epsilon_{\rm B}(q)^2 
                                / \hbar^2} \right) \times \nonumber \\
        &&\qquad \qquad \times \left( \delta_{m',n'} + {2 \over \beta
        \epsilon_{\rm B}(q')} \, \tanh {\beta \epsilon_{\rm B}(q') 
        \over 2} \, {\omega_{m'} \omega_{n'} - \epsilon_{\rm B}(q')^2
        /\hbar^2 \over \omega_{n'}^2 + \epsilon_{\rm B}(q')^2/\hbar^2} 
        \right) \Delta({\bf q},\omega_n;{\bf q}',\omega_{n'}) \ ,
 \label{dothpi} \\
        &&\overline{\langle \pi({\bf q},\omega_m) 
        \pi({\bf q}',\omega_{m'}) \rangle} = \nonumber \\ &&\quad =
        {n_0 \hbar q^2 /m \over \omega_m^2 + \epsilon_{\rm B}(q)^2 / 
        \hbar^2} \, \left( \beta\hbar \delta_{m,-m'} - {2 \hbar \over 
        \epsilon_{\rm B}(q)} \, \tanh {\beta \epsilon_{\rm B}(q) 
        \over 2} \, {\omega_m \omega_{m'} + \epsilon_{\rm B}(q)^2 / 
        \hbar^2 \over \omega_{m'}^2 + \epsilon_{\rm B}(q)^2 / \hbar^2}
        \right) (2 \pi)^d \delta({\bf q}+{\bf q}') \nonumber \\
        &&+ {n_0^2 q^2 {q'}^2 / m^2 \over [\omega_m^2 + 
        \epsilon_{\rm B}(q)^2 / \hbar^2] [\omega_{m'}^2 + 
        \epsilon_{\rm B}(q')^2 / \hbar^2]} \sum_{n,n'} \left(
        \delta_{m,n} + {2 \over \beta \epsilon_{\rm B}(q)} \, \tanh
        {\beta \epsilon_{\rm B}(q) \over 2} \, {\omega_m \omega_n - 
        \epsilon_{\rm B}(q)^2 / \hbar^2 \over \omega_n^2 + 
        \epsilon_{\rm B}(q)^2 / \hbar^2} \right) \times \nonumber \\
        &&\qquad \qquad \times \left( \delta_{m',n'} + {2 \over \beta
        \epsilon_{\rm B}(q')} \, \tanh {\beta \epsilon_{\rm B}(q') 
        \over 2} \, {\omega_{m'} \omega_{n'} - \epsilon_{\rm B}(q')^2
        /\hbar^2 \over \omega_{n'}^2 + \epsilon_{\rm B}(q')^2/\hbar^2} 
        \right) \Delta({\bf q},\omega_n;{\bf q}',\omega_{n'}) \ .
 \label{dopipi}
\end{eqnarray}
These results should be contrasted with
Eqs.~(\ref{dsthth})--(\ref{dspipi}), which are valid for periodic
boundary conditions, and may be recovered from
(\ref{dothth})--(\ref{dopipi}) by taking into account only those terms
in the brackets containing the Kronecker symbols $\delta_{m,-m'}$, 
$\delta_{m,n}$, and $\delta_{m',n'}$.

From Eq.~(\ref{dopipi}), we can deduce the equal--``time'' density
correlation function,
\begin{eqnarray}
        &&\overline{S({\bf q},\tau;{\bf q}',\tau)} = 
        {n_0 \hbar^2 q^2 \over 2 m \epsilon_{\rm B}(q)} \,
        \tanh {\beta \epsilon_{\rm B}(q) \over 2} \, 
        (2 \pi)^d \delta({\bf q}+{\bf q}') \nonumber \\ &&\qquad 
        + {n_0^2 \hbar^2 q^2 {q'}^2 \over m^2 \epsilon_{\rm B}(q) 
        \epsilon_{\rm B}(q')} \, {1 \over (\beta \hbar)^2} \sum_{n,n'}
        {\Delta({\bf q},\omega_n;{\bf q}',\omega_{n'}) \over 
                [\omega_n^2 + \epsilon_{\rm B}(q)^2 / \hbar^2]
        [\omega_{n'}^2 + \epsilon_{\rm B}(q')^2 / \hbar^2]} \times
                        \nonumber \\ &&\qquad \quad \times \left[ 
        {\epsilon_{\rm B}(q) \over \hbar} \left( e^{-i\omega_n \tau} -
        {\cosh [(\beta\hbar-2\tau) \epsilon_{\rm B}(q)/2\hbar] \over 
        \cosh [\beta \epsilon_{\rm B}(q)/2]} \right) - i\omega_n \,
        {\sinh [(\beta\hbar-2\tau) \epsilon_{\rm B}(q)/2\hbar] \over 
        \cosh [\beta \epsilon_{\rm B}(q)/2]} \right] \times\nonumber\\ 
        &&\quad \times \left[ {\epsilon_{\rm B}(q')\over\hbar}
        \left( e^{-i\omega_{n'}\tau} - 
        {\cosh [(\beta\hbar-2\tau) \epsilon_{\rm B}(q')/2\hbar] \over
        \cosh [\beta\epsilon_{\rm B}(q')/2]} \right) - i\omega_{n'} \,
        {\sinh [(\beta\hbar-2\tau) \epsilon_{\rm B}(q') / 2 \hbar]
        \over \cosh [\beta \epsilon_{\rm B}(q')/2]} \right] \ ,
 \label{dopstf}
\end{eqnarray}
generalizing both Eqs.~(\ref{dostst}) and (\ref{opstrf}). The
$d$--dimensional structure factor is now a function of $\tau$,
implying that the mean--square variations of the vortex positions
depend on the depth of the considered cross--section perpendicular to
the magnetic field in the sample. On the surfaces at $\tau = 0$ and
$\tau = \beta \hbar$, the fluctuations vanish. Similarly, for the
leading term of the current correlation function (i.e., the one
originating in the phase fluctuations) we find
\begin{eqnarray}
        &&\overline{C_{ij}({\bf q},\tau;{\bf q}',\tau)} = 
        n_0 m \, {\epsilon_{\rm B}(q) \over 2} \,
        \tanh {\beta \epsilon_{\rm B}(q) \over 2} \, P_{ij}^L({\bf q})
        \, (2 \pi)^d \delta({\bf q}+{\bf q}') \nonumber \\
        &&\qquad + n_0^2 q_i q_j' \, {1 \over (\beta \hbar)^2}
        \sum_{n,n'} {\Delta({\bf q},\omega_n;{\bf q}',\omega_{n'})
        \over [\omega_n^2 + \epsilon_{\rm B}(q)^2 / \hbar^2]
        [\omega_{n'}^2 + \epsilon_{\rm B}(q')^2 / \hbar^2]} \times
                        \nonumber \\ &&\qquad \quad \times \left[ 
        i \omega_n \left( e^{-i\omega_n \tau} - 
        {\cosh [(\beta\hbar-2\tau) \epsilon_{\rm B}(q)/2\hbar] \over 
        \cosh [\beta \epsilon_{\rm B}(q)/2]} \right) - 
        {\epsilon_{\rm B}(q) \over \hbar} \,
        {\sinh [(\beta\hbar-2\tau) \epsilon_{\rm B}(q)/2\hbar] \over 
        \cosh [\beta \epsilon_{\rm B}(q)/2]} \right] \times\nonumber\\ 
        &&\quad\times \left[ i\omega_{n'} \left( e^{-i\omega_{n'}\tau}
        - {\cosh [(\beta\hbar-2\tau) \epsilon_{\rm B}(q')/2\hbar] 
        \over \cosh [\beta\epsilon_{\rm B}(q')/2]} \right) - 
        {\epsilon_{\rm B}(q') \over \hbar} \,
        {\sinh [(\beta\hbar-2\tau) \epsilon_{\rm B}(q') / 2 \hbar]
        \over \cosh [\beta \epsilon_{\rm B}(q')/2]} \right] \ .
 \label{dopcrf}
\end{eqnarray}
Eq.~(\ref{tilcor}) then yields the tilt correlations.

In the center of the sample, $\tau = \beta \hbar / 2$, these
expressions simplify considerably; upon assuming translational
invariance in the $d$ transverse ``spatial'' directions, the density
and current correlators read
\begin{eqnarray}
        &&\overline{S({\bf q},\tau = \beta \hbar / 2)} = 
        {n_0 \hbar^2 q^2 \over 2 m \epsilon_{\rm B}(q)} \,
        \tanh {\beta \epsilon_{\rm B}(q) \over 2}
        + {1 \over (\beta\hbar)^2 \Omega} \sum_{n,n'} 
        {(n_0 q^2/m)^2 \,\Delta({\bf q},\omega_n;-{\bf q},\omega_{n'})
        \over [\omega_n^2 + \epsilon_{\rm B}(q)^2 / \hbar^2] 
        [\omega_{n'}^2 + \epsilon_{\rm B}(q)^2/\hbar^2]} \nonumber \\
        &&\qquad \qquad \qquad \times 
                        \left( e^{- i \omega_n \beta \hbar / 2} - 
        {1 \over \cosh [\beta \epsilon_{\rm B}(q) / 2]} \right)
        \left( e^{- i \omega_{n'} \beta \hbar / 2} - 
        {1 \over \cosh [\beta \epsilon_{\rm B}(q) / 2]} \right) \ ,
 \label{domsst} \\
        &&\overline{C_{ij}({\bf q},\tau = \beta \hbar / 2)} = 
        \Biggl[ n_0 m \, {\epsilon_{\rm B}(q) \over 2} \, 
                \tanh {\beta \epsilon_{\rm B}(q) \over 2} + 
                {1 \over (\beta\hbar)^2 \Omega} \sum_{n,n'} 
        {n_0^2 q^2 \, \Delta({\bf q},\omega_n;-{\bf q},\omega_{n'}) 
                                \, \omega_n \omega_{n'} \over 
        [\omega_n^2 + \epsilon_{\rm B}(q)^2 / \hbar^2]
        [\omega_{n'}^2 + \epsilon_{\rm B}(q)^2/\hbar^2]} \nonumber \\
        &&\qquad \qquad \qquad \times 
                        \left( e^{- i \omega_n \beta \hbar / 2} - 
        {1 \over \cosh [\beta \epsilon_{\rm B}(q) / 2]} \right)
        \left( e^{- i \omega_{n'} \beta \hbar / 2} - 
        {1 \over \cosh [\beta \epsilon_{\rm B}(q) / 2]} \right) 
                                        \Biggr] P_{ij}^L({\bf q}) \ .
 \label{domscr}
\end{eqnarray}
Note that $e^{- i \omega_n \beta \hbar / 2} = (-1)^n$. For static
defects (\ref{statdf}), the disorder contribution to
Eq.~(\ref{domscr}) vanishes, while the static structure factor becomes 
\begin{equation}
        \overline{S({\bf q},\tau = \beta \hbar / 2)} = 
        {n_0 \hbar^2 q^2 \over 2 m \epsilon_{\rm B}(q)} \,
        \tanh {\beta \epsilon_{\rm B}(q) \over 2} + \Delta({\bf q}) 
        \left( {n_0 \hbar^2 q^2 \over m \epsilon_{\rm B}(q)^2} \right)^2
        \left( 1- {1 \over \cosh [\beta \epsilon_{\rm B}(q)/2]}
                                                        \right)^2 \ .
 \label{opdost}
\end{equation}

\subsection{Depletion and vortex tilt modulus} 
 \label{obtilt}

Within the Gaussian approximation (\ref{expgre}),
(\ref{dothth})--(\ref{dopipi})we can readily obtain the disorder
contribution to the Green's function for open boundary conditions,
supplementing the pure part (\ref{opgref}),
\begin{eqnarray}
        &&\overline{G_\Delta({\bf q},\omega_m;{\bf q}',\omega_m')} =
        {n_0/\hbar^2 \over [\omega_m^2 +\epsilon_{\rm B}(q)^2/\hbar^2]
        [\omega_{m'}^2 + \epsilon_{\rm B}(q')^2/\hbar^2]} \sum_{n,n'}
        \Delta({\bf q},\omega_n;{\bf q}',\omega_{n'}) \times \nonumber
        \\ &&\Biggl[ \left( {\hbar q^2 \over 2m} + i \omega_m \right) 
        \delta_{m,n} + {\tanh [\beta \epsilon_{\rm B}(q) / 2] \over 
                \beta \epsilon_{\rm B}(q) / 2} {[\omega_m \omega_n
                - \epsilon_{\rm B}(q)^2 / \hbar^2] \hbar q^2 / 2m 
        - i (\omega_m + \omega_n) \epsilon_{\rm B}(q)^2 / \hbar^2
        \over \omega_n^2 + \epsilon_{\rm B}(q)^2/\hbar^2} \Biggr] 
                \! \times \nonumber \\ &&\qquad \times \Biggl[ \left( 
        {\hbar {q'}^2 \over 2m} - i \omega_{m'} \right) \delta_{m',n'} 
        + {\tanh [\beta \epsilon_{\rm B}(q')/2] \over 
        \beta \epsilon_{\rm B}(q') / 2} \times \nonumber \\ 
        &&\qquad \qquad \qquad \qquad \qquad \times 
        {[\omega_{m'} \omega_{n'} - \epsilon_{\rm B}(q')^2 / \hbar^2]
                \hbar {q'}^2 / 2m + i (\omega_{m'} + \omega_{n'}) 
                        \epsilon_{\rm B}(q')^2 / \hbar^2 \over 
        \omega_{n'}^2 + \epsilon_{\rm B}(q')^2 / \hbar^2} \Biggr] \ .
 \label{dopgrf}
\end{eqnarray}
Upon assuming as usual translational invariance in ${\bf r}$, this
leads to the disorder--induced depletion, i.e., a disentanglement of
pinned vortex lines,
\begin{eqnarray}
        \overline{n_\Delta(\tau)} &&= {n_0 \over \hbar^2 \Omega}
        \int \! {d^dq \over (2 \pi)^d} {1 \over (\beta \hbar)^2}
        \sum_{n,n'} {\Delta({\bf q},\omega_n;-{\bf q},\omega_n')
        \over [\omega_n^2 + \epsilon_{\rm B}(q)^2 / \hbar^2]
        [\omega_{n'}^2 + \epsilon_{\rm B}(q)^2 / \hbar^2]} \times
        \nonumber \\ &&\qquad \qquad \times \Biggl[ 
        \left( {\hbar q^2 \over 2 m} + i \omega_n \right) 
        \left( e^{-i \omega_n \tau} - 
        {\cosh [(\beta \hbar - 2 \tau) \epsilon_{\rm B}(q) / 2 \hbar] 
        \over \cosh [\beta \epsilon_{\rm B}(q) / 2]} \right) \nonumber
        \\ &&\qquad \qquad \qquad \qquad \qquad - 
                        \left( {\epsilon_{\rm B}(q) \over \hbar} + 
        {i \omega_n \hbar^2 q^2 \over 2 m \epsilon_{\rm B}(q)} \right) 
        {\sinh [(\beta \hbar - 2 \tau) \epsilon_{\rm B}(q) / 2 \hbar]
        \over \cosh [\beta \epsilon_{\rm B}(q) / 2]} \Biggr] \times
        \nonumber \\ &&\qquad \qquad \times \Biggl[ 
        \left( {\hbar q^2 \over 2 m} - i \omega_{n'} \right) 
        \left( e^{-i \omega_{n'} \tau} - 
        {\cosh [(\beta \hbar - 2 \tau) \epsilon_{\rm B}(q) / 2 \hbar] 
        \over \cosh [\beta \epsilon_{\rm B}(q) / 2]} \right) \nonumber
        \\ &&\qquad \qquad \qquad \qquad \qquad + 
                        \left( {\epsilon_{\rm B}(q) \over \hbar} -
        {i\omega_{n'} \hbar^2 q^2 \over 2m\epsilon_{\rm B}(q)} \right)
        {\sinh [(\beta \hbar - 2 \tau) \epsilon_{\rm B}(q) / 2 \hbar]
        \over \cosh [\beta \epsilon_{\rm B}(q) / 2]} \Biggr] \ ,
 \label{dopdpl}
\end{eqnarray}
compare Eq.~(\ref{dodepl}) for periodic boundary conditions. It is a
straightforward, but rather tedious task to compute the
disorder renormalization of the normal--fluid density from
Eqs.~(\ref{curexp}) and (\ref{mdengr}), following Landau's arguments
(see Sec.~\ref{pbcofr}); besides expanding to second order in 
${\bf q}\cdot{\bf v}$, this involves a number of Matsubara frequency
sums for which the list in App.~\ref{bosmat} is helpful. The final
result reads
\begin{eqnarray}
        \overline{\rho_{n \, ij \, \Delta}(\tau)} &&= 
        - {n_0 \over \hbar \Omega} \int \! {d^dq \over (2 \pi)^d} \, 
        q_i q_j \, {1 \over (\beta \hbar)^2} \sum_{n,n'}
        {\Delta({\bf q},\omega_n;-{\bf q},\omega_n') \over 
        [\omega_n^2 + \epsilon_{\rm B}(q)^2 / \hbar^2] [\omega_{n'}^2 
        + \epsilon_{\rm B}(q)^2 / \hbar^2]} \times \nonumber \\ 
        &&\times \Biggl\{ \Biggl[ \left( {\hbar q^2 \over 2 m} 
                + i \omega_n \right) \left( e^{-i \omega_n \tau} - 
        {\cosh [(\beta \hbar - 2 \tau) \epsilon_{\rm B}(q) / 2 \hbar] 
        \over \cosh [\beta \epsilon_{\rm B}(q)/2]} \right)\nonumber\\ 
        &&\qquad \qquad \qquad \qquad 
        - \left( {\epsilon_{\rm B}(q) \over \hbar} + {i \omega_n 
        \hbar^2 q^2 \over 2 m \epsilon_{\rm B}(q)} \right) {\sinh 
        [(\beta \hbar - 2 \tau) \epsilon_{\rm B}(q) / 2 \hbar] \over 
        \cosh [\beta \epsilon_{\rm B}(q)/2]} \Biggr] \times\nonumber\\
        &&\quad \times \Biggl[ {\omega_{n'}^2 - 
        \epsilon_{\rm B}(q)^2 / \hbar^2 + i \omega_{n'} \hbar q^2 / m 
        \over \omega_{n'}^2 + \epsilon_{\rm B}(q)^2 / \hbar^2} \,
                e^{-i \omega_{n'} \tau} \nonumber \\ 
        &&\qquad \qquad + \left( {\hbar q^2 \over 2m} - i \omega_{n'}
        \right) \left( {\beta \hbar \sinh [\epsilon_{\rm B}(q) \tau /
                \hbar] \over \sinh [\beta \epsilon_{\rm B}(q)]} - 
        {\tau \cosh [(\beta\hbar-2\tau) \epsilon_{\rm B}(q)/2\hbar]
        \over \cosh [\beta \epsilon_{\rm B}(q)/2]} \right) \nonumber\\
        &&\qquad + \left( {\epsilon_{\rm B}(q) \over \hbar} -
        {i\omega_{n'} \hbar^2 q^2 \over 2m\epsilon_{\rm B}(q)} \right)
        \left( {\beta \hbar \cosh [\epsilon_{\rm B}(q) \tau / \hbar]
                \over \sinh [\beta \epsilon_{\rm B}(q)]} - {\tau 
        \sinh [(\beta\hbar-2\tau) \epsilon_{\rm B}(q)/2\hbar] \over 
        \cosh [\beta \epsilon_{\rm B}(q)/2]} \right)\Biggr]\nonumber\\ 
        &&\quad + \Biggl[ {\omega_n^2 -\epsilon_{\rm B}(q)^2/\hbar^2 - 
        i\omega_n \hbar q^2/m \over \omega_n^2 + \epsilon_{\rm B}(q)^2
        / \hbar^2} \, e^{-i \omega_n \tau} \nonumber \\ 
        &&\qquad \qquad - \left( {\hbar q^2 \over 2 m} + i \omega_n
        \right) \left( {\beta \hbar \sinh [\epsilon_{\rm B}(q) \tau /
                \hbar] \over \sinh [\beta \epsilon_{\rm B}(q)]} - 
        {\tau \cosh [(\beta\hbar-2\tau) \epsilon_{\rm B}(q)/2\hbar]
        \over \cosh [\beta \epsilon_{\rm B}(q)/2]} \right) \nonumber\\
        &&\qquad + \left( {\epsilon_{\rm B}(q) \over \hbar} +
        {i\omega_n \hbar^2 q^2 \over 2m\epsilon_{\rm B}(q)} \right)
        \left( {\beta \hbar \cosh [\epsilon_{\rm B}(q) \tau / \hbar]
                \over \sinh [\beta \epsilon_{\rm B}(q)]} - {\tau 
        \sinh [(\beta\hbar-2\tau) \epsilon_{\rm B}(q)/2\hbar] \over 
        \cosh [\beta\epsilon_{\rm B}(q)/2]} \right)\Biggr] \nonumber\\
        &&\quad \times \Biggl[ \left( {\hbar q^2 \over 2 m} 
        - i \omega_{n'} \right) \left( e^{-i \omega_{n'} \tau} - 
        {\cosh [(\beta \hbar - 2 \tau) \epsilon_{\rm B}(q) / 2 \hbar] 
        \over \cosh [\beta \epsilon_{\rm B}(q)/2]} \right)\nonumber\\ 
        &&\qquad \qquad \qquad \qquad + \left( 
        {\epsilon_{\rm B}(q) \over \hbar} - 
        {i\omega_{n'} \hbar^2 q^2 \over 2m\epsilon_{\rm B}(q)} \right)
        {\sinh [(\beta \hbar - 2 \tau) \epsilon_{\rm B}(q) / 2 \hbar]
        \over \cosh [\beta \epsilon_{\rm B}(q)/2]} \Biggr] \Biggr\}\ ,
 \label{dopnfl}
\end{eqnarray}
and is to be contrasted with Eq.~(\ref{dorhon}). Note that as a
consequence of the open boundary conditions (which suppress
fluctuations in the boson order parameter at the boundaries), both
results (\ref{dopdpl}) and (\ref{dopnfl}) vanish for $\tau = 0$ and
$\tau = \beta \hbar$, respectively. At $\tau = \beta \hbar / 2$, these
expressions simplify, and the depletion and normal--fluid density in
the center of the bulk become
\begin{eqnarray}
        &&\overline{n(\tau = \beta \hbar / 2)} - n_0 = 
        \int \! {d^dq \over (2 \pi)^d} \left[ v(q)^2 - {1 + 2 v(q)^2 
        \over e^{\beta \epsilon_{\rm B}(q)} + 1} \right] \nonumber\\
        &&\quad + {n_0 \over \hbar^2 \Omega} \int \! 
        {d^dq \over (2 \pi)^d} \, {1 \over (\beta \hbar)^2}
        \sum_{n,n'} {\Delta({\bf q},\omega_n;-{\bf q},\omega_n') \over
                [\omega_n^2 + \epsilon_{\rm B}(q)^2 / \hbar^2] 
        [\omega_{n'}^2 + \epsilon_{\rm B}(q)^2 / \hbar^2]} \left( 
        {\hbar q^2 \over 2 m} + i \omega_n \right) \times \nonumber \\ 
        &&\qquad \times \left( {\hbar q^2 \over 2 m} - i \omega_{n'}
        \right) \left( e^{- i \omega_n \beta \hbar / 2} - 
                {1 \over \cosh [\beta \epsilon_{\rm B}(q)/2]} \right) 
        \left( e^{- i \omega_{n'} \beta \hbar / 2} - {1 \over 
        \cosh [\beta \epsilon_{\rm B}(q)/2]} \right) \ ,
 \label{dodbde} \\
        &&\overline{\rho_{n \, ij}(\tau = \beta \hbar / 2)} = 
                - {\beta \hbar^2 \over 4 d} \, \delta_{ij} \int \! 
                        {d^dq \over (2 \pi)^d} \left( {q \over 
        \sinh [\beta \epsilon_{\rm B}(q)/2]} \right)^2 \nonumber \\ 
        &&\quad - {n_0 \over \hbar \Omega} \int \! 
        {d^dq \over (2 \pi)^d} \, q_i q_j \, {1 \over (\beta \hbar)^2}
        \sum_{n,n'} {\Delta({\bf q},\omega_n;-{\bf q},\omega_n') \over 
        [\omega_n^2 + \epsilon_{\rm B}(q)^2 / \hbar^2] [\omega_{n'}^2 
        + \epsilon_{\rm B}(q)^2 / \hbar^2]} \times \nonumber \\ 
        &&\qquad \times \Biggl\{ 
        \left( {\hbar q^2 \over 2 m} + i \omega_n \right) 
        \left( e^{- i \omega_{n'} \beta \hbar / 2} - {1 \over 
        \cosh [\beta \epsilon_{\rm B}(q)/2]} \right) \times\nonumber\\ 
        &&\qquad \qquad \times 
        \left[ {\omega_{n'}^2 - \epsilon_{\rm B}(q)^2 / \hbar^2 
        + i \omega_{n'} \hbar q^2 / m \over \omega_{n'}^2 + 
        \epsilon_{\rm B}(q)^2 / \hbar^2} \, e^{-i \omega_{n'} \tau} +
        {\beta \hbar / 2 \over \sinh [\beta \epsilon_{\rm B}(q) / 2]}
        \left( {\epsilon_{\rm B}(q) \over\hbar} - {i\omega_{n'}\hbar^2
        q^2 \over 2 m \epsilon_{\rm B}(q)} \right) \right] \nonumber\\ 
        &&\qquad + \left( {\hbar q^2 \over 2m} - i \omega_{n'} \right)
        \left( e^{- i \omega_n \beta \hbar / 2} - {1 \over 
        \cosh [\beta \epsilon_{\rm B}(q)/2]} \right) \times\nonumber\\
        &&\qquad \times \left[ {\omega_n^2 - \epsilon_{\rm B}(q)^2 /
        \hbar^2 - i \omega_n \hbar q^2 / m \over \omega_n^2 + 
        \epsilon_{\rm B}(q)^2 / \hbar^2} \, e^{-i \omega_n \tau} +
        {\beta \hbar / 2 \over \sinh [\beta \epsilon_{\rm B}(q) / 2]}
        \left( {\epsilon_{\rm B}(q) \over \hbar} + {i\omega_n \hbar^2
        q^2 \over 2 m \epsilon_{\rm B}(q)} \right) \right] \Biggr\}\ .
 \label{dodbnf}
\end{eqnarray}

For static disorder (\ref{statdf}), we find the following
comparatively simple expressions,
\begin{eqnarray}
        \overline{n_\Delta(\tau = \beta \hbar /2)} &&= 
        {n_0 \over 4 m^2} \int \! {d^dq \over (2 \pi)^d} \, 
        \left( {\hbar q \over \epsilon_{\rm B}(q)} \right)^4 
        \Delta({\bf q}) \left( 1 - {1 \over 
                \cosh [\beta \epsilon_{\rm B}(q)/2]} \right)^2 \ ,
 \label{dosddp} \\
        \overline{\rho_{n \, ij \,\Delta}(\tau = \beta \hbar / 2)} &&=
        {n_0 \over m} \int \! {d^dq \over (2 \pi)^d} \, q_i q_j \,
        \left( {\hbar^2 q \over \epsilon_{\rm B}(q)^2} \right)^2 
        \Delta({\bf q}) \times \nonumber \\
        &&\quad \times \left( 1 - 
                {1 \over \cosh [\beta \epsilon_{\rm B}(q)/2]} -
                {\beta \epsilon_{\rm B}(q) / 2 \over 
                        \sinh [\beta \epsilon_{\rm B}(q) / 2]} +
                {\beta \epsilon_{\rm B}(q) \over 
                        \sinh [\beta \epsilon_{\rm B}(q)]} \right) \ .
 \label{dosdrn}
\end{eqnarray}
In contrast to Eqs.~(\ref{stdode}) and (\ref{stdorn}), these results
do depend on $\beta \hbar$. With the defects being ``oriented'' along
the $\tau$ direction, it is plausible that there are no
finite--thickness corrections to the disorder contributions for 
quantities such as the depletion and the normal--fluid density,
provided periodic boundary conditions are applied; for all points
along the defect (world) line are then fully equivalent. This explains
in physical terms why the effects of static disorder on Bose
superfluids turned out to be independent on temperature, see
Sec.~\ref{disbos}. With open boundary conditions, on the other hand,
the enhanced thermal wandering near the surfaces comes into play
(compare Sec.~\ref{flines}), which renders the defect influence
dependent on the sample thickness. As is apparent from 
Eqs.~(\ref{domsst})--(\ref{opdost}) and
(\ref{dodbde})--(\ref{dosdrn}), these finite--size contributions
typically decay exponentially with growing sample thickness 
$\beta \hbar$, the characteristic length (the analog of the thermal
de Broglie--wavelength for superfluids) being $\propto \beta \hbar
c_1$ at small wavenumbers. Note in addition that there appears no
simple relation such as (\ref{rnnrel}) or (\ref{ednrnr}) between the
disorder contributions to the depletion and the normal--fluid density
at any $\beta \hbar < \infty$ any more. We defer the explicit
evaluation of Eqs.~(\ref{dodbde}) and (\ref{dodbnf}) for several types
of defects to the following section.


\section{Explicit results for various kinds of disorder}
 \label{expdis}

We now specialize the cumbersome expressions obtained for general
disorder in the previous section to specifically interesting types of
defects, and in some cases list the corresponding results for both
periodic and open boundary conditions along the $\tau \sim z$
direction. Although we have the application to magnetic flux liquids
in high--temperature superconductors in mind, we express our results
in the boson superfluid notation in $d$ ``transverse''
dimensions. Only in the final subsection \ref{expcon}, when we
summarize a number of experimentally relevant consequences of our
investigations, we put $d=2$ explicitly and translate to the flux line
language.

\subsection{Uncorrelated disorder -- point defects}
 \label{uncdis}

We start with disorder completely uncorrelated in ${\bf r}$ {\it and}
$\tau$,
\begin{equation}
        \Delta({\bf q},\omega_m;{\bf q}',\omega_m') = 
                        \Delta \, (2\pi)^d \delta({\bf q}+{\bf q}') \,
                                        \beta \hbar \delta_{m,-m'} \ ;
 \label{unccor}
\end{equation}
the quenched disorder average restores translational invariance in
``space'' and ``time''. In the flux line picture, Eq.~(\ref{unccor})
describes point defects, e.g. oxygen vacancies in the cuprate
planes. (For ``true'' bosons, a realization of this space-- and
time--dependent disorder is more difficult to conceive.) For this
highly ``dynamical'' disorder (\ref{unccor}), we find the following
defect contributions to the structure factor and the leading component
of the ``static'' current correlation function (stemming from the
phase fluctuations only), first for periodic boundary conditions [see
Eqs.~(\ref{dostst}) and (\ref{dsthth})],
\begin{eqnarray}
        {\rm p.b.c.} \, : \quad \overline{S_\Delta({\bf q})} &&= 
        {\Delta \over \hbar \epsilon_{\rm B}(q)} \left( 
        {n_0 \hbar^2 q^2 \over 2 m \epsilon_{\rm B}(q)} \right)^2 
                \coth {\beta \epsilon_{\rm B}(q) \over 2} 
                \left( 1 + {\beta \epsilon_{\rm B}(q) \over 
                        \sinh [\beta \epsilon_{\rm B}(q)]} \right) \ ,
 \label{ptdpst} \\
        \overline{C_{ij \, \Delta}({\bf q})} &&=
        - {\Delta \over \hbar \epsilon_{\rm B}(q)} 
        \left( {n_0 \hbar q \over 2} \right)^2 
                \coth {\beta \epsilon_{\rm B}(q) \over 2} 
                \left( 1 - {\beta \epsilon_{\rm B}(q) \over 
        \sinh [\beta \epsilon_{\rm B}(q)]} \right) P_{ij}^L({\bf q}) \ ,
 \label{ptdpcr}
\end{eqnarray}
while the results for for open boundary conditions
[Eqs.~(\ref{domsst}),(\ref{domscr})], taken in the center of the
sample ($\tau = \beta \hbar / 2$), read
\begin{eqnarray}
        {\rm o.b.c.} \, : \quad \overline{S_\Delta({\bf q})} &&=
        {\Delta \over \hbar \epsilon_{\rm B}(q)} \left( 
        {n_0 \hbar^2 q^2 \over 2 m \epsilon_{\rm B}(q)} \right)^2 
                \tanh {\beta \epsilon_{\rm B}(q) \over 2} 
                \left( 1 - {\beta \epsilon_{\rm B}(q) \over 
                        \sinh [\beta \epsilon_{\rm B}(q)]} \right) \ ,
         \label{ptdost} \\
        \overline{C_{ij \, \Delta}({\bf q})} &&=
        - {\Delta \over \hbar \epsilon_{\rm B}(q)} 
        \left( {n_0 \hbar q \over 2} \right)^2 
                \tanh {\beta \epsilon_{\rm B}(q) \over 2} 
                \left( 1 + {\beta \epsilon_{\rm B}(q) \over 
        \sinh [\beta \epsilon_{\rm B}(q)]} \right) P_{ij}^L({\bf q}) \ .
 \label{ptdocr}
\end{eqnarray}
As in the pure case, the limit $\beta \hbar \to \infty$ is thus
approached from different sides for periodic and open boundary
conditions, respectively.

In order to compute the depletion from Eqs.~(\ref{trdepl}) and
(\ref{dodbde}) for periodic and open boundary conditions, we use the
Matsubara frequency sums tabulated in App.~\ref{bosmat}. The results
are
\begin{eqnarray}
        {\rm p.b.c.} \, : \quad \overline{n_\Delta} &&= 
        {n_0 \over 4} \int \! {d^dq \over (2 \pi)^d} \,
        {\Delta \over \hbar \epsilon_{\rm B}(q)}
        \coth {\beta \epsilon_{\rm B}(q) \over 2} \times \nonumber \\ 
        &&\qquad \qquad \times \left[ 
        \left( {\hbar^2 q^2 \over 2 m \epsilon_{\rm B}(q)} \right)^2
        \left( 1 + {\beta \epsilon_{\rm B}(q) \over 
        \sinh [\beta \epsilon_{\rm B}(q)]} \right) - 
        \left( 1 - {\beta \epsilon_{\rm B}(q) \over 
        \sinh [\beta \epsilon_{\rm B}(q)]} \right) \right] \ ,
 \label{ptdpdp} \\
        {\rm o.b.c.} \, : \quad \overline{n_\Delta} &&= 
        {n_0 \over 4} \int \! {d^dq \over (2 \pi)^d} \,
        {\Delta \over \hbar \epsilon_{\rm B}(q)}
        \tanh {\beta \epsilon_{\rm B}(q) \over 2} \times \nonumber \\ 
        &&\qquad \qquad \times \left[  
        \left( {\hbar^2 q^2 \over 2 m \epsilon_{\rm B}(q)} \right)^2
        \left( 1 - {\beta \epsilon_{\rm B}(q) \over 
        \sinh [\beta \epsilon_{\rm B}(q)]} \right) - 
        \left( 1 + {\beta \epsilon_{\rm B}(q) \over 
        \sinh [\beta \epsilon_{\rm B}(q)]} \right) \right] \ .
 \label{ptdodp}
\end{eqnarray}
In the limit $\beta \hbar \to \infty$, these expressions coincide, of
course,
\begin{equation}
        \overline{n_\Delta} = 
        {n_0 \over 4} \int \! {d^dq \over (2 \pi)^d} \, {\Delta \over 
        \hbar \epsilon_{\rm B}(q)} \left[ \left( {\hbar^2 q^2 \over 
        2 m \epsilon_{\rm B}(q)} \right)^2 - 1 \right] = 
        - \Delta \, {n_0^2 \hbar \over 4 m} \int \! {d^dq \over 
        (2 \pi)^d} \, {q^2 V(q) \over \epsilon_{\rm B}(q)^3} \ .
 \label{ptdldp}
\end{equation}
With $x = \hbar q / 4 \sqrt{\pi n_0 a}$, we find for a delta--function
repulsive potential [$V(q) \approx V_0 = 4 \pi a / m$]
\begin{equation}
        \overline{n_\Delta} = - \Delta \, {m n_0^{d/2} a^{(d/2)-1} 
                        \over 2^{4-d} \pi \Gamma(d/2) \hbar^{d+1}}
        \int_0^\infty \! {x^{d-2} \over (1 + x^2)^{3/2}} \, dx \ ;
 \label{ptdedp}
\end{equation}
in two dimensions, this yields
\begin{equation}
        d = 2 \, : \quad \overline{n_\Delta} = 
                                - \Delta \, m n_0 / 4 \pi \hbar^3 \ ,
 \label{ptd2dp}
\end{equation}
and in the three--dimensional case
\begin{equation}
        d = 3 \, : \quad \overline{n_\Delta} = 
        - \Delta \, m n_0^{3/2} a^{1/2} / \pi^{3/2} \hbar^4 \ . 
 \label{ptd3dp}
\end{equation}
By exploiting the identities 
$\int_0^\infty [x^{d-2} (\coth x - 1) - x^{d-1} / \sinh^2 x] dx = 
- \delta_{d,2} - [(d-2)/(d-1)]$ $\int_0^\infty x^{d-1} / \sinh^2 x dx$
and $\int_0^\infty [x^{d-2} (1 - \tanh x) - x^{d-1} / \cosh^2 x] dx = 
-[(d-2)/(d-1)]$ $\int_0^\infty x^{d-1} / \cosh^2 x dx$ for $d \geq 2$,
we can derive the leading finite--temperature / sample thickness
corrections in the phonon approximation, 
$\epsilon_{\rm B}(q) \approx \hbar c_1 q$ ($x = \beta \hbar c_1 q/2$),  
\begin{eqnarray}
        {\rm p.b.c.} \, : \quad &&\overline{\Delta n_\Delta(T)} =
        - \Delta \, {n_0 \over 4 \hbar} \int \! {d^dq \over (2 \pi)^d}
        \, \Biggl\{ {n_0 \hbar^2 q^2 V(q) \over 
                                        m \epsilon_{\rm B}(q)^3}
        \left( \coth {\beta \epsilon_{\rm B}(q) \over 2} - 1 \right)
        \nonumber \\ &&\qquad \qquad \qquad \qquad \qquad \qquad - 
        {\beta / 2 \over \sinh^2 [\beta \epsilon_{\rm B}(q)/2]} \left[
        1 + \left( {\hbar^2 q^2 \over 2 m \epsilon_{\rm B}(q)} 
                                        \right)^2 \right] \Biggr\}
 \label{ptpidp} \\
        &&\approx {\Delta n_0 (k_{\rm B} T)^{d-1} \over 
        4 \pi^{d/2} \Gamma(d/2) \hbar^{d+1} c_1^d} \left[ \delta_{d,2}
        + {d-2 \over d-1}\int_0^\infty\!{x^{d-1} \over \sinh^2 x}\, dx
        + {(k_{\rm B}T)^2 \over m^2 c_1^4} \int_0^\infty{x^{d+1} \over
                        \sinh^2 x} \, dx \right] \ , \nonumber \\
        {\rm o.b.c.} \, : \quad &&\overline{\Delta n_\Delta(T)} =
        \Delta \, {n_0 \over 4 \hbar} \int \! {d^dq \over (2 \pi)^d}
        \, \Biggl\{ {n_0 \hbar^2 q^2 V(q) 
                \over m \epsilon_{\rm B}(q)^3} \left( 1 - 
        \tanh {\beta \epsilon_{\rm B}(q) \over 2} \right) \nonumber \\
        &&\qquad \qquad \qquad \qquad \qquad \qquad - 
        {\beta / 2 \over \cosh^2 [\beta \epsilon_{\rm B}(q)/2]} \left[
        1 + \left( {\hbar^2 q^2 \over 2 m \epsilon_{\rm B}(q)}
                                        \right)^2 \right] \Biggr\}  
 \label{ptoidp} \\
        &&\approx - {\Delta n_0 (k_{\rm B} T)^{d-1} \over 
                4 \pi^{d/2} \Gamma(d/2) \hbar^{d+1} c_1^d} \left[ 
        {d-2 \over d-1} \int_0^\infty\!{x^{d-1} \over \cosh^2 x} \, dx
        + {(k_{\rm B}T)^2 \over m^2 c_1^4} \int_0^\infty{x^{d+1} \over
                                \cosh^2 x} \, dx \right] \ ; \nonumber
\end{eqnarray}
with Eqs.~(\ref{integ3}) and (\ref{integ4}) of App.~\ref{momint} we
finally find in two dimensions 
\begin{eqnarray}
        d = 2 \, : \quad 
        {\rm p.b.c.} \, : \quad \overline{\Delta n_\Delta(T)} &&=
                {\Delta n_0 k_{\rm B} T \over 4 \pi \hbar^3 c_1^2} 
                                                + {\cal O}(T^3) \ ,
 \label{ptp2dp} \\
        {\rm o.b.c.} \, : \quad \overline{\Delta n_\Delta(T)} &&=
                                                - {\cal O}(T^3) \ ,
 \label{pto2dp}
\end{eqnarray}
while for $d > 2$
\begin{eqnarray}
        {\rm p.b.c.} \, : \quad \overline{\Delta n_\Delta(T)} &&=
        {(d-2) \Gamma(d-1) \zeta(d-1) \over 2^d \pi^{d/2} \Gamma(d/2)}
        \, {\Delta n_0 (k_{\rm B} T)^{d-1} \over \hbar^{d+1} c_1^d}\ ,
 \label{ptpddp} \\
        {\rm o.b.c.} \, : \quad \overline{\Delta n_\Delta(T)} &&=
                                - \left( 1 - {1 \over 2^{d-2}} \right)
        {(d-2) \Gamma(d-1) \zeta(d-1) \over 2^d \pi^{d/2} \Gamma(d/2)}
        \, {\Delta n_0 (k_{\rm B} T)^{d-1} \over \hbar^{d+1} c_1^d}\ ,
 \label{ptoddp}
\end{eqnarray}
e.g., in three dimensions,
\begin{eqnarray}
        d = 3 \, : \quad 
        {\rm p.b.c.} \, : \quad \overline{\Delta n_\Delta(T)} &&=
                \Delta n_0 (k_{\rm B} T)^2 / 24 \hbar^4 c_1^3 \ ,
 \label{ptp3dp} \\
        {\rm o.b.c.} \, : \quad \overline{\Delta n_\Delta(T)} &&=
                - \Delta n_0 (k_{\rm B} T)^2 / 48 \hbar^4 c_1^3 \ .
 \label{pto3dp}
\end{eqnarray}
As expected, the sign of the finite--temperature / size corrections is
opposite for periodic and open boundary conditions, respectively,
again reflecting the fact that the enhanced thermal wandering near the
surfaces increases the tendency for entanglement of the (world) lines
in the case of open boundary conditions. In $d=2$ the leading
correction $\propto k_{\rm B} T$ vanishes for open boundary 
conditions. 

With somewhat more effort, the disorder renormalization of the
normal--fluid density / tilt modulus may be obtained evaluating
(\ref{donfld}) and (\ref{dodbnf}), respectively; as the disorder
correlator is isotropic, one finds
\begin{eqnarray}
        {\rm p.b.c.} \, : \quad \overline{\rho_{n \, \Delta}} &&= {n_0
        \beta^2 \hbar^4 \over 8 d m} \int \! {d^dq \over (2 \pi)^d} \,
                        {\Delta q^4 \over \hbar \epsilon_{\rm B}(q)} 
                        {\cosh [\beta \epsilon_{\rm B}(q) / 2] \over 
                        \sinh^3 [\beta \epsilon_{\rm B}(q) / 2]} \ ,
 \label{ptdprn} \\
        {\rm o.b.c.} \, : \quad \overline{\rho_{n \,\Delta}} &&= -{n_0
        \beta^2 \hbar^4 \over 8 d m} \int \! {d^dq \over (2 \pi)^d} \,
                        {\Delta q^4 \over \hbar \epsilon_{\rm B}(q)} 
                        {1 \over \sinh [\beta \epsilon_{\rm B}(q)]} \ ;
 \label{ptdorn}
\end{eqnarray}
note that these results for periodic and open boundary conditions are
not simply related, because the normal--fluid density arises from the
next--to--leading order in a long--wavelength expansion. However, in
the limit $\beta \hbar \to \infty$ one has 
\begin{equation}
        \overline{\rho_{n \, \Delta}(T = 0)} = 0 \ ,
 \label{ptdlrn}
\end{equation}
implying that mere point disorder cannot alter the tilt modulus of a
flux liquid in thick samples; see also App.~\ref{galtil}. For finite
``temperatures'', the leading corrections become for periodic and open
boundary conditions, respectively [using $\int_0^\infty x^{d+2} 
(\cosh x / \sinh^3 x) dx = [(d+2)/2] \int_0^\infty x^{d+1}/\sinh^2 x$
and Eq.~(\ref{integ3})]
\begin{eqnarray}
        {\rm p.b.c.} \, : \quad \overline{\rho_{n \, \Delta}(T)} &&= 
        {\Gamma(d+3)\zeta(d+1) \over 2^{d+1}\pi^{d/2}\Gamma(1+d/2)} \,
        {\Delta n_0(k_{\rm B}T)^{d+1} \over m \hbar^{d+1}c_1^{d+4}}\ ,
 \label{ptpdrn} \\
        {\rm o.b.c.} \, : \quad \overline{\rho_{n \, \Delta}(T)} &&= 
                                - \left( 1 - {1 \over 2^{d+3}} \right)
        {\Gamma(d+3)\zeta(d+3) \over 2^{d+2}\pi^{d/2}\Gamma(1+d/2)} \,
        {\Delta n_0(k_{\rm B}T)^{d+1} \over m \hbar^{d+1}c_1^{d+4}}\ ,
 \label{ptodrn}
\end{eqnarray}
which in two dimensions becomes
\begin{eqnarray}
        d = 2 \, : \quad        
        {\rm p.b.c.} \, : \quad \overline{\rho_{n \, \Delta}(T)} &&= 
        3\zeta(3) \Delta n_0 (k_{\rm B} T)^3 / \pi m \hbar^3 c_1^6 \ ,
 \label{ptp2rn} \\
        {\rm o.b.c.} \, : \quad \overline{\rho_{n \, \Delta}(T)} &&= -
        93\zeta(5)\Delta n_0 (k_{\rm B} T)^3/ 64\pi m\hbar^3 c_1^6 \ ,
 \label{pto2rn}
\end{eqnarray}
while in three dimensions
\begin{eqnarray}
        d = 3 \, : \quad 
        {\rm p.b.c.} \, : \quad \overline{\rho_{n \, \Delta}(T)} &&= 
        \pi^2 \Delta n_0 (k_{\rm B} T)^4 / 9 m \hbar^4 c_1^7 \ ,
 \label{ptp3rn} \\
        {\rm o.b.c.} \, : \quad \overline{\rho_{n \, \Delta}(T)} &&= 
        - \pi^4 \Delta n_0 (k_{\rm B} T)^4 / 192 m \hbar^4 c_1^7 \ .
 \label{pto3rn}
\end{eqnarray}
Again, these finite--temperature / sample thickness corrections
display opposite signs for the two different boundary conditions.

\subsection{Correlated random disorder -- nearly isotropic splay}
 \label{aisspl}

For magnetic flux lines in high--temperature superconductors, a
deliberate introduction of splayed columnar defects has been suggested
as a means to enhance flux pinning \cite{hwaled,lednel}. In the
flux line notation, the disorder correlator for such randomly tilted
columnar defects, each described by a trajectory 
${\bf r}_i(z) = {\bf R}_i + {\bf v}_i z$, with a Gaussian
distribution of the tilts ${\bf v}_i$, $P[{\bf v}_i] \propto \prod_i 
e^{- v_i^2 / 2 v_D^2}$, reads in Fourier space \cite{hwaled,devsca}
\begin{equation}
        \Delta({\bf q},q_z) = {\Delta_1 \over \sqrt{2 \pi} v_D q} \,
                                        e^{- q_z^2 / 2 v_D^2 q^2} \ .
 \label{splcor}
\end{equation}
Upon taking the limit $v_D \to \infty$, with $\Delta_1 / v_D$ held
fixed, one has $\Delta({\bf q},q_z) \propto 1/q$, i.e., an ``almost''
isotropic situation; as this correlator was obtained by starting from
a distribution centered around the $z$ axis, which thus constitutes a
preferred direction, this limit does not quite lead to the ``truly''
isotropic limit 
$\Delta({\bf q},q_z) \propto 1 / ({\bf q}^2 + q_z^2)^{-1/2}$ [compare
the two--dimensional correlator (\ref{rnlncr})]. The additional $q_z$
dependence here is not expected to affect the physical implications
drastically, but would render the evaluation of the depletion and tilt
modulus considerably more cumbersome. 

We thus study the simpler case of ``neaarly isotropic'' splay, namely
\begin{equation}
        \Delta({\bf q},\omega_m;{\bf q}',\omega_m') = 
        {\Delta \over q} \, (2\pi)^d \delta({\bf q}+{\bf q}') \,
                                        \beta \hbar \delta_{m,-m'} \ .
 \label{rspcor}
\end{equation}
With this correlator, we can immediately take over the results for
point defects from the previous subsection, because none of the
Matsubara frequency sums is changed; we simply have to divide $\Delta$
by $q$ everywhere. Hence
\begin{eqnarray}
        {\rm p.b.c.} \, : \quad \overline{S_\Delta({\bf q})} &&= 
        {\Delta \over \hbar q \epsilon_{\rm B}(q)} \left( 
        {n_0 \hbar^2 q^2 \over 2 m \epsilon_{\rm B}(q)} \right)^2 
                \coth {\beta \epsilon_{\rm B}(q) \over 2} 
                \left( 1 + {\beta \epsilon_{\rm B}(q) \over 
                        \sinh [\beta \epsilon_{\rm B}(q)]} \right) \ ,
 \label{rsdpst} \\
        \overline{C_{ij \, \Delta}({\bf q})} &&=
        - {\Delta \over \hbar q \epsilon_{\rm B}(q)} 
        \left( {n_0 \hbar q \over 2} \right)^2 
                \coth {\beta \epsilon_{\rm B}(q) \over 2} 
                \left( 1 - {\beta \epsilon_{\rm B}(q) \over 
        \sinh [\beta \epsilon_{\rm B}(q)]} \right) P_{ij}^L({\bf q})
 \label{rsdpcr}
\end{eqnarray}
for periodic boundary conditions, while
\begin{eqnarray}
        {\rm o.b.c.} \, : \quad \overline{S_\Delta({\bf q})} &&=
        {\Delta \over \hbar q \epsilon_{\rm B}(q)} \left( 
        {n_0 \hbar^2 q^2 \over 2 m \epsilon_{\rm B}(q)} \right)^2 
                \tanh {\beta \epsilon_{\rm B}(q) \over 2} 
                \left( 1 - {\beta \epsilon_{\rm B}(q) \over 
                        \sinh [\beta \epsilon_{\rm B}(q)]} \right) \ ,
 \label{rsdost} \\
        \overline{C_{ij \, \Delta}({\bf q})} &&=
        - {\Delta \over \hbar q \epsilon_{\rm B}(q)} 
        \left( {n_0 \hbar q \over 2} \right)^2 
                \tanh {\beta \epsilon_{\rm B}(q) \over 2} 
                \left( 1 + {\beta \epsilon_{\rm B}(q) \over 
        \sinh [\beta \epsilon_{\rm B}(q)]} \right) P_{ij}^L({\bf q})
 \label{rsdocr}
\end{eqnarray}
for open boundary conditions, evaluated at $\tau = \beta \hbar / 2$.

In the same manner, dividing the integrands of Eqs.~(\ref{ptdpdp}) and
(\ref{ptdodp}) by $q$ yields the defect contribution to the depletion 
\begin{eqnarray}
        {\rm p.b.c.} \, : \quad \overline{n_\Delta} &&= 
        {n_0 \over 4} \int \! {d^dq \over (2 \pi)^d} \,
        {\Delta \over \hbar q \epsilon_{\rm B}(q)}
        \coth {\beta \epsilon_{\rm B}(q) \over 2} \times \nonumber \\ 
        &&\qquad \qquad \times \left[ 
        \left( {\hbar^2 q^2 \over 2 m \epsilon_{\rm B}(q)} \right)^2
        \left( 1 + {\beta \epsilon_{\rm B}(q) \over 
        \sinh [\beta \epsilon_{\rm B}(q)]} \right) - 
        \left( 1 - {\beta \epsilon_{\rm B}(q) \over 
        \sinh [\beta \epsilon_{\rm B}(q)]} \right) \right] \ ,
 \label{rsdpdp} \\
        {\rm o.b.c.} \, : \quad \overline{n_\Delta} &&= 
        {n_0 \over 4} \int \! {d^dq \over (2 \pi)^d} \,
        {\Delta \over \hbar q \epsilon_{\rm B}(q)}
        \tanh {\beta \epsilon_{\rm B}(q) \over 2} \times \nonumber \\ 
        &&\qquad \qquad \times \left[  
        \left( {\hbar^2 q^2 \over 2 m \epsilon_{\rm B}(q)} \right)^2
        \left( 1 - {\beta \epsilon_{\rm B}(q) \over 
        \sinh [\beta \epsilon_{\rm B}(q)]} \right) - 
        \left( 1 + {\beta \epsilon_{\rm B}(q) \over 
        \sinh [\beta \epsilon_{\rm B}(q)]} \right) \right] \ .
 \label{rsdodp}
\end{eqnarray}
In the limit $\beta \hbar \to \infty$, both expressions lead to
\begin{equation}
        \overline{n_\Delta} = 
        {n_0 \over 4} \int \! {d^dq \over (2 \pi)^d} \, {\Delta \over 
        \hbar q \epsilon_{\rm B}(q)} \left[ \left( {\hbar^2 q^2 \over 
                2 m \epsilon_{\rm B}(q)} \right)^2 - 1 \right] =
        - \Delta \, {n_0^2 \hbar \over 4 m} \int \! {d^dq \over 
        (2 \pi)^d} \, {q V(q) \over \epsilon_{\rm B}(q)^3} \ .
 \label{rsdldp}
\end{equation}
For $V(q) \approx V_0 = 4 \pi a / m$ ($x = \hbar q/4\sqrt{\pi a n_0}$)
we thus find for the disorder contribution to 
$n - \overline{n_0} = n - \overline{|\langle \psi \rangle|^2}$,
\begin{equation}
        \overline{n_\Delta} = - \Delta \, {m n_0^{(d-1)/2} a^{(d-3)/2} 
                \over 2^{6-d} \pi^{3/2} \Gamma(d/2) \hbar^d}  
        \int_0^\infty \! {x^{d-3} \over (1 + x^2)^{3/2}} \, dx \ ,
 \label{rsdedp}
\end{equation}
which is {\it negative} and logarithmically {\it divergent} in two
dimensions, suggesting that $\overline{|\langle \psi \rangle|^2}$
becomes unbounded above! Evidently the flux lines are easily trapped
and swept along by the disorder; thermal wandering presumably becomes
superdiffusive, leading to the divergence of $\overline{n_\Delta}$ (see
Ref.~\cite{kamnel} for a related case). The mapping onto
nonrelativistic bosons is inadequate to describe completely this
``superentangled'' state of interacting lines. In three dimensions,
the corrections are finite,
\begin{equation}
        \overline{n_\Delta} = - \Delta m n_0 / 4 \pi^2 \hbar^3 \ ,
 \label{rsi3dp}
\end{equation}
and the wandering remains diffusive. Upon dividing the expressions of
the previous subsection by $q$, we find for the finite--temperature /
sample thickness corrections for $d > 2$,
\begin{eqnarray}
        &&{\rm p.b.c.}\, :\nonumber \\ &&\overline{\Delta n_\Delta(T)}
        \approx {\Delta n_0 (k_{\rm B} T)^{d-2} \over 8 \pi^{d/2} 
        \Gamma(d/2) \hbar^d c_1^{d-1}} \left[ \delta_{d,3}
        + {d-3 \over d-2} \int_0^\infty\!{x^{d-2} \over \sinh^2 x} dx
        + {(k_{\rm B}T)^2 \over m^2 c_1^4} 
        \int_0^\infty \! {x^d \over \sinh^2 x} dx \right] \ ,
 \label{rspidp} \\
        &&{\rm o.b.c.}\, :\nonumber \\ &&\overline{\Delta n_\Delta(T)} 
        \approx - {\Delta n_0 (k_{\rm B} T)^{d-2} \over 
                8 \pi^{d/2} \Gamma(d/2) \hbar^d c_1^{d-1}} \left[ 
        {d-3 \over d-2} \int_0^\infty \!{x^{d-2} \over \cosh^2 x} \, dx
        + {(k_{\rm B}T)^2 \over m^2 c_1^4} \
        \int_0^\infty \! {x^d \over \cosh^2 x} \, dx \right] \ ,
 \label{rsoidp}
\end{eqnarray}
which for $d = 3$ yields
\begin{eqnarray}
        d = 3 \, : \quad 
        {\rm p.b.c.} \, : \quad \overline{\Delta n_\Delta(T)} &&=
                {\Delta n_0 k_{\rm B} T \over 4 \pi^2 \hbar^3 c_1^2} 
                                                + {\cal O}(T^3) \ ,
 \label{rsp2dp} \\
        {\rm o.b.c.} \, : \quad \overline{\Delta n_\Delta(T)} &&=
                                                - {\cal O}(T^3) \ ,
 \label{rso2dp}
\end{eqnarray}
while for $d > 3$
\begin{eqnarray}
        {\rm p.b.c.} \, : \quad \overline{\Delta n_\Delta(T)} &&=
        {(d-3) \Gamma(d-2) \zeta(d-2) \over 2^d \pi^{d/2} \Gamma(d/2)}
        \, {\Delta n_0 (k_{\rm B} T)^{d-2} \over \hbar^d c_1^{d-1}}\ ,
 \label{rspddp} \\
        {\rm o.b.c.} \, : \quad \overline{\Delta n_\Delta(T)} &&=
                                - \left( 1 - {1 \over 2^{d-3}} \right)
        {(d-3) \Gamma(d-2) \zeta(d-2) \over 2^d \pi^{d/2} \Gamma(d/2)}
        \, {\Delta n_0 (k_{\rm B} T)^{d-2} \over \hbar^d c_1^{d-1}}\ .
 \label{rsoddp}
\end{eqnarray}

In the same manner, we obtain for the normal--fluid density,
\begin{eqnarray}
        {\rm p.b.c.} \, : \quad \overline{\rho_{n \, \Delta}} &&= {n_0 
        \beta^2 \hbar^4 \over 8 d m} \int \! {d^dq \over (2 \pi)^d} \,
                        {\Delta q^3 \over \hbar \epsilon_{\rm B}(q)} 
                        {\cosh [\beta \epsilon_{\rm B}(q) / 2] \over 
                        \sinh^3 [\beta \epsilon_{\rm B}(q) / 2]} \ ,
 \label{rsdprn} \\
        {\rm o.b.c.} \, : \quad \overline{\rho_{n \,\Delta}} &&= -{n_0
        \beta^2 \hbar^4 \over 8 d m} \int \! {d^dq \over (2 \pi)^d} \,
                        {\Delta q^3 \over \hbar \epsilon_{\rm B}(q)} 
                        {1 \over \sinh [\beta \epsilon_{\rm B}(q)]} \ ,
 \label{rsdorn}
\end{eqnarray}
which approaches zero as $\beta \hbar \to \infty$, as is shown to be
the case for any disorder correlator $\Delta({\bf q},\omega_m)$
independent of $\omega_m$ in App.~\ref{galtil}. For finite
temperatures or sample thickness,
\begin{eqnarray}
        {\rm p.b.c.} \, : \quad \overline{\rho_{n \, \Delta}(T)} &&= 
        {\Gamma(d+2) \zeta(d) \over 2^{d+1} \pi^{d/2} \Gamma(1+d/2)} 
        \, {\Delta n_0 (k_{\rm B} T)^d \over m \hbar^d c_1^{d+3}} \ ,
 \label{rspdrn} \\
        {\rm o.b.c.} \, : \quad \overline{\rho_{n \, \Delta}(T)} &&= 
                                - \left( 1 - {1 \over 2^{d+2}} \right)
        {\Gamma(d+2) \zeta(d+2) \over 2^{d+2} \pi^{d/2} \Gamma(1+d/2)} 
        \, {\Delta n_0 (k_{\rm B} T)^d \over m \hbar^d c_1^{d+3}} \ ;
 \label{rsodrn}
\end{eqnarray}
in two dimensions this becomes
\begin{eqnarray}
        d = 2 \, : \quad        
        {\rm p.b.c.} \, : \quad \overline{\rho_{n \, \Delta}(T)} &&= 
        \pi \Delta n_0 (k_{\rm B} T)^2 / 8 m \hbar^2 c_1^5 \ ,
 \label{rsp2rn} \\
        {\rm o.b.c.} \, : \quad \overline{\rho_{n \, \Delta}(T)} &&= 
        - \pi^3 \Delta n_0 (k_{\rm B} T)^2 / 256 m \hbar^2 c_1^5 \ ,
 \label{rso2rn}
\end{eqnarray}
and for $d=3$
\begin{eqnarray}
        d = 3 \, : \quad 
        {\rm p.b.c.} \, : \quad \overline{\rho_{n \, \Delta}(T)} &&= 
        2\zeta(3) \Delta n_0 (k_{\rm B} T)^3 /\pi^2 m\hbar^3 c_1^6 \ ,
 \label{rsp3rn} \\
        {\rm o.b.c.} \, : \quad \overline{\rho_{n \, \Delta}(T)} &&= -
        31\zeta(5)\Delta n_0(k_{\rm B}T)^3 /32\pi^2 m\hbar^3 c_1^6 \ .
 \label{rso3rn}
\end{eqnarray}
We see that {\it only} for $d=3$ and with open boundary conditions
there exists a simple relation between the finite--size corrections to
the disorder renormalization of the depletion and the normal--fluid
density, namely
\begin{equation}
        d = 3 \, , \; {\rm o.b.c.} \, : \quad 
        \overline{\rho_{n \, \Delta}(T)} = [31 \zeta(5) / 9 \zeta(3)] 
                                \, m \overline{\Delta n_\Delta(T)} \ .
 \label{rso3dr}
\end{equation}

Finally, we remark that in the above calculations we have not taken
into account the possibility of a {\it direct} coupling of the defects
to the tangent field (i.e., the momentum or current density in the
boson language), but have exclusively studied the effects of disorder
which couples to the particle density, i.e., changes the chemical
potential locally. The latter mechanism should remain the most
important effect of the disorder even in the case of splayed
correlated defects; however, it is conceivable that in addition a term
of the form ${\bf t}_D({\bf r},\tau)\cdot{\bf g}({\bf r},\tau)$, see
Eq.~(\ref{momden}), with ${\bf t}_D({\bf r},\tau)$ representing
another quenched random variable, may have to be included at least in
an ``effective'' theory for which one further step of some
course--graining procedure has been carried out. We neglect this
scenario here, as well as in Secs.~\ref{cortlt} and \ref{twofam} 
below.

\subsection{Correlated disorder -- parallel untilted extended defects}
 \label{cordis}

We now turn to the analog of static disorder for bosons, namely
untilted extended defects in flux liquids. We can use the correlator
(\ref{boedcr}), with the understanding that $d_\parallel = 0$
describes point defects in the boson representation, i.e., linear
(columnar) defects in the flux line picture, and similarly
$d_\parallel = 1$ represents defect lines / planes, respectively,
etc. The results may then simply be expressed in terms of the
transverse dimensionality $d_\perp = d - d_\parallel$.

The disorder contribution to the static structure factor with periodic
[see Eq.~(\ref{doedst})] and open boundary conditions 
[cf.~(\ref{opdost})] read
\begin{eqnarray}
        {\rm p.b.c.} \, : \quad \overline{S_\Delta({\bf q})} &&= 
        \Delta \left( {n_0 \hbar^2 q_\perp^2 \over 
                        m \epsilon_{\rm B}(q_\perp)^2} \right)^2 
                (2 \pi)^{d_\parallel} \delta({\bf q}_\parallel) \ ,
 \label{exdpst} \\
        {\rm o.b.c.} \, : \quad \overline{S_\Delta({\bf q})} &&= 
        \Delta \left( {n_0 \hbar^2 q_\perp^2 \over 
                        m \epsilon_{\rm B}(q_\perp)^2} \right)^2 
        \left( 1 - {1 \over \cosh [\beta \epsilon_{\rm B}(q_\perp)/2]}
        \right)^2 (2 \pi)^{d_\parallel} \delta({\bf q}_\parallel) \ ,
 \label{exdost}
\end{eqnarray}
whereas the leading part of the current (tilt) correlators vanish in
this case in the static limit.

The results for the depletion and the normal--fluid density in the
limit $\beta \hbar \to \infty$ have already been explicitly given in 
Secs.~\ref{pbcofr} and \ref{pbexdo}, where we had also seen that for
periodic boundary conditions the finite--temperature / sample
thickness corrections vanish entirely, 
\begin{equation}
        {\rm p.b.c.} \, : \quad \overline{\Delta n_\Delta(T)} = 0 \, ,
                \quad \overline{\rho_{n \, ij \, \Delta}(T)} = 0 \ .
 \label{exdpdp}
\end{equation}
However, for open boundary conditions there are finite--size
corrections, induced by the stronger thermal fluctuations in the
vicinity of the sample surfaces. Eq.~(\ref{dosddp}) yields (in the
sample center, $\tau = \beta \hbar / 2$)
\begin{equation}
        {\rm o.b.c.} \, : \; \overline{\Delta n_\Delta(T)} = 
        - \Delta \, {n_0 \over 2 m^2} 
        \int \! {d^{d_\perp} q_\perp \over (2 \pi)^{d_\perp}} \, 
        \left({\hbar q_\perp \over \epsilon_{\rm B}(q_\perp)}\right)^4
        \left( {1 \over \cosh [\beta \epsilon_{\rm B}(q)/2]} -
        {1 \over 2 \cosh^2 [\beta \epsilon_{\rm B}(q)/2]} \right) \ ,
 \label{exdodp}
\end{equation}
which can be evaluated in the phonon approximation using
Eqs.~(\ref{integ6}) and (\ref{integ4}) of App.~\ref{momint}, where
also the function $\xi(d)$ is defined,
\begin{equation}
        {\rm o.b.c.} \, : \; \overline{\Delta n_\Delta(T)} = 
        - {2 \Gamma(d_\perp) \over \pi^{d_\perp/2} \Gamma(d_\perp/2)}
        \left[ \xi(d_\perp) -2^{-d_\perp}\left( 1-2^{2-d_\perp}\right) 
        \zeta(d_\perp-1) \right] {\Delta n_0 (k_{\rm B} T)^{d_\perp} 
                        \over m^2 \hbar^{d_\perp} c_1^{d_\perp+4}} \ ;
 \label{exoddp}
\end{equation}
for the relevant transverse dimensionalities this reduces to 
[$\xi(2) = G \approx 0.916$ denotes Catalan's constant]
\begin{eqnarray}
        d_\perp = 1 \, : \quad \overline{\Delta n_\Delta(T)} &&= 
        - (\pi - 1) \Delta n_0 k_{\rm B} T / 2 \pi m^2 \hbar c_1^5 \ ,
 \label{exo1dp} \\
        d_\perp = 2 \, : \quad \overline{\Delta n_\Delta(T)} &&= - 
        (4G-\ln 2) \Delta n_0 (k_{\rm B}T)^2/2\pi m^2 \hbar^2 c_1^6\ ,
 \label{exo2dp} \\
        d_\perp = 3 \, : \quad \overline{\Delta n_\Delta(T)} &&= - 
        (3\pi-1) \Delta n_0 (k_{\rm B}T)^3 / 12 m^2 \hbar^3 c_1^7 \ .
 \label{exo3dp}
\end{eqnarray}

The normal--fluid density along the direction of the extended defects
remains unaffected by the disorder,
\begin{equation}
        {\rm o.b.c.} \, : \quad 
        \overline{\rho_{n \, \parallel \, \Delta}(T)} = 0 \ .
 \label{exparn}
\end{equation}
In the perpendicular directions, Eq.~(\ref{dosdrn}) leads to
\begin{eqnarray}
        {\rm o.b.c.} \, : \quad 
        \overline{\rho_{n \, \perp \,\Delta}(T)} &&= 
        - {\Delta n_0 \over d_\perp m} \int \! 
        {d^{d_\perp} q_\perp \over (2 \pi)^{d_\perp}} \left( 
        {\hbar q_\perp \over \epsilon_{\rm B}(q_\perp)} \right)^4
        \times \nonumber \\ &&\qquad \times 
        \left( {1 \over \cosh [\beta \epsilon_{\rm B}(q_\perp)/2]} +
                {\beta \epsilon_{\rm B}(q_\perp) / 2 \over 
                        \sinh [\beta \epsilon_{\rm B}(q_\perp) / 2]} -
                {\beta \epsilon_{\rm B}(q_\perp) \over 
                \sinh [\beta \epsilon_{\rm B}(q_\perp)]} \right) \ ,
 \label{exdirn}
\end{eqnarray}
which in the phonon approximation is readily evaluated using
Eqs.~(\ref{integ5}) and (\ref{integ6}),
\begin{eqnarray}
        \overline{\rho_{n \, \perp \,\Delta}(T)} &&= -
        {2 \Gamma(d_\perp) \over \pi^{d_\perp/2} \Gamma(1+d_\perp/2)}
        \left[ \xi(d_\perp) + \left( 1 - 2^{-d_\perp} \right) \left( 
        1 - 2^{-(d_\perp+1)} \right) d_\perp \zeta(d_\perp+1) \right]
        \times \nonumber\\ &&\qquad \qquad \qquad \qquad \qquad \times
        {\Delta n_0 (k_{\rm B} T)^{d_\perp} \over 
                                m \hbar^{d_\perp} c_1^{d_\perp+4}} \ .
 \label{exodrn}
\end{eqnarray}
Explicitly one gets
\begin{eqnarray}
        d_\perp = 1 \, : \quad 
        \overline{\rho_{n \, \perp \, \Delta}(T)} &&= 
        - (\pi+4) \Delta n_0 k_{\rm B} T / 4 m \hbar c_1^5 \nonumber\\ 
        &&= [\pi(\pi+4)/2(\pi-1)] \, m \overline{\Delta n_\Delta(T)}\ ,
 \label{exo1rn} \\
        d_\perp = 2 \, : \quad 
        \overline{\rho_{n \, \perp \, \Delta}(T)} &&= 
        - \left[ 16 G + 21 \zeta(3) \right] \Delta n_0 (k_{\rm B} T)^2
        / 8 \pi m \hbar^2 c_1^6 \nonumber \\
        &&= [\left( 16 G + 21 \zeta(3) \right) / 
        (16 G - 4 \ln 2)] \, m \overline{\Delta n_\Delta(T)} \ ,
 \label{exo2rn} \\
        d_\perp = 3 \, : \quad 
        \overline{\rho_{n \, \perp \, \Delta}(T)} &&= - \pi (7\pi+8) 
        \Delta n_0 (k_{\rm B} T)^3 / 48 m \hbar^3 c_1^7 \nonumber \\
        &&=[\pi(7\pi+8)/4(3\pi-1)]\, m\overline{\Delta n_\Delta(T)}\ ;
 \label{exo3rn}
\end{eqnarray}
thus the finite--temperature / size corrections to the disorder
renormalization of the depletion and the normal--fluid density are
proportional to each other for these parallel untilted extended
defects, albeit with a complicated numerical factor.

\subsection{Correlated disorder -- parallel tilted extended defects}
 \label{cortlt}

More generally, we can also discuss ``moving'' or tilted parallel
extended defects, say, along the x direction; the corresponding
disorder correlator then becomes
\begin{eqnarray}
        \Delta({\bf r},\tau;{\bf r}',\tau') &&= \Delta \, \delta
        \Bigl( {\bf r}_\perp-{\bf r}_\perp'+v(\tau-\tau'){\bf e}_x
        \Bigr) \ , \nonumber \\ \Delta({\bf q},\omega_m) &&= 
        \Delta \, (2 \pi)^{d_\parallel} \delta({\bf q}_\parallel) \, 
        \beta \hbar \delta_{\omega_m,q_x v} \ ,
 \label{tedcor}
\end{eqnarray}
generalizing Eq.~(\ref{boedcr}). Recall that ${\bf r_\perp}$ refers to
directions perpendicular to a set of $d_\parallel$--dimensional
defects; $(d_\parallel,d_\perp) = (1,d-1)$ means line defects, while
$(d_\parallel,d_\perp) = (2,d-2)$ corresponds to planar disorder. For
magnetic flux liquids, the tilt parameter is simply the tangent of the
tilt angle, $v = \tan \alpha$, see also Eq.~(\ref{flplds}) in
App.~\ref{galtil}. We shall restrict ourselves for simplicity to thick
samples in this and the following subsection. If one wanted to apply
periodic boundary conditions, one would have to keep the transverse
momenta discrete as well, $q_{x,k} = 2 \pi k / L_\perp$ and ensure
that $\beta \hbar v / L_\perp = m / k$ is a rational number.

In the limit $\beta \hbar \to \infty$, Eqs.~(\ref{dostst}) and
(\ref{domsst}) yield the static structure factor
\begin{equation}
        \overline{S({\bf q})} = 
        {n_0 \hbar^2 q^2 \over 2 m \epsilon_{\rm B}(q)}
        + \Delta \left( {n_0 q_\perp^2 / m \over q_x^2 v^2 
                + \epsilon_{\rm B}(q_\perp)^2 / \hbar^2} \right)^2 
                (2 \pi)^{d_\parallel} \delta({\bf q}_\parallel) \ ,
 \label{texdst}
\end{equation}
while the leading contribution to the static current correlation
function [see, e.g., Eq.~(\ref{domscr})] reads
\begin{equation}
        \overline{C_{ij}({\bf q})} = 
        \left[ {n_0 m \epsilon_{\rm B}(q) \over 2} 
        - \Delta \left( {n_0 q_\perp q_x v \over q_x^2 v^2 
                + \epsilon_{\rm B}(q_\perp)^2 / \hbar^2} \right)^2 
        (2 \pi)^{d_\parallel} \delta({\bf q}_\parallel) \right]
                                                P_{ij}^L({\bf q}) \ .
 \label{tlexst}
\end{equation}
For $v \to 0$, these results of course reduce to those of the previous
subsection. In real time, the density and phase correlation functions
display singularities when $v \to c_1$, i.e., when the defects move at
the speed of sound of the superfluid. 

The disorder contribution to the depletion (\ref{trdepl}) becomes
\begin{equation}
        \overline{n_\Delta} = \Delta \, {n_0 \over \hbar^2} 
        \int \! {d^{d_\perp} q_\perp \over (2 \pi)^{d_\perp}} \, 
        {(\hbar q_\perp^2 / 2 m)^2 - q_x^2 v^2 \over 
        [\epsilon_{\rm B}(q_\perp)^2 / \hbar^2 + q_x^2 v^2]^2} \ ;
 \label{tlexdp}
\end{equation}
this can be easily seen to diverge for any nonzero tilt $v$ in the
cases $d_\perp = 1$ and $d_\perp = 2$, which we again interpret as
indicating that the world lines are ``dragged'' along by the tilted
extended defects, leading to superdiffusive behavior. Only for $v=0$
does the depletion remain finite, see Eq.~(\ref{exoddp}). For 
$d_\perp = 3$ (point disorder in a three--dimensional superfluid
moving at constant speed $v$) one finds for delta--function
interaction with strength $V_0 = 4 \pi a / m$,
\begin{equation}
        d_\perp = 3 \, : \quad \overline{n_\Delta} =
        {\Delta m^2 n_0^{1/2} \over 8 \pi^{3/2} \hbar^3 a^{1/2}} \,
        {1 \over \left( 1 + m^2 v^2 / 4 \pi n_0 a \right)^{1/2}} \ .
 \label{tex3dp}
\end{equation}

The disorder renormalization of the normal--fluid density and tilt
modulus perpendicular to the extended defects can be inferred from
Eq.~(\ref{dotilt}), see also App.~\ref{galtil},
\begin{equation}
        \overline{\rho_{n \,\perp i \,\Delta}} = \Delta {n_0 \over m}  
        \int \! {d^{d_\perp} q_\perp \over (2 \pi)^{d_\perp}} \,
        q_\perp^2 q_{\perp i}^2 \, 
        {\epsilon_{\rm B}(q_\perp)^2 / \hbar^2 - 3 q_x^2 v^2 \over 
        [\epsilon_{\rm B}(q_\perp)^2 / \hbar^2 + q_x^2 v^2]^3} \ ,
 \label{ttexrn}
\end{equation}
while $\overline{\rho_{n \, \parallel \, \Delta}} = 0$, of course. In
the case $d_\perp = 1$, there is only one perpendicular direction, and
\begin{equation}
        d_\perp = 1 \, : \quad \overline{\rho_{n \, \perp \, \Delta}}
        = {\Delta m^3 \over 16 \pi^{3/2} \hbar n_0^{1/2} a^{3/2}} \,
        {1 - m^2 v^2 / 2 \pi n_0 a \over 
                \left( 1 + m^2 v^2 / 4 \pi n_0 a \right)^{5/2}} \ ;
 \label{tex1rn}
\end{equation}
for $d_\perp = 2$, on the other hand, we have to distinguish between
the direction along the tilt and perpendicular to it,
\begin{eqnarray}
        d_\perp = 2 \, : \quad\overline{\rho_{n \, \perp x \, \Delta}}
        &&= {\Delta m n_0 \over \pi \hbar^2 v^2}
        \left[ {1 + m^2 v^2 / 2 \pi n_0 a \over 
        \left( 1 + m^2 v^2 / 4\pi n_0 a \right)^{3/2}} - 1 \right] \ , 
 \label{tex2rn} \\
\overline{\rho_{n \, \perp y \, \Delta}} &&= 
        {\Delta m n_0 \over \pi \hbar^2 v^2} \left[ 1 - {1 \over 
        \left( 1 + m^2 v^2 / 4 \pi n_0 a \right)^{1/2}} \right] \ ,
 \label{tey2rn}
\end{eqnarray}
and finally, for $d_\perp = 3$ one gets
\begin{eqnarray}
        d_\perp = 3 \, : \quad \overline{\rho_{n \, \perp x \,\Delta}}
        &&= {4 \Delta n_0^2 a \over \hbar^3 v^3} \left[ {\rm arsinh}
        \, {mv \over (4 \pi n_0 a)^{1/2}} - {m v / (4 \pi n_0 a)^{1/2}
        \over \left( 1 + m^2 v^2/4\pi n_0 a \right)^{1/2}} \right] \ ,
 \label{tex3rn} \\
\overline{\rho_{n \, \perp y \, \Delta}} &&= 
        {2 \Delta n_0^2 a \over \hbar^3 v^3} 
        \left[ {m v \over (4 \pi n_0 a)^{1/2}}
                \left(1 + {m^2 v^2 \over 4 \pi n_0 a} \right)^{1/2} 
        - {\rm arsinh} \, {mv \over (4 \pi n_0 a)^{1/2}} \right] \ .
 \label{tey3rn}
\end{eqnarray}
These results show that for general nonzero tilts quite complicated
formulas arise; and it is only in the very special situation $v = 0$
that simple relations of the form 
$\overline{\rho_{n \, \perp \, \Delta}} \propto \overline{n_\Delta}$
emerge.

It is interesting to compute the fraction of particles / lines
$\overline{n_v}$ that are actually ``dragged'' along by the defects,
\begin{equation}
        m v {\overline n_v} = {1 \over \beta \hbar \Omega} \Big| 
        \langle g_x({\bf q} = {\bf 0},\omega_m  = 0) \rangle \Big| \ .
 \label{dragln}
\end{equation}
With Eq.~(\ref{traped}) we find
\begin{equation}
        {\overline n_v} = \Delta {n_0 \over m^2} 
                \int {d^{d_\perp}q_\perp \over (2 \pi)^{d_\perp}} 
        \left( q_\perp q_x \over \epsilon_{\rm B}(q_\perp)^2/\hbar^2 
                                        + q_x^2 v^2 \right)^2 \ .
 \label{pinned}
\end{equation}
For $v \to 0$ this precisely coincides with the disorder--induced
normal--fluid density [Eqs.~(\ref{doedfd}),(\ref{ednrnr})]; for static
disorder, the superfluid density is thus reduced by exactly the
average fraction of particles that are pinned to the defects.

\subsection{Two families of symmetrically tilted extended defects}
 \label{twofam}

As a final step of generalization, we now introduce two families of
extended defects, which are symmetrically ``moving'' (tilted) along
the $x$ direction \cite{hwaled}. The disorder correlator in this
situation reads
\begin{eqnarray}
        \Delta({\bf r},\tau;{\bf r}',\tau') &&= {\Delta \over 4} \, 
        \Bigl[ \delta \bigl( {\bf r}_\perp-{\bf r}_\perp' + 
        v (\tau-\tau') {\bf e}_x \bigr) + \delta \bigl( 
        {\bf r}_\perp-{\bf r}_\perp' - v (\tau-\tau') {\bf e}_x \bigr) 
        \nonumber \\ &&\qquad \qquad
        + \delta \bigl( {\bf r}_\perp-{\bf r}_\perp' + v (\tau+\tau')
        {\bf e}_x \bigr) + \delta \bigl( {\bf r}_\perp-{\bf r}_\perp' 
        - v (\tau+\tau') {\bf e}_x \bigr) \Bigr] \ , \nonumber \\
        \Delta({\bf q},\omega_m;{\bf q}',\omega_{m'}) &&= 
        {\Delta \over 4} \, (2 \pi)^d \delta({\bf q}+{\bf q}') \, 
        (2 \pi)^{d_\parallel} \delta({\bf q}_\parallel) \times
        \nonumber \\ &&\qquad \qquad \times (\beta \hbar)^2 
        \Bigl( \delta_{m,-m'} + \delta_{m,m'} \Bigr) \Bigl( 
        \delta_{\omega_m,q_x v} + \delta_{\omega_m,-q_x v} \Bigr) \ .
 \label{tfedcr}
\end{eqnarray}
It is important to realize that translational invariance along the
$\tau$ direction is broken here, which originates in the fact that
there is a strong correlation between defects ``moving'' in the
positive and negative $x$ directions; or, equivalently, of defects
``moving'' forward and backward in imaginary time. As a consequence,
we have to resort to our most general formulas in Secs.~\ref{disbos}
and \ref{disfll}, and Eq.~(\ref{afistm}), for example, is not
applicable; but again, we shall only discuss the limit 
$\beta \hbar \to \infty$ here. Note that in the limit $v \to 0$, the
results of Sec.~\ref{cordis} for untilted parallel defects are
recovered.

Interestingly, with this correlator the static structure function, see
Eq.~(\ref{dopstf}), for example, becomes identical with the result
(\ref{texdst}) for a single family of tilted extended disorder. This
must be the case, as in a fixed cross section at $\tau$, those two
cases look the same. However, in the leading contribution to the
current correlation function, the effects of the two oppositely
``moving'' defect families precisely cancel each other, leaving behind
just the pure result (this is actually true for finite $\beta \hbar$
as well, with any boundary condition applied). This implies that in
the Green's function, from which the depletion and the normal--fluid
density are to be determined, we shall face a partial cancellation of
the disorder influence, namely in those contributions stemming from
the phase correlations, respectively.

Again because of cancellations, the disorder--induced depletion,
obtained from Eq.~(\ref{dodbde}), turns out to contain only the first
term in the numerator of Eq.~(\ref{tlexdp}), 
\begin{equation}
        \overline{n_\Delta} = \Delta \, {n_0 \over 4 m^2} 
        \int \! {d^{d_\perp} q_\perp \over (2 \pi)^{d_\perp}} \, 
        \left( {q_\perp^2 \over 
        \epsilon_{\rm B}(q_\perp)^2/\hbar^2 + q_x^2 v^2} \right)^2 \ ;
 \label{tfexdp}
\end{equation}
i.e., the leading contribution (in a long--wavelength expansion),
which had caused the divergences discussed in the previous subsection,
actually has disappeared. Therefore the depletion remains finite even
for nonzero tilt. The explicit results read
\begin{eqnarray}
        d_\perp = 1 \, : \quad \overline{n_\Delta} &&= 
        {\Delta m^2 \over 64 \pi^{3/2} \hbar n_0^{1/2} a^{3/2}} \, 
        {1 \over \left( 1 + m^2 v^2 / 4 \pi n_0 a \right)^{3/2}} \ ,
 \label{tfe1dp} \\
        d_\perp = 2 \, : \quad \overline{n_\Delta} &&= 
        {\Delta m^2 \over 16 \pi^2 \hbar^2 a} \, 
        {1 \over \left( 1 + m^2 v^2 / 4 \pi n_0 a \right)^{1/2}} \ ,
 \label{tfe2dp} \\
        d_\perp = 3 \, : \quad \overline{n_\Delta} &&= 
        {\Delta m n_0 \over 4 \pi \hbar^3 v} \, 
        {\rm arsinh} \, {m v \over (4 \pi n_0 a)^{1/2}} \ .
 \label{tfe3dp} 
\end{eqnarray}

In a similar manner, the disorder renormalization of the normal--fluid
density is reduced, and Eq.~(\ref{dodbnf}) leads to
\begin{equation}
        \overline{\rho_{n \,\perp i \,\Delta}} = \Delta {n_0 \over m}  
        \int \! {d^{d_\perp} q_\perp \over (2 \pi)^{d_\perp}} \,
        q_\perp^2 q_{\perp i}^2 \, 
        {\epsilon_{\rm B}(q_\perp)^2 / \hbar^2 - q_x^2 v^2 \over 
        [\epsilon_{\rm B}(q_\perp)^2 / \hbar^2 + q_x^2 v^2]^3} \ ,
 \label{tfexrn}
\end{equation}
to be contrasted with Eq.~(\ref{ttexrn}). For $d_\perp = 1$ this
becomes 
\begin{equation}
        d_\perp = 1 \, : \quad \overline{\rho_{n \, \perp \,\Delta}}
        = {\Delta m^3 \over 16 \pi^{3/2} \hbar n_0^{1/2} a^{3/2}} \, 
        {1 - m^2 v^2 / 8 \pi n_0 a \over 
                \left( 1 + m^2 v^2 / 4 \pi n_0 a \right)^{5/2}} =
        4 m \overline{n_\Delta} \, {1 - m^2 v^2 / 8 \pi n_0 a 
                                \over 1 + m^2 v^2 / 4 \pi n_0 a} \ .
 \label{tfe1rn} 
\end{equation}
For $d_\perp = 2$ we find
\begin{eqnarray}
        d_\perp = 2 \, : \quad \overline{\rho_{n \, \perp x \, \Delta}}
        &&= {\Delta m^3 \over 8 \pi^2 \hbar^2 a} \, 
        {1 \over \left( 1 + m^2 v^2 / 4 \pi n_0 a \right)^{3/2}} 
        = {2 m \overline{n_\Delta} \over 1 + m^2 v^2 / 4 \pi n_0 a}  \ ,
 \label{tfx2rn} \\
        \overline{\rho_{n \, \perp y \, \Delta}}
        &&= {\Delta m^3 \over 8 \pi^2 \hbar^2 a} \, 
        {1 \over \left( 1 + m^2 v^2 / 4 \pi n_0 a \right)^{1/2}} 
        = 2 m \overline{n_\Delta} \ ;
 \label{tfy2rn}
\end{eqnarray}
and finally in three ``transverse'' dimensions,
\begin{eqnarray}
        &&d_\perp = 3\, : \quad\overline{\rho_{n \, \perp x \,\Delta}}
        = {\Delta n_0^2 a \over \hbar^3 v^3} \left[
        {\rm arsinh} \, {m v \over (4 \pi n_0 a)^{1/2}} - 
\bigskip
        {m v \over (4 \pi n_0 a)^{1/2}} \, {1 - m^2 v^2 / 4 \pi n_0 a
        \over \left( 1 + m^2 v^2/4 \pi n_0 a \right)^{1/2}} \right]\ ,
 \label{tfx3rn} \\
        &&\quad \overline{\rho_{n \, \perp y \, \Delta}}
                = {\Delta n_0^2 a \over 4 \hbar^3 v^3} 
        \left[ {m v \over (4 \pi n_0 a)^{1/2}} 
        \left( 1 + {m^2 v^2 \over 4 \pi n_0 a} \right)^{1/2} - 
        \left( 1 - {m^2 v^2 \over 2 \pi n_0 a} \right) 
        {\rm arsinh} \, {m v \over (4 \pi n_0 a)^{1/2}} \right] \ .
 \label{tfy3rn} 
\end{eqnarray}

\subsection{Experimental consequences for vortex physics}
 \label{expcon}

We conclude by translating the results of the previous subsections
into the flux line language, specializing to $d=2$ dimensions, and
discussing some of the experimental consequences of our results. As
noted in Sec.~\ref{flines}, we approximate the inter--vortex repulsion
by $V(q) = V_0 / (1 + \lambda^2 q^2)$, with $V_0 = \phi_0^2 / 4 \pi$
and the magnetic flux quantum $\phi_0 = h c / 2 e$. As shown in
Table~\ref{analog}, the particle mass is to be replaced by the
effective line tension 
${\tilde \epsilon}_1 = (m_\perp / m_z) \epsilon_0 \ln (\lambda / \xi)$,
where the energy scale is set by 
$\epsilon_0 = (\phi_0 / 4 \pi \lambda)^2$; we shall also need the
scattering length 
\begin{equation}
        a = {\tilde \epsilon}_1 V_0 / 4 \pi = 
                {\tilde \epsilon}_1 (\phi_0 / 4 \pi)^2 \ , 
 \label{scalen}
\end{equation}
and the characteristic velocity 
\begin{equation}
        c_1 = (n_0 V_0 / {\tilde \epsilon}_1)^{1/2} = 
                (n_0 \phi_0^2 / 4 \pi {\tilde \epsilon}_1)^{1/2} \ .
 \label{souspe}
\end{equation}
With these substitutions, and the replacements $\hbar \to k_{\rm B} T$
and $\beta \hbar \to L$, the structure factor and other correlation
functions are readily obtained from our results in the previous
sections. In the following, we shall review the explicit consequences
for the fraction of entangled lines as well as the renormalization of
the tilt modulus by disorder and through modification of the boundary
conditions.

For a {\it pure system}, we found in Sec.~\ref{pspebc} that the
mean ``boson'' order parameter squared 
$n_0 = |\langle \psi \rangle|^2$ as function of temperature in the
limit of thick samples $L \to \infty$ is 
\begin{equation}
        n_0(T) = n \left[ 1 + a / ( k_{\rm B} T)^2 \right]^{-1}
        = n \left[ 1 + {\tilde \epsilon}_1 
                (\phi_0 / 4 \pi k_{\rm B} T)^2 \right]^{-1} \ .
 \label{entaln}
\end{equation}
$\langle \psi \rangle$ measures the degree of entanglement, which
increases with increasing temperature. The vortex contribution to the
tilt modulus [see Eq.~(\ref{rentil})] is 
${c^v_{44}}^{-1} = (n_0 {\tilde \epsilon}_1)^{-1}$. The leading
finite--size correction for sample thicknesses $L < \infty$ to the
fraction of disentangled lines diverges logarithmically, if 
{\it periodic} boundary conditions are employed along the
magnetic--field direction, implying that in fact the ``boson'' order
parameter vanishes in this case. On the other hand, for {\it open}
boundary conditions, which are more realistic for most samples, the
reduction in the ``boson'' order parameter is finite, 
cf. Eq.~(\ref{2opdep}),
\begin{equation}
        \Delta n(L) = - (\ln 2) {\tilde \epsilon}_1 
                                        / 2 \pi k_{\rm B} T L \ .
 \label{fsentl}
\end{equation}
To leading order in the sample thickness corrections,
Eqs.~(\ref{2Tnfld}) and (\ref{2opfld}) yield for the tilt modulus with
periodic and open boundary conditions, respectively,
\begin{equation}
        {c^v_{44}}^{-1}(L) = (n {\tilde \epsilon}_1)^{-1} 
        \left[ 1 \mp 3 \zeta(3) k_{\rm B} T / 2 \pi n
        {\tilde \epsilon}_1 c_1^4 L^3 \right] \quad \left\{ \quad
        \begin{array}{c} {\rm p.b.c.} \\ {\rm o.b.c.} \end{array} 
                                                        \right. \ ,
 \label{fstilt}
\end{equation}
with $\zeta(3) \approx 1.202$. As already discussed, the different
signs for the finite--size corrections reflect the fact that for open
boundary conditions thermal flux line wandering is effectively
enhanced near the sample surfaces, and therefore both entanglement and
response to external tilt are facilitated; the firm restrictions of
periodic boundary conditions, on the other hand, reduce the number of
entangled lines, and stiffen the system towards external tilting. 
According to Eq.~(\ref{anisot}), any renormalization of the tilt
modulus may also be viewed as a changing the mass anisotropy 
$m_\perp / m_z < 1$, with an increasing ${c^v_{44}}^{-1}$ implying a
smaller effective mass ratio and vice versa. Any enhancement of the
normal--fluid density, and correspondingly a reduction of
${c^v_{44}}^{-1}$, either through finite--size effects as in
Eq.~(\ref{fstilt}) for periodic boundary conditions, or through the
influence of (correlated) disorder, will therefore render the
superconducting system effectively more three--dimensional.

Uncorrelated {\it point defects}, e.g., oxygen vacancies in the
copper--oxide planes, {\it increase} the fraction of entangled
lines (hence the minus sign) by [Eq.~(\ref{ptd2dp})]
\begin{equation}
        \overline{n_\Delta} = - \Delta n_0 {\tilde \epsilon}_1 
                                        / 4 \pi (k_{\rm B} T)^3 \ ,
 \label{ptentf}
\end{equation}
clearly as a result of {\it enhanced} wandering of each flux line
searching for an optimal path through the sample while trying to
accomodate most profitably with the uncorrelated pinning centers. In
very thick samples, the tilt modulus, however, remains unaffected by
this kind of isotropically distributed disorder. The leading
finite--thickness corrections to the fraction of pinned vortices
remarkably display {\it different} power--law thickness dependences
for periodic and open boundary conditions, respectively, see
Eqs.~(\ref{ptp2dp}) and (\ref{pto2dp}),
\begin{equation}
        \overline{\Delta n_\Delta(L)} = \left\{ \begin{array}{ll} 
        \Delta n_0 / 4 \pi c_1^2 (k_{\rm B} T)^2 L & $\qquad$
                                                {\rm p.b.c.} \\
        - 9 \zeta(3) \Delta n_0 / 32 \pi {\tilde \epsilon}_1^2 
        c_1^6 L^3 & $\qquad$ {\rm o.b.c.} \end{array} \right. \ ;
 \label{ptfsen}
\end{equation}
the corresponding expressions for the tilt modulus read
\begin{equation}
        \overline{{c^v_{44}}^{-1}(L)} = \left\{ \begin{array}{ll} 
        (n {\tilde \epsilon}_1)^{-1} \left[ 1 - 3 \zeta(3) \Delta / 
        \pi {\tilde \epsilon}_1^2 c_1^6 L^3 \right] & $\qquad$
                                                {\rm p.b.c.} \\ 
        (n {\tilde \epsilon}_1)^{-1} \left[ 1 + {\cal O}(1/L^3)
        \right] & $\qquad$ {\rm o.b.c.} \end{array} \right. \ .
 \label{ptfstm}
\end{equation}

For an ensemble of {\it randomly tilted linear defects}, i.e.,
nearly isotropic splay, we found that the disorder contribution to
the fraction of entangled lines actually diverges
[Eq.~(\ref{rsdedp})], because the thermal flux line wandering becomes
superdiffusive as the vortices are ``convected'' along with the tilted
defects. However, in thick samples the tilt modulus is unaffected by
this isotropic disorder, and only in a finite sample does one get the
following tilt modulus renormalization,
\begin{equation}
        \overline{{c^v_{44}}^{-1}(L)} = \left\{ \begin{array}{ll} 
        (n {\tilde \epsilon}_1)^{-1} \left[ 1 - \pi \Delta / 
        8 {\tilde \epsilon}_1^2 c_1^5 L^2 \right] & $\qquad$
                                                {\rm p.b.c.} \\ 
        (n {\tilde \epsilon}_1)^{-1} \left[ 1 + \pi^3 \Delta / 
        256 {\tilde \epsilon}_1^2 c_1^5 L^2 \right] & $\qquad$
                                {\rm o.b.c.} \end{array} \right. \ .
 \label{rsfstm}
\end{equation}

The logarithmic divergence of the disorder--induced depletion 
$\overline{n_\Delta}$ also occurs for {\it parallel columnar
defects} ($d_\perp = 2$), as long as the tilt angle $\alpha$ remains
nonzero. (For this case, the divergence can presumably be removed by
redefining the time--like direction to account for the vortex
``drift''.) In the spetial situation $\alpha = 0$, Eq.~(\ref{dopdf2})
becomes
\begin{equation}
        \overline{n_\Delta} = 
        \Delta {\tilde \epsilon}_1^2 / 16 \pi^2 a (k_{\rm B} T)^2 \ ,
 \label{colden}
\end{equation}
and the corresponding finite--size corrections are nonzero only for
open boundary conditions and read, according to (\ref{exo2dp}),
\begin{equation}
        \overline{\Delta n_\Delta(L)} = - (4 G - \ln 2) \Delta n_0 /
                        2 \pi {\tilde \epsilon}_1^2 c_1^6 L^2 \ ,
 \label{cdfsen}
\end{equation}
with $G \approx 0.916$. For a tilt in the $x$ direction with angle
$\alpha > 0$, the non--vanishing tilt modulus tensor components in a
thick sample are [Eqs.~(\ref{tex2rn}), (\ref{tey2rn})]
\begin{eqnarray}
        \overline{{c^v_{xx}}^{-1}} &&= 
        {1 \over n {\tilde \epsilon}_1} \left[ 1 - {\Delta \over 
                \pi (k_{\rm B} T)^2 (\tan \alpha)^2} \left( {1 + 
                {\tilde \epsilon}_1^2 (\tan \alpha)^2 / 2 \pi n_0 a
                \over \left[ 1 + {\tilde \epsilon}_1^2 (\tan \alpha)^2
                / 4 \pi n_0 a \right]^{3/2}} - 1 \right) \right] \ ,
 \label{colxtm} \\
        \overline{{c^v_{yy}}^{-1}} &&= 
        {1 \over n {\tilde \epsilon}_1} \left[ 1 - {\Delta \over 
                \pi (k_{\rm B} T)^2 (\tan \alpha)^2} \left( 1 - 
        {1 \over \left[ 1 + {\tilde \epsilon}_1^2 (\tan \alpha)^2 / 
                        4 \pi n_0 a \right]^{1/2}} \right) \right] \ .
 \label{colytm}
\end{eqnarray}
We remark again that this reduction of the tilt modulus by the
correlated linear disorder can be reinterpreted as an increase of the
mass anisotropy ratio $m_\perp / m_z$; the vortices are thus
effectively rendered more three--dimensional by their interaction with
the columnar defects, as should be expected. This mechanism increases
the decoupling field in the presence of correlated disorder, above
which the vortices change from lines, as they have been treated
throughout this paper), to point--like ``pancakes''. For 
$\alpha \to 0$, both the expressions (\ref{colxtm}) and (\ref{colytm})
reduce to
\begin{equation}
        \overline{{c^v_{44}}^{-1}} = (n {\tilde \epsilon}_1)^{-1} 
                \left[ 1 - \Delta {\tilde \epsilon}_1^2 / 
                        8 \pi^2 n_0 a (k_{\rm B} T)^2 \right] \ ,
 \label{col0tm}
\end{equation}
which remains unchanged even for finite sample thickness in the case
of periodic boundary conditions (toroidal geometry), while we find for
open boundary conditions
\begin{equation}
        \overline{{c^v_{44}}^{-1}(L)} = (n {\tilde \epsilon}_1)^{-1}
        \left[ 1 - \Delta {\tilde \epsilon}_1^2 / 
        8 \pi^2 n_0 a (k_{\rm B} T)^2 + [16 G + 21 \zeta(3)] \Delta / 
                8 \pi {\tilde \epsilon}_1^2 c_1^6 L^2 \right] \ .
 \label{cdfstm}
\end{equation}
For the infinitely thick sample, Eq.~(\ref{col0tm}) leads to an
estimate for the Bose glass transition temperature from the
high--temperature flux liquid phase, complementary to the estimate in
Ref.~\cite{nelvin}. We note that the average disorder potential for 
columnar defects is $\overline{V_D} = - U_0 b^2 / d^2$, where 
$U_0 \approx \phi_0^2 / 2 (4 \pi \lambda)^2$ denotes the defect
potential well strength, $b$ its width, and $d$ the average distance
between the columns \cite{nelvin}. Accordingly, the defect correlator
becomes $\Delta_0 = U_0^2 b^4 / d^2$. One important modification has
to be taken into account near the localization transition (more
precisely, for temperatures above the depinning temperature $T_{dp}$),
however, namely the effective thermal renormalization of both the
disorder strength $U(T)$ and the replacement of one factor of $b^2$ by
the effective transverse localization length $l_\perp(T)$, i.e.:
\begin{equation}
        \Delta_0 \longrightarrow 
                        \Delta(T) = U(T)^2 l_\perp(T)^2 b^2 / d^2 \ ,
 \label{disren}
\end{equation}
where
\begin{equation}
        U(T) = U_0 (T^* / T)^2 \quad , \qquad
        l_\perp(T) = d (T / T^*)^2 \ ,
 \label{theren}
\end{equation}
and the characteristic temperature is given by
$k_{\rm B} T^* = b ({\tilde \epsilon}_1 U_0)^{1/2}$ \cite{nelvin}. 
Upon inserting these results, $n_0 = B / \phi_0$, and
Eq.~(\ref{scalen}) into Eq.~(\ref{col0tm}), and collecting terms, we
find that
\begin{equation}
        \overline{{c^v_{44}}^{-1}(T)} = (n {\tilde \epsilon}_1)^{-1}
        \Bigl[ 1 - \left( T_{\rm BG} / T \right)^4 \Bigr] \ .
 \label{tilins}
\end{equation}
So $c_{44}^v$ diverges as expected at the Bose glass transition
temperature \cite{nelvin},
\begin{equation}
        T_{\rm BG} = T^* [\phi_0 / (4 \pi \lambda)^2 B]^{1/4} \ ,
 \label{bgltem}
\end{equation}
in agreement with a similar analysis in Ref.~\cite{hwaled}. Although
one should not trust the exponent for the vanishing of
${c_{44}^v}^{-1}$ predicted by the approximate formula (\ref{tilins}),
this analysis suffices to estimate the transition temperature. Upon
generalizing to tilted linear disorder, and utilizing
Eqs.~(\ref{colxtm}) and (\ref{colytm}) we can locate the instabilities
for tilt in the $x$ and $y$ directions, respectively, as a function of
the tilt angle $\alpha$, or instead introducing the dimensionless
parameter 
\begin{equation}
        u(\alpha) = 
        {\tilde \epsilon}_1^2 (\tan \alpha)^2 / 4 \pi n_0 a \ ,
 \label{angpar}
\end{equation}
and find
\begin{eqnarray}
        \left[ {T_x(\alpha) \over T_{\rm BG}} \right]^4 &&= 
        {2 \over u(\alpha)} \left( {1+2 u(\alpha) \over 
                                [1+u(\alpha)]^{3/2}} - 1 \right) \ , 
 \label{cdxins} \\
        \left[ {T_y(\alpha) \over T_{\rm BG}} \right]^4 &&= 
        {2 \over u(\alpha)} 
                \left( 1 - {1 \over [1+u(\alpha)]^{1/2}} \right) \ .
 \label{cdyins}
\end{eqnarray}
The temperatures $T_x$ and $T_y$, where the instabilities occur, are
plotted in Fig.~\ref{cdtxty} as a function of the parameter $u$. 
Notice that while always $T_y(\alpha) > 0$, $T_x(\alpha_c) = 0$ for
$u_c = (1 + \sqrt{5})/2$, suggesting that the Bose glass remains
stable towards tilt up to this critical tilt angle $\alpha_c$. For 
$\alpha < \alpha_c$, the system will thus not respond to transverse
magnetic fields, i.e., display a {\it transverse Meissner effect}
\cite{nelvin}. The curve $T_x(\alpha)$ can therefore be viewed as
describing the phase boundary (localization transition line) in a
phase diagram with the thermodynamic variables temperature and
transverse magnetic field. However, within the Bogoliubov (Gaussian)
approximation used here, we do not find the cusp in $T_x(\alpha)$ that
is predicted by the Bose glass scaling theory \cite{nelvin}.

For the case of {\it two families of symmetrically tilted} (angles 
$\pm \alpha$) {\it columnar defects}, the fraction of pinned lines
remains finite because the leading logarithmic divergences for both
directions precisely cancel; Eq.~(\ref{tfe2dp}) yields
\begin{equation}
        \overline{n_\Delta} = \Delta {\tilde \epsilon}_1^2 / 
        16 \pi^2 a (k_{\rm B} T)^2 \left[ 1 + {\tilde \epsilon}_1^2 
                        (\tan \alpha)^2 / 4 \pi n_0 a \right]^{1/2} \ ;
 \label{stcden}
\end{equation}
and the tilt modulus components (for $L \to \infty$) deriving from 
Eqs.~(\ref{tfx2rn}) and (\ref{tfy2rn}) are
\begin{eqnarray}
        \overline{{c^v_{xx}}^{-1}} &&= 
        {1 \over n {\tilde \epsilon}_1} \left( 1 - {\Delta 
        {\tilde \epsilon}_1^2 \over 8 \pi^2 n_0 a (k_{\rm B} T)^2} \,
        {1 \over \left[ 1 + {\tilde \epsilon}_1^2 (\tan \alpha)^2 /
                        4 \pi n_0 a \right]^{3/2}} \right) \ ,
 \label{stcxtm} \\
        \overline{{c^v_{yy}}^{-1}} &&= 
        {1 \over n {\tilde \epsilon}_1} \left( 1 - {\Delta 
        {\tilde \epsilon}_1^2 \over 8 \pi^2 n_0 a (k_{\rm B} T)^2} \,
        {1 \over \left[ 1 + {\tilde \epsilon}_1^2 (\tan \alpha)^2 /
                        4 \pi n_0 a \right]^{1/2}} \right) \ .
 \label{stcytm}
\end{eqnarray}
As before, we can investigate the instabilities of this system towards
external tilt, and find
\begin{eqnarray}
        \left[ T_x(\alpha) / T_{\rm BG} \right]^4 
                                        &&= [1+u(\alpha)]^{-3/2} \ ,
 \label{scxins} \\
        \left[ T_y(\alpha) / T_{\rm BG} \right]^4 
                                        &&= [1+u(\alpha)]^{-1/2} \ ;
 \label{scyins}
\end{eqnarray}
note that both $T_x(\alpha)$ are always $T_y(\alpha)$ positive (see
Fig.~\ref{sctxty}), and the Bose glass remains stable irrespective of
the tilt angle $\alpha$. 

Turning to {\it parallel defect planes} ($d_\perp = 1$), oriented in
the $yz$ plane (i.e., $\alpha = 0$), the reduction in the boson order
parameter (\ref{doedf1}) is
\begin{equation}
        \overline{n_\Delta} = \Delta {\tilde \epsilon}_1^2 / 
                64 \pi^{3/2} n_0^{1/2} a^{3/2} k_{\rm B} T \ .
 \label{plnden}
\end{equation}
For any nonzero tilt angle, we again find a divergence, as in the case
of linear defects. According to Eq.~(\ref{exo1dp}), the leading
finite--$L$ corrections to (\ref{plnden}) for $\alpha = 0$ are in the
case of open boundary conditions 
\begin{equation}
        \overline{\Delta n_\Delta(L)} = - (\pi - 1) \Delta n_0 / 
                        2 \pi {\tilde \epsilon}_1^2 c_1^5 L \ ,
 \label{pdfsen}
\end{equation}
while no such finite--thickness contributions occur if periodic
boundary conditions are applied. Tilting the defect planes in the $x$
direction with angle $\alpha$, the components of the tilt modulus 
[see Eq.~\ref{tex1rn}] read
\begin{eqnarray}
        \overline{{c^v_{xx}}^{-1}} &&= {1 \over n {\tilde \epsilon}}
        \left[ 1 - {\Delta {\tilde \epsilon}_1^2 \over 
                                16 (\pi n_0 a)^{3/2} k_{\rm B} T} \, 
        {1 - {\tilde \epsilon}_1^2 (\tan \alpha)^2 / 2 \pi n_0 a \over
                \left[ 1 + {\tilde \epsilon}_1^2 (\tan \alpha)^2 / 
                                4 \pi n_0 a \right]^{5/2}} \right] \ ,
 \label{plnxtm} \\
        \overline{{c^v_{yy}}^{-1}} &&= (n {\tilde \epsilon})^{-1} \ ,
 \label{plnytm} 
\end{eqnarray}
i.e., the tilt modulus in the $y$ direction transverse to both the
tilt and the magnetic field remains unaffected by the disorder. When
$\alpha \to 0$, Eq.~(\ref{plnxtm}) becomes
\begin{equation}
        \overline{{c^v_{xx}}^{-1}} = (n {\tilde \epsilon})^{-1}
        \left( 1 - \Delta {\tilde \epsilon}_1^2 / 
                        16 (\pi n_0 a)^{3/2} k_{\rm B} T \right) \ ;
 \label{pln0tm}
\end{equation}
only for open boundary conditions will there be corrections for finite
sample thickness, which we have worked out in Eq.~(\ref{exo1rn}) to be
\begin{equation}
        \overline{{c^v_{xx}}^{-1}(L)} = (n {\tilde \epsilon})^{-1}
        \left[ 1 - \Delta {\tilde \epsilon}_1^2 / 
                                16 (\pi n_0 a)^{3/2} k_{\rm B} T 
        + (\pi + 4) / 4 {\tilde \epsilon}_1^2 c_1^5 L \right] \ .
 \label{plfstm}
\end{equation}
As for the linear defects, we can estimate a critical temperature from 
Eq.~(\ref{pln0tm}). The ``bare'' disorder correlator for planar
defects is $U_0^2 b^2 / d$, where $b$ denotes the diameter of the
defect potential, and $d$ the average defect distance. The thermal
renormalizations for $T > T_{dp}$ now read \cite{nelvin}
\begin{eqnarray}
        &&\Delta_0 \longrightarrow \Delta(T) = U(T)^2 b^2 / d \ ,
 \label{pldren} \\
        U(T) = U_0 (b/d)^{2/3} (T^*/T)^{2/3} &&\quad , \qquad
        l_\perp(T) = d (b/d)^{2/3} (T/T^*)^{4/3} ,
 \label{plthrn}
\end{eqnarray}
implying that $\Delta(T) = U_0^2 b^3 / d^2$, independent of $T$. This
yields
\begin{equation}
        \overline{{c^v_{xx}}^{-1}(T)} = (n {\tilde \epsilon}_1)^{-1}
        \Bigl[ 1 - {\tilde T}_{\rm BG} / T \Bigr] \ ,
 \label{pldins}
\end{equation}
and the instabilitiy occurs at the Bose glass transition temperature
marking the onset of localization to the defect planes 
\begin{equation}
        {\tilde T}_{\rm BG} = T^* (b / 4 d)^2 
                                (\phi_0 / 2 \pi \lambda^2 B)^{3/2} \ .
 \label{pldbgt}
\end{equation}
As a function of tilt angle $\alpha$, or equivalently, the parameter
$u$ introduced in Eq.~(\ref{angpar}), we furthermore find
\begin{equation}
        {T_x \over {\tilde T}_{\rm BG}} = 
                        {1-2 u(\alpha) \over [1+u(\alpha)]^{5/2}} \ ,
 \label{pdxins}
\end{equation}
see Fig.~\ref{pdsptx}. In this approximation, the critical angle
beyond which the Bose glass phase of localized vortices becomes
unstable towards tilt is given by $u_c = 1/2$.

Upon generalizing to an ensemble of {\it two families of symmetrically
tilted planar defects} (in thick samples), the divergences present for
each of the families with positive and negative tilt angle,
respectively, again cancel, and the resulting average density of
pinned flux lines (\ref{tfe1dp}) is
\begin{equation}
        \overline{n_\Delta} = \Delta {\tilde \epsilon}_1^2 / 64 
        \pi^{3/2} n_0^{1/2} a^{3/2} k_{\rm B} T \left[ 1 + {\tilde 
        \epsilon}_1^2 (\tan \alpha)^2 / 4 \pi n_0 a \right]^{3/2} \ ,
 \label{stpden}
\end{equation}
and the tilt modulus [from Eq.~(\ref{tfe1rn})] along the tilt
direction reads 
\begin{equation}
        \overline{{c^v_{xx}}^{-1}} = 
        {1 \over n {\tilde \epsilon}_1} \left( 1 - {\Delta 
        {\tilde \epsilon}_1^2 \over 16 (\pi n_0 a)^{3/2} k_{\rm B} T} 
        \, {1 - {\tilde \epsilon}_1^2 (\tan \alpha)^2 / 8 \pi n_0 a
        \over \left[ 1 + {\tilde \epsilon}_1^2 (\tan \alpha)^2 /
                        4 \pi n_0 a \right]^{5/2}} \right) \ ,
 \label{stpxtm}
\end{equation}
while of course 
$\overline{{c^v_{yy}}^{-1}} = (n {\tilde \epsilon}_1)^{-1}$. The angle
dependence of the instability in $\overline{{c^v_{xx}}^{-1}}$ is
readily found to be 
\begin{equation}
        {T_x \over {\tilde T}_{\rm BG}} = 
                        {1-u(\alpha)/2 \over [1+u(\alpha)]^{5/2}} \ ,
 \label{sdxins}
\end{equation}
compare Eq.~(\ref{pdxins}). As is shown in Fig.~\ref{pdsptx}, in a
system with two families of symmetrically tilted planar defects the
stability region of the Bose glass in the phase diagram is
enhanced, the critical angle now being determined by $u_c = 2$.
        

\section{Summary and discussion}
 \label{sumdis}

In this paper, we have investigated the influence of weak uncorrelated
and correlated disorder on superfluid bosons as well as magnetic flux
liquids in high--$T_c$ superconductors, exploiting the formal analogy
of the statistical mechanics of directed lines in $d+1$ dimensions
with the path integral representation of the quantum mechanics of
$d$--dimensional particles. We have specifically addressed the
important issue of how different boundary conditions in the
directed--polymer case modify the results in systems with finite 
thickness $L$.

For disordered boson superfluids, we have generalized the previous
studies by Huang and Meng \cite{khuang}, and Giorgini, Pitaevskii, and
Stringari \cite{giorgi} on the effect of point disorder on the
depletion as well as the superfluid density to the case of correlated
disorder, i.e., linear or planar defects. In Sec.~\ref{disbos}, we
have derived general results for the density, current, and vorticity
correlation functions, as well as the condensate fraction and the
normal--fluid density. Explicit formulas for the contributions by
pointlike, linear, and planar disorder, as well as randomly positioned
and oriented line defects, to the depletion and normal--fluid density
in two and three dimensions are summarized in Sec.~\ref{pbexdo}. We
have also found that these results for static disorder are independent
of temperature (at least to first order in the defect correlator).

In the case of magnetic flux line liquids in high--temperature
superconductors, we saw that different boundary conditions can
actually change the effects of disorder qualitatively. Even in the
absence of disorder, we found that a homogeneous condensate fraction
persists at finite sample thicknesses if open boundary conditions are
applied, contrary to the case of periodic (``bosonic'') boundary
conditions, where long--range order is destroyed by ``thermal'' 
fluctuations in this ($2+1$)--dimensional system. We have physically
related the qualitatively and quantitatively different results for
periodic and open boundary conditions to the enhanced thermal
wandering of flux lines near the sample boundaries in the latter case,
see Fig.~\ref{wavfun}. In Sec.~\ref{disfll}, we have studied the
situation for open boundary conditions in some detail, looking at
precisely those quantities which are the analogs to those studied for
the Bose superfluid, namely density and tilt correlation functions,
the ``boson order parameter'', which is measure of the fraction of
entangled lines, and the tilt modulus. In Sec.~\ref{expdis}, we have
compared our findings in two and three dimensions for the two kinds of
boundary conditions we explored, specializing to (in the flux line
language) point defects, nearly isotropically splayed linear disorder,
parallel untilted and tilted extended defects (lines and planes), as
well as two families of symmetrically tilted extended defects. We have
thus significantly generalized earlier work on weak point disorder in
infinitely thick systems \cite{nelled} in two important respects,
namely by (i) studying correlated disorder as well, and (ii) taking
carefully into account the role of the more realistic open boundary
conditions.

In Sec.~\ref{expcon}, we provide explicit results for the average
fraction of entangled flux lines and the tilt modulus, and elucidate
some of their experimental consequences. E.g., we have commented on
how the defect influence on the tilt modulus effected by anisotropic
(correlated) disorder may be reinterpreted as a renormalization of the
effective mass anisotropy, illustrating how columnar and planar
defects render the system effectively more three--dimensional; this
renormalization clearly increases the decoupling field, above which
the vortices degenerate from three--dimensional objects (lines) to
weakly coupled two--dimensional ones (pancakes). Furthermore we have
discussed how the renormalized tilt modulus may be used to estimate
the location of the instability towards a localized phase from the
high--temperature side of the phase diagram, and to study the
stability of this Bose glass with respect to transverse magnetic
fields (``transverse Meissner effect'').

Our entire investigation was based on the Gaussian approximation, and
nonlinear effects have been neglected. Analysis of these
nonlinearities, as well as disorder which couples directly to the tilt
field (corresponding to a boson current) are interesting avenues for
future investigation. We have restricted our study to local (in $z$)
and isotropic repulsive vortex interactions, which is a fair
approximation for low magnetic fields. It would be interesting to
extend our results to the high--field regime, perhaps along the lines
suggested in the Introduction.


\acknowledgments

We benefitted from discussions with V.B.~Geshkenbein, E.~Frey, and
S.~Teitel. One of us (D.R.N.) is indebted to T.~Hwa, P.~Le~Doussal,
H.S.~Seung, and V.M.~Vinokur for the collaborations represented in 
Refs.~\cite{nelseu,nelvin,hwaled} and \cite{nelled}. Many ideas
developed jointly with these authors have been incorporated into this
article. This research was supported by the National Science
Foundation, in part by the MRSEC Program through Grant DMR-9400396,
and through Grant DMR-9417047. U.C.T. acknowledges support from the
Deutsche Forschungsgemeinschaft (DFG) under Contract Ta.~177/1-1,2 and
from the Engineering and Physical Sciences Research Council (EPSRC)
through Grant GR/J78327.

\newpage


\appendix

\section{Derivation of the disorder effects on the tilt modulus using
         affine transformations}
 \label{galtil}

In this first Appendix, we provide yet another derivation of the
disorder--induced renormalization of the tilt modulus of a flux line
liquid, utilizing an affine transformation \cite{hwaled}, which
is applicable in the (thermodynamic) limit of thick samples $L \to
\infty$, and for disordered systems in which translational invariance
is statistically restored. Upon applying a uniform tilt
(\ref{afftra}), the transformed action in polar representation reads 
[see Eqs.~(\ref{enetra}),(\ref{chmtra})]
\begin{equation}
        S'[\pi',\Theta'] = {\tilde S}[\pi,\Theta] + \int_0^\infty\! dz
        \int \! d^dr \left[ i {k_{\rm B} T \over {\tilde \epsilon}_1}
                {\bf h} \cdot \bbox{\nabla} \Theta({\bf r},z) - 
        {h^2 \over {\tilde \epsilon}_1^2} - \delta V_D({\bf r},z) 
                \right] \Bigl( n_0 + \pi({\bf r},z) \Bigr) \ ,
 \label{aftrac}
\end{equation}
where ${\tilde S}[\pi,\Theta]$ denotes the pure part (in the
flux line notation) of the action (\ref{haract}). This can also be
written as
\begin{equation}
        S'[\pi',\Theta'] = 
        {\tilde S}'[\pi',\Theta'] - \int \! dz' \int \! d^dr' \left[ 
        \delta V_D({\bf r}'-{\bf h} z' / {\tilde \epsilon}_1,z') + 
                        {h^2 \over 2 {\tilde \epsilon}_1} \right] 
                        \Bigl( n_0 + \pi'({\bf r}',z') \Bigr) \ .
 \label{afshac}
\end{equation}
The partition function for the tilted system therefore becomes (we
henceforth omit the prime)
\begin{equation}
        \ln Z_{\rm gr} = \ln {\tilde Z}_{\rm gr} + \ln \Bigg \langle
        \exp \left\{ {1 \over k_{\rm B} T} \int \! dz \int \! d^dr
        \left[ \delta V_D({\bf r}-{\bf h} z / {\tilde \epsilon}_1,z)
        + {h^2 \over 2 {\tilde \epsilon}_1} \right] \Bigl( n_0 + 
        \pi({\bf r},z) \Bigr) \right\} \Bigg \rangle_{{\tilde S}} \ ,
 \label{aftrfz}
\end{equation}
where the thermodynamic average $\langle \ldots \rangle$ is to be
performed using the pure action, and the vortex contribution to the
tilt modulus [see Eq.~(\ref{rentil})] can then be inferred from
\begin{equation}
        {c^v_{ij}}^{-1} = {k_{\rm B} T \over n^2 L \Omega} \,
        {\delta^2 \ln Z_{\rm gr} \over \delta h_i \delta h_j}
                                \Bigg \vert_{{\bf h} = {\bf 0}} \ .
 \label{tilder}
\end{equation}

In order to compute the thermodynamic and quenched disorder averages,
we perform a double cumulant expansion, which yields
\begin{equation}
        \overline{\ln Z_{\rm gr}} = \ln {\tilde Z} +
        {n L \Omega \over {\tilde \epsilon}_1 k_{\rm B} T} \, h^2 +
        {1 \over 2 (k_{\rm B} T)^2} \int \! {d^dq \over (2 \pi)^d}
                                \int \! {d q_z \over 2 \pi} \left[      
        \Delta({\bf q},q_z;-{\bf q},-q_z)_{\bf h} + \left( {h^2 \over
        2 {\tilde \epsilon}_1} \right)^2 \right] S({\bf q},q_z) \ ;
 \label{cumexp}
\end{equation}
here, $S({\bf q},q_z)$ denotes the density correlation function of the
pure system (\ref{fouden}). Assuming now that the quenched disorder
average restores translational invariance statistically, 
\begin{equation}
        \Delta({\bf q},q_z;{\bf q}',q_z') = \Delta({\bf q},q_z) \,
        \Omega \delta_{{\bf q},-{\bf q}'} \, L \delta_{q_z,-q_z'} \ ,
 \label{fltrdo}
\end{equation}
and expanding the disorder correlator to second order in the tilt, one
obtains with Eq.~(\ref{tilder}) 
\begin{eqnarray}
        \overline{{c^v_{ij}}^{-1}}
        &&= {1 \over n {\tilde \epsilon}_1} \left[ \delta_{ij}
                + {1 \over 2 n_0 {\tilde \epsilon}_1 k_{\rm B} T} 
        \int \! {d^dq \over (2 \pi)^d} \int \! {d q_z \over 2 \pi} 
                                \, q_i q_j \, S({\bf q},q_z) \,
        {\partial^2 \over \partial q_z^2} \Delta({\bf q},q_z) \right]
 \label{aftlt1} \\
        &&= {1 \over n {\tilde \epsilon}_1} \left[ \delta_{ij}
                + {1 \over 2 n_0 {\tilde \epsilon}_1 k_{\rm B} T}
        \int \! {d^dq \over (2 \pi)^d} \int \! {d q_z \over 2 \pi} 
                                \, q_i q_j \, \Delta({\bf q},q_z) \,
        {\partial^2 \over \partial q_z^2} S({\bf q},q_z) \right] \ .
 \label{aftlt2}
\end{eqnarray}
If the system is statistically isotropic (with respect to the $d$
transverse dimensions) as well, Eq.~(\ref{aftlt2}) reduces to 
\cite{hwaled}
\begin{equation}
        \overline{{c^v_{44}}^{-1}} = 
                                {1 \over n {\tilde \epsilon}_1} 
        \left[ 1 + {1 \over 2 d n_0 {\tilde \epsilon}_1 k_{\rm B} T} 
        \int \! {d^dq \over (2 \pi)^d} \int \! {d q_z \over 2 \pi} \,
        q^2 \Delta({\bf q},q_z) {\partial^2 \over \partial q_z^2} 
                                        S({\bf q},q_z) \right] \ .
 \label{afitlt}
\end{equation}
Inserting
\begin{equation}
        {\partial^2 \over \partial q_z^2} S({\bf q},q_z) = 
        - {2 k_{\rm B} T n_0 q^2 \over {\tilde \epsilon}_1} \,
        {\epsilon_{\rm B}(q)^2 / (k_{\rm B}T)^2 - 3 q_z^2 \over \left[
        \epsilon_{\rm B}(q)^2 / (k_{\rm B}T)^2 + q_z^2 \right]^3} \ ,
 \label{sqderv}
\end{equation}
we finally arrive at
\begin{equation}
        \overline{{c^v_{ij}}^{-1}} = 
        {1 \over n {\tilde \epsilon}_1} \left[ \delta_{ij} - 
        {1 \over {\tilde \epsilon}_1^2} \int \! {d^dq \over (2 \pi)^d}
                        \int \! {d q_z \over 2 \pi} \, q_i q_j q^2 \, 
        {\epsilon_{\rm B}(q)^2 / (k_{\rm B}T)^2 - 3 q_z^2 \over \left[
        \epsilon_{\rm B}(q)^2 / (k_{\rm B} T)^2 + q_z^2 \right]^3}
                                \, \Delta({\bf q},q_z) \right] \ .
 \label{afistm}
\end{equation}
compare Eq.~(\ref{dotilt}) in Sec.~\ref{pbcorr}. 

Eq.~(\ref{aftlt1}) immediately shows that uncorrelated point defects,
$\Delta({\bf q},q_z) = \Delta$, and ``almost isotropically'' splayed
columnar defects, $\Delta({\bf q},q_z) = \Delta / q$ \cite{hwaled},
have no influence on the tilt modulus in thick samples at all, which
is true for any disorder whose correlator is independent of $q_z$. For
parallel extended defects, tilted in the $x$ direction, 
\begin{equation}
        \Delta({\bf q}_\parallel,{\bf q}_\perp,q_z) = 
        \Delta \, \Omega_\parallel \delta_{{\bf q}_\parallel,{\bf 0}} 
                \, L \delta_{q_z,q_x \tan \alpha} \ ,
 \label{flplds}
\end{equation}
one sees that the disorder does not change the tilt modulus along the
defect directions, 
\begin{equation}
        \overline{c^v_\parallel} = n {\tilde \epsilon}_1 \ ,
 \label{aftlpa}
\end{equation}
while in the perpendicular directions, compare Eq.~(\ref{ttexrn}),
\begin{equation}
        \overline{{c^v_{\perp i}}^{-1}} = 
                {1 \over n {\tilde \epsilon}_1} \left[ 1 - 
                        {\Delta \over {\tilde \epsilon}_1^2} \int \! 
        {d^{d_\perp}q_\perp \over (2 \pi)^{d_\perp}} \, q_{\perp i}^2 
        q_\perp^2 \, {\epsilon_{\rm B}(q_\perp)^2 / (k_{\rm B}T)^2 - 
        3 q_x^2 \tan^2 \alpha \over \left[ \epsilon_{\rm B}(q_\perp)^2
        / (k_{\rm B} T)^2 + q_x^2 \tan^2 \alpha \right]^3} \right] \ .
 \label{aftlpe}
\end{equation}
As discussed in Sec.~\ref{expcon}, a divergence of the transverse tilt
modulus in a certain direction, $\overline{{c^v_{\perp i}}^{-1}} = 0$,
marks a localization transition to a (generalized) Bose glass phase;
Eq.~(\ref{aftlpe}) can thus be employed to obtain a
``high--temperature'' estimate for the Bose glass transition
temperature \cite{hwaled}.

\newpage

\section{Phase--only approximation}
 \label{poappr}

Considering the two types of fluctuations in our system, described by
the harmonic action (\ref{haract}), it may be tempting to integrate
out the density fluctuations, which acquire a mass $V_0$, see 
Eqs.~(\ref{dshars}) and (\ref{matrix}), and only retain the massless
phase modes; certainly this should correctly describe the leading
terms in a long--wavelength expansion. In this Appendix, we study this
simple phase--only approximation (with periodic boundary conditions),
and explain its deficiencies.

We thus define the effective action $S_{\rm eff}[\Theta]$ by
\begin{equation}
        e^{- S_{\rm eff}[\Theta] / \hbar} = 
        \int \! {\cal D}[\pi] \, e^{- S_0[\pi,\Theta] / \hbar} \ .
 \label{poeffs}
\end{equation}
The purely quadratic functional integral is readily evaluated, with
the result
\begin{eqnarray}
        S_{\rm eff}[\Theta] = {\rm const.} &&+ 
        {1 \over \beta \hbar \Omega} \sum_{{\bf q},\omega_m} 
        \Biggl\{ {n_0 \hbar^2 \over 2 m} \, 
                {\hbar^2 q^2 \over \epsilon_{\rm B}(q)^2} \,
        \Bigl [\omega_m^2 + \epsilon_{\rm B}(q)^2 / \hbar^2 \Bigr] 
        \Theta(-{\bf q},-\omega_m)\Theta({\bf q},\omega_m) \nonumber\\
        &&-{n_0 \over 2 m} \, {\hbar^2q^2 \over \epsilon_{\rm B}(q)^2}
        \, \Bigl[ 2 \hbar \omega_m \Theta(-{\bf q},-\omega_m) + 
                        \delta V_D(-{\bf q},-\omega_m) \Bigr] 
                        \delta V_D({\bf q},\omega_m) \Biggr\} \ ;
 \label{effact}
\end{eqnarray}
shifting the $\Theta$ field as in Eq.~(\ref{shflds}), one then finds
for the phase correlation function precisely the result
(\ref{dsthth}). However, as density fluctuations have now been
neglected entirely, the Green's function is simply
\begin{equation}
        \overline{G({\bf q},\omega_m;{\bf q}',\omega{m'})} \approx
        n_0 \overline{\langle \Theta({\bf q},\omega_m)
                        \Theta({\bf q'},\omega_{m'}) \rangle} \ ,
 \label{pogref}
\end{equation}
and similarly the current correlation function is merely given
by the first term in Eq.~(\ref{gaumcr}),
\begin{eqnarray}
        \overline{C_{ij}({\bf q},\omega_m'{\bf q}',\omega_{m'})} &&=
        {n_0 m \epsilon_{\rm B}(q)^2 / \hbar \over \omega_m^2 + 
        \epsilon_{\rm B}(q)^2 / \hbar^2} \, P_{ij}^L({\bf q}) \, 
        (2\pi)^d \delta({\bf q}+{\bf q}') \, \beta\hbar \delta_{m,-m'}
        \nonumber \\ &&\quad - {n_0^2 q_i q_j' \omega_m \omega_{m'}
        \over [\omega_m^2 + \epsilon_{\rm B}(q)^2 / \hbar^2] 
                [\omega_{m'}^2 + \epsilon_{\rm B}(q')^2 / \hbar^2]} \, 
                \Delta({\bf q},\omega_m;{\bf q}',\omega_{m'}) \ ,
 \label{pocurr}
\end{eqnarray}
where, in order to be consistent with our long--wavelength expansion,
we should use $\epsilon_{\rm B}(q) \approx \hbar c_1 q$. Thus all the
nonlinear terms, i.e., the vorticity contributions (\ref{fouvor})
present in the ``full'' theory (\ref{dshars}) have disappeared, and as
the current correlation function is therefore purely longitudinal, see
Eq.~(\ref{momvor}), the normal--fluid density will necessarily be zero
at all temperatures in the phase--only approximation,
\begin{equation}
        \overline{\rho_n(T)} = 0 \ .
 \label{ponfld}
\end{equation}

Using Eq.~(\ref{parnum}), we can also compute the average depletion,
\begin{eqnarray}
        \overline{n} - n_0 &&= {m c_1 \over 2} \int \! 
                {d^dq \over (2 \pi)^d} \, {1 \over q} \left( 1 + 
        {2 \over e^{\beta \hbar c_1 q} - 1} \right) \nonumber \\
        &&+ {n_0 \over \hbar^2 \Omega} \int \! {d^dq \over (2 \pi)^d}
        \, {1 \over (\beta \hbar)^2} \sum_{m,m'} {\omega_m \omega_{m'}
        \Delta({\bf q},\omega_m;-{\bf q},\omega_{m'}) \over 
        (\omega_m^2 + c_1^2 q^2) (\omega_{m'}^2 + c_1^2 q^2)} \ .
 \label{podepl}
\end{eqnarray}
For the pure system, this yields the at least qualitatively correct
result that the zero--temperature depletion diverges logarithmically
in one dimension; however, for $d \geq 2$, $n(T=0)$ cannot be
determined from Eq.~(\ref{podepl}). On the other hand, the
finite--temperature corrections $\Delta n(T)$ turn out to be identical
to the leading contributions (\ref{findep}) obtained for the ``full''
theory in the phonon approximation [i.e., neglect any terms of order
$q^2$ in the integrand of Eq.~(\ref{dodepl})]. In the same manner, the
disorder contribution to (\ref{podepl}) yields the leading terms in a
long--wavelength expansion for the depletion, whenever these do not
vanish. E.g., for uncorrelated (``point'') disorder in space and
imaginary time, one finds [see (\ref{ptdpdp})]
\begin{equation}
        \overline{n_\Delta(T)} = 
        - \Delta {n_0 \over 4 \hbar} \int \! {d^dq \over (2 \pi)^d} \,
        {\coth [\beta\epsilon_{\rm B}(q)/2] \over \epsilon_{\rm B}(q)}
                \left( 1 - {\beta \epsilon_{\rm B}(q) \over 
                        \sinh [\beta \epsilon_{\rm B}(q)]} \right) \ ,
 \label{pouncd}
\end{equation}
and for the case of ``almost isotropic splay'' accordingly,
cf. Eq.~(\ref{rsdpdp}), 
\begin{equation}
        \overline{n_\Delta(T)} = 
        - \Delta {n_0 \over 4 \hbar} \int \! {d^dq \over (2 \pi)^d} \,
        {\coth [\beta\epsilon_{\rm B}(q)/2]\over q\epsilon_{\rm B}(q)}
                \left( 1 - {\beta \epsilon_{\rm B}(q) \over 
                        \sinh [\beta \epsilon_{\rm B}(q)]} \right) \ ,
 \label{poispl}
\end{equation}
while for ``tilted extended defects'' the result is
temperature--independent [see Eq.~(\ref{tlexdp})], 
\begin{equation}
        \overline{n_\Delta} = - \Delta {n_0 \over \hbar^2} \int \! 
        {d^dq_\perp \over (2 \pi)^{d_\perp}} \, {q_x^2 v^2 \over 
        [\epsilon_{\rm B}(q_\perp)^2 / \hbar^2 + q_x^2 v^2]^2} \ .
 \label{potilt}
\end{equation}
All of these expressions correctly represent the leading contributions
of the results obtained in the ``full'' theory where the density
fluctuations are taken into as well. However, at $T=0$ the integrals
(\ref{pouncd}) and (\ref{poispl}) become ill--defined when the phonon
approximation is employed, as is required for the sake of consistency
here. Note in addition that for the case of ``untilted'' correlated
disorder (along the $\tau$ direction), and for two families of
``symmetrically tilted extended defects'' the lowest--wavenumber 
contributions actually cancel, leaving the very poor result 
$\overline{n_\Delta(T)} = 0$.

It is actually this very mechanism which leads to a vanishing
normal--fluid density in the phase--only approximation, namely a
cancellation of the leading--order terms in the long--wavelength
expansion. This can be made explicit by actually calculating the
average momentum density in a system with walls moving relative to the
superfluid; a quick computation shows that even with disorder
\begin{equation}
        \overline{\langle {\bf g}({\bf r},\tau) \rangle} = 
                        0 + {\cal O}[({\bf q}\cdot{\bf v})^2] \ ,
 \label{pomomd}
\end{equation}
quite in accord with the fact that within the framework of the
phase--only approximation there is no vorticity in the system, and
hence $\overline{\rho_n(T)} = 0$ for all temperatures.

Summarizing our experiences, we have realized that the phase--only
approximation (\ref{effact}) is sufficient {\it only} for the
evaluation of the longitudinal current correlations, and typically
also for the leading finite--temperature corrections to the condensate
fraction. It produces quite unsatisfactory results for the transverse
current and vorticity correlations, however, which are crucially
affected by their coupling to density fluctuations, and hence fails to
detect any nonzero normal--fluid density at finite temperatures.

\newpage

\section{Open boundary conditions:
         correlation functions and disorder contributions}
 \label{obcder}

In this Appendix, we derive the density and phase correlation
functions in Gaussian approximation for a disordered flux line system
described by an action of the form (\ref{haract}), with open boundary
conditions. This calculation follows and generalizes
Refs.~\cite{nelseu} and \cite{nelled}, i.e., the disorder
contributions will be treated to first order in the disorder
correlator $\Delta$, and the open--boundary correlators will be
expressed in terms of matrix products of the corresponding correlation
functions obtained with periodic boundary conditions. Although the
physical application we have in mind is that for magnetic flux
liquids, we retain the boson notation here in order to facilitate
comparison between the two cases.

Our starting point is Eq.~(\ref{cnpath}), where the boundary
conditions (\ref{constr}) have been explicitly taken into account
through the insertion of a static constraint field
$\lambda({\bf r})$. Introducing the vector
\begin{equation}
        u({\bf q},\omega_m) = 
        \left( \begin{array}{c} 0 \\ \delta V_D({\bf q},\omega_m) 
                \end{array} \right) + i \hbar \lambda({\bf q}) \ ,
 \label{auxfld}
\end{equation}
we can write the new effective harmonic action as
\begin{equation}
        {\tilde S_0}[\lambda,\Upsilon] = 
        {1 \over 2 \beta \hbar \Omega} \sum_{{\bf q},\omega_m} \left[ 
        \Upsilon^T(-{\bf q},-\omega_m) {\bf A}({\bf q},\omega_m) 
        \Upsilon({\bf q},\omega_m) - 2 \Upsilon^T(-{\bf q},-\omega_m) 
                                u({\bf q},\omega_m) \right] \ ,
 \label{opefac}
\end{equation}
where ${\bf A}({\bf q},\omega_m)$ is the matrix (\ref{matrix}),
obeying ${\bf A}^T(-{\bf q},-\omega_m) = {\bf A}({\bf q},\omega_m)$.
Transforming (with Jacobian 1) to new fields 
${\tilde \Upsilon}({\bf q},\omega_m) = \Upsilon({\bf q},\omega_m) - 
{\bf A}^{-1}({\bf q},\omega_m) u({\bf q},\omega_m)$ 
[see Eq.~(\ref{shflds})] , the effective action 
${\tilde S_0}[\lambda,{\tilde \Upsilon}]$ is purely quadratic in 
${\tilde \Upsilon}$, and hence the functional integral over these
variables is readily performed, leading to
\begin{eqnarray}
        \langle \Upsilon({\bf q},\omega_m)
                \Upsilon^T({\bf q}',\omega_{m'}) \rangle &&= 
        \hbar {\bf A}^{-1}({\bf q},\omega_m) \, \beta \hbar \Omega \,
                \delta_{{\bf q},-{\bf q}'} \delta_{m,-m'} \nonumber \\
        &&+ {\bf A}^{-1}({\bf q},\omega_m) \, 
                \langle u({\bf q},\omega_m) u^T({\bf q}',\omega_{m'}) 
        \rangle_\lambda \, {\bf A}^{-1}(-{\bf q}',-\omega_{m'}) \ ,
 \label{intder}
\end{eqnarray}
with the inverse matrix ${\bf A}^{-1}({\bf q},\omega_m)$ given in
Eq.~(\ref{matinv}), and
\begin{equation}
        \langle u({\bf q},\omega_m) 
                u^T({\bf q}',\omega_{m'}) \rangle_\lambda = 
        {\int \! {\cal D}[\lambda({\bf q})] \, u({\bf q},\omega_m) 
        u^T({\bf q}',\omega_{m'}) \, e^{- S_u[\lambda] / \hbar}
                        \over \int \! {\cal D}[\lambda({\bf q})] \, 
                                e^{- S_u[\lambda] / \hbar}} \ ,
 \label{uucorr}
\end{equation}
where
\begin{equation}
        S_u[\lambda] = - {1 \over 2 \beta \hbar \Omega} 
                \sum_{{\bf q},\omega_m} u^T(-{\bf q},-\omega_m) 
        {\bf A}^{-1}({\bf q},\omega_m) u({\bf q},\omega_m) \ .
 \label{uactio}
\end{equation}
At this point we caution the reader to note that Eq.~(\ref{uucorr})
does {\it not} simply reduce to the negative of the first line in
Eq.~(\ref{intder}), for the functional integral is over the 
{\it static} field $\lambda({\bf q})$ only, and not over the dynamic
variable $u({\bf q},\omega_m)$; the fluctuations will only vanish at
the special points $\tau = 0$ and $\tau = \beta \hbar$.

We thus have to evaluate Eq.~(\ref{uucorr}) very carefully. Upon
defining
\begin{equation}
        {\bf A}^{-1}({\bf q}) = {\bf A}^{-1}({\bf q},\tau = 0) =
        {1 \over \beta\hbar} \sum_m {\bf A}^{-1}({\bf q},\omega_m) \ , 
 \label{statam}
\end{equation}
the inverse of which is explicitly written down in Eq.~(\ref{stamat}),
and furthermore
\begin{equation}
        a({\bf q}) = {1 \over \beta \hbar} \sum_m 
        {\bf A}^{-1}({\bf q},\omega_m) \left( \begin{array}{c} 0 \\
                \delta V_D({\bf q},\omega_m) \end{array} \right) \ ,
 \label{avectr}
\end{equation}
and then again shifting (with Jacobian 1) the integration variable 
${\tilde \lambda}({\bf q}) = \lambda({\bf q}) - i {\bf A}({\bf q})
a({\bf q}) / \hbar$, it becomes a straightforward task to perform the
purely quadratic path integral over ${\tilde \lambda}$, with the
result
\begin{eqnarray}
        &&\langle u({\bf q},\omega_m) 
                        u^T({\bf q}',\omega_{m'}) \rangle_\lambda = 
        - \hbar {\bf A}({\bf q}) \Omega \delta_{{\bf q},-{\bf q}'} + 
        \delta V_D({\bf q},\omega_m) \delta V_D({\bf q}',\omega_{m'})
        \left( \begin{array}{cc} 0 & 0 \\ 0 & 1 \end{array} \right) 
                                        \nonumber \\ &&\qquad \qquad 
        - {\bf A}({\bf q}) {1 \over \beta \hbar} \sum_n
        {\bf A}^{-1}({\bf q},\omega_n) \delta V_D({\bf q},\omega_n)
        \left( \begin{array}{cc} 0 & 0 \\ 0 & 1 \end{array} \right) 
                \delta V_D({\bf q}',\omega_{m'}) \nonumber \\
        &&\qquad \qquad - \delta V_D({\bf q},\omega_n) \left(
        \begin{array}{cc} 0 & 0 \\ 0 & 1 \end{array} \right) {1 \over 
        \beta \hbar} \sum_{n'} {\bf A}^{-1}(-{\bf q}',-\omega_{n'}) 
        \delta V_D({\bf q}',\omega_{n'}) {\bf A}({\bf q}') \nonumber\\
        &&+ {\bf A}({\bf q}) {1 \over \beta \hbar} \sum_n
        {\bf A}^{-1}({\bf q},\omega_n) \delta V_D({\bf q},\omega_n)
        \left( \begin{array}{cc} 0 & 0 \\ 0 & 1 \end{array} \right) 
        {1 \over \beta \hbar} \sum_{n'} 
        {\bf A}^{-1}(-{\bf q}',-\omega_{n'}) 
        \delta V_D({\bf q}',\omega_{n'}) {\bf A}({\bf q}') \ .
 \label{ucorre}
\end{eqnarray}
Finally performing the quenched--disorder average, and then collecting
the various terms, we find for the matrix of the averaged density and
phase correlation functions
\begin{eqnarray}
        &&\overline{\langle \Upsilon({\bf q},\omega_m)
                \Upsilon^T({\bf q}',\omega_{m'}) \rangle} = (2 \pi)^d 
        \delta({\bf q}+{\bf q}') \hbar {\bf A}^{-1}({\bf q},\omega_m)
        \left[ \beta \hbar \delta_{m,-m'} - {\bf A}({\bf q})
        {\bf A}^{-1}({\bf q},-\omega_{m'}) \right] \nonumber \\ 
        &&\qquad + {\bf A}^{-1}({\bf q},\omega_m) 
        {1 \over (\beta \hbar)^2} \sum_{n,n'} \Biggl\{ 
        \left[ \beta \hbar \delta_{m,n} - 
        {\bf A}({\bf q}) {\bf A}^{-1}({\bf q},\omega_n) \right] 
        \left( \begin{array}{cc} 0 & 0 \\ 0 & 1 \end{array} \right) 
                        \times \nonumber \\ &&\qquad \qquad \times 
        \left[ \beta \hbar \delta_{m',n'} - {\bf A}({\bf q}') 
                        {\bf A}^{-1}({\bf q}',\omega_{n'}) \right]^T 
        \Delta({\bf q},\omega_n;{\bf q}',\omega_{n'}) \Biggr\} 
                {\bf A}^{-1}(-{\bf q}',-\omega_{m'}) \nonumber \\ 
        &&\quad = (2 \pi)^d \delta({\bf q}+{\bf q}') \left[ 
        \beta \hbar \delta_{m,-m'} - {\bf A}^{-1}({\bf q},\omega_m)     
                                        {\bf A}({\bf q}) \right] 
                \hbar {\bf A}^{-1}({\bf q},-\omega_{m'}) \nonumber \\ 
        &&\qquad + {1 \over (\beta \hbar)^2} \sum_{n,n'} \left[ 
        \beta \hbar \delta_{m,n} - {\bf A}^{-1}({\bf q},\omega_m) 
        {\bf A}({\bf q}) \right] {\bf A}^{-1}({\bf q},\omega_n) 
        \left( \begin{array}{c} 0 \\ 1 \end{array} \right) \times 
        \nonumber \\ &&\qquad \qquad \times \Biggl\{ \left[ \beta
        \hbar \delta_{m',n'} - {\bf A}^{-1}({\bf q}',\omega_{m'})
        {\bf A}({\bf q}') \right] {\bf A}^{-1}({\bf q}',\omega_{n'}) 
        \left( \begin{array}{c} 0 \\ 1 \end{array} \right) \Biggr\}^T 
                \Delta({\bf q},\omega_n;{\bf q}',\omega_{n'}) \ .
 \label{opdisc}
\end{eqnarray}
It can be easily checked that indeed the fluctuations vanish at the
boundaries,
\begin{eqnarray}
        \overline{\langle \Upsilon({\bf q},\tau = 0) 
                \Upsilon^T({\bf q}',\tau = 0) \rangle} &&=
        \overline{\langle \Upsilon({\bf q},\tau = \beta \hbar) 
        \Upsilon^T({\bf q}',\tau = \beta \hbar) \rangle} \nonumber \\
        &&= {1 \over (\beta \hbar)^2} \sum_{m,m'}
        \overline{\langle \Upsilon({\bf q},\omega_m) 
                \Upsilon^T({\bf q}',\omega_{m'}) \rangle} = 0 \ ,
 \label{boucon}
\end{eqnarray}
as required by Eqs.~(\ref{constr}). Also, if all the second terms in
the square brackets, which break translational invariance in $\tau$,
are disregarded, we recover the two--point correlation functions
(\ref{dsthth})--(\ref{dspipi}) for periodic boundary conditions
obtained in Sec.~\ref{pbcorr}.

\newpage

\section{Matsubara frequency sums and momentum integrals}
 \label{sumsin}

For the reader's convenience, we finally provide a list of useful
explicit formulas for the (bosonic) Matsubara frequency sums required
in the above calculations, as well as some definite integrals needed
for the momentum integrations.

\subsection{Bosonic Matsubara frequency sums}
 \label{bosmat}

The following fundamental formula for bosonic and fermionic Matsubara
frequency sums ($n = 0, \pm 1, \pm2, \ldots$) is readily obtained by
performing an appropriate contour integration in the complex plane
(for details see Ref.~\cite{fetwal}, p.~248 {\it ff.})
\begin{equation}
        \tau \geq 0 \, : \quad {1 \over \beta \hbar}
        \sum_n {e^{i \omega_n \tau} \over i \omega_n - \epsilon/\hbar}
        = \mp \, {e^{\tau \epsilon/\hbar} \over 
                        e^{\beta \epsilon} \mp 1} \; {\rm for} \; 
        \left\{ \begin{array}{ll} 
        {\rm bosons \; with} & \omega_n = 2 n \pi / \beta \hbar \ , \\
        {\rm fermions \; with} & 
        \omega_n = (2 n + 1) \pi / \beta \hbar \ . \end{array} \right.
 \label{bfmat1}
\end{equation}
Consequently,
\begin{equation}
        {1 \over \beta \hbar} \sum_n {e^{i \omega_n \tau} \over 
        \omega_n^2 + \epsilon^2/\hbar^2} = {\hbar \over 2 \epsilon} \,
        {e^{(\beta - \tau/\hbar) \epsilon} \pm e^{\tau \epsilon/\hbar}
        \over e^{\beta \epsilon} \mp 1} \longrightarrow
        {\hbar \over 2 \epsilon} \left\{ \begin{array}{ll}
        \coth \left( \beta \epsilon / 2 \right) & {\rm (bosons)} \\
        \tanh \left( \beta \epsilon / 2 \right) & {\rm (fermions)}
        \end{array} \right. {\rm for} \; \tau \to 0 \ .
 \label{bfmat2}
\end{equation}

Now specializing to bosons, i.e., periodic boundary conditions in
imaginary time and therefore Matsubara frequencies with even integers,
we may write down the following sequence of formulas (for 
$\tau \geq 0$): 
\begin{eqnarray}
        {1 \over \beta \hbar} \sum_n 
        {e^{- i \omega_n \tau} \over i \omega_n + \epsilon/\hbar} &&=
        {- 1 \over \beta \hbar} \sum_n 
        {e^{i \omega_n \tau} \over i \omega_n - \epsilon/\hbar} =
        {e^{\tau \epsilon/\hbar} \over e^{\beta \epsilon} - 1} \ ,
 \label{bsmat1} \\
        {- 1 \over \beta \hbar} \sum_n 
        {e^{- i \omega_n \tau} \over i \omega_n - \epsilon/\hbar}&&=
        {1 \over \beta \hbar} \sum_n 
        {e^{i \omega_n \tau} \over i \omega_n + \epsilon/\hbar} =
        {e^{(\beta - \tau/\hbar) \epsilon} \over 
                                        e^{\beta \epsilon} - 1} \ ,
 \label{bsmat2} \\
        {1 \over \beta \hbar} \sum_n {e^{- i \omega_n \tau} \over
        \omega_n^2 + \epsilon^2/\hbar^2} &&= {\hbar
        \cosh \left[ (\beta \hbar - 2 \tau) \epsilon / 2\hbar \right]
        \over 2 \epsilon \sinh \left( \beta \epsilon / 2 \right)} \ ,
 \label{bsmat3} \\
        {1 \over \beta \hbar} \sum_n {i \omega_n e^{- i \omega_n \tau}
        \over \omega_n^2 + \epsilon^2/\hbar^2} &&=
        {\sinh \left[ (\beta \hbar - 2 \tau) \epsilon / 2\hbar \right]
                \over 2 \sinh \left( \beta \epsilon / 2 \right)} \ .
 \label{bsmat4}
\end{eqnarray}
One then immediately sees that ($m,n = 0, \pm 1, \pm2, \ldots$)
\begin{equation}
        {1 \over (\beta \hbar)^2} \sum_{m,n}
        {i (\omega_m - \omega_n) e^{- i (\omega_m + \omega_n) \tau}
                        \over (\omega_m^2 + \epsilon^2/\hbar^2)
                        (\omega_n^2 + \epsilon^2/\hbar^2)} = 0 \ ,
 \label{dmsum1}  
\end{equation}
and hence
\begin{equation}
        {1 \over (\beta \hbar)^2} \sum_{m,n}
        {(\omega_m \omega_n + \epsilon^2/\hbar^2) 
                e^{- i (\omega_m + \omega_n) \tau} \over 
                                (\omega_m^2 + \epsilon^2/\hbar^2) 
                                (\omega_n^2 + \epsilon^2/\hbar^2)} =
        {e^{\beta \epsilon} \over 
                        \left( e^{\beta \epsilon} - 1 \right)^2} = 
        {1 \over 4 
        \left[ \sinh \left( \beta \epsilon / 2 \right) \right]^2} \ ,
 \label{dmsum2}
\end{equation}
independent of $\tau$.

Differentiation of Eqs.~(\ref{bsmat3}) and (\ref{bsmat4}) with respect
to $\epsilon$ furthermore yields
\begin{eqnarray}
        {1 \over \beta \hbar} \sum_n {e^{- i \omega_n \tau}
                        \over (\omega_n^2 + \epsilon^2/\hbar^2)^2} &&= 
        {\hbar^3 \over 4 \epsilon^3 
        \sinh \left( \beta \epsilon / 2 \right)} \Biggl\{ \cosh 
        \left[ {(\beta \hbar - 2 \tau) \epsilon \over 2 \hbar} \right]
        \nonumber \\ &&\qquad \qquad \qquad \qquad
        - {\tau \epsilon \over \hbar} \sinh 
        \left[ {(\beta \hbar - 2 \tau) \epsilon \over 2 \hbar} \right]
        + {\beta \epsilon \over 2} \,
        {\cosh \left( \tau \epsilon / \hbar \right) \over
                \sinh \left( \beta \epsilon / 2 \right)} \Biggr\} \ ,
 \label{dfmsm1} \\
        {1 \over \beta \hbar} \sum_n {i \omega_n e^{- i \omega_n \tau}
                        \over (\omega_n^2 + \epsilon^2/\hbar^2)^2} &&=
        {\hbar \over 4 \epsilon 
        \sinh \left( \beta \epsilon / 2 \right)} \left\{ \tau \cosh
        \left[ {(\beta \hbar - 2 \tau) \epsilon \over 2 \hbar} \right]
        - {\beta \hbar \over 2} \,
        {\sinh \left( \tau \epsilon / \hbar \right) \over
                \sinh \left( \beta \epsilon / 2 \right)} \right\} \ ,
 \label{dfmsm2}
\end{eqnarray}
and combining these with Eq.~(\ref{bsmat3}) gives
\begin{eqnarray}
        {1 \over \beta \hbar} \sum_n {\omega_n^2 e^{- i \omega_n \tau}
                        \over (\omega_n^2 + \epsilon^2/\hbar^2)^2} &&=
        {\hbar \over 4 \epsilon 
        \sinh \left( \beta \epsilon / 2 \right)} \Biggl\{ \cosh 
        \left[ {(\beta \hbar - 2 \tau) \epsilon \over 2 \hbar} \right]
        \nonumber \\ 
        &&\; \qquad \qquad \qquad + {\tau \epsilon \over \hbar} \sinh 
        \left[ {(\beta \hbar - 2 \tau) \epsilon \over 2 \hbar} \right]
        - {\beta \epsilon \over 2} \,
        {\cosh \left( \tau \epsilon / \hbar \right) \over
                \sinh \left( \beta \epsilon / 2 \right)} \Biggr\} \ ,
 \label{dfmsm3} \\
        {1 \over \beta \hbar} \sum_n 
        {(\omega_n^2 - \epsilon^2/\hbar^2) e^{- i \omega_n \tau}
                        \over (\omega_n^2 + \epsilon^2/\hbar^2)^2} &&=
        {1 \over 2 \sinh \left( \beta \epsilon / 2 \right)} \left\{ 
        \tau \sinh 
        \left[ {(\beta \hbar - 2 \tau) \epsilon \over 2 \hbar} \right]
        - {\beta \hbar \over 2} \,
        {\cosh \left( \tau \epsilon / \hbar \right) \over
                \sinh \left( \beta \epsilon / 2 \right)} \right\} \ .
 \label{dfmsm4}
\end{eqnarray}

\subsection{Momentum integrals}
 \label{momint}

We finally list some useful formulas for some of the momentum
integrals required for Bose systems in $d$ dimensions, see 
Refs.~\cite{fetwal,graryz}:
\begin{eqnarray}
        \int_0^\infty \! {x^{d-2} \over e^x - 1} \, dx &&=
                                        \Gamma(d-1) \, \zeta(d-1) \ ,
 \label{integ1} \\
        \int_0^\infty \! {x^{d-2} \over e^x + 1} \, dx &&= \left\{
        \begin{array}{cc} \ln 2 & \; {\rm for} \quad d = 2 \ , \\
        \left( 1 - 2^{-(d-2)} \right) \Gamma(d-1) \, \zeta(d-1) 
                & \; {\rm for} \quad d > 2 \ , \end{array} \right.
 \label{integ2} \\
        \int_0^\infty \! {x^{d-1} \over \sinh^2 x} \, dx &&= 
                        2^{-(d-2)} \, \Gamma(d) \, \zeta(d-1) \ ,
 \label{integ3} \\
        \int_0^\infty \! {x^{d-1} \over \cosh^2 x}\, dx &&= 2^{-(d-2)} 
                \left( 1 - 2^{-(d-2)} \right) \Gamma(d) \, \zeta(d-1) \ ,
 \label{integ4} \\
        \int_0^\infty \! {x^{d-1} \over \sinh x} \, dx &&= 
                2 \left( 1 - 2^{-d} \right) \Gamma(d) \, \zeta(d) \ ,
 \label{integ5} \\
        \int_0^\infty \! {x^{d-1} \over \cosh x} \, dx &&= 
                        2 \, \Gamma(d) \, \xi(d) \ .
 \label{integ6}
\end{eqnarray}
Here, $\zeta(n) = \sum_{k=1}^\infty 1 / k^n$ and 
$\xi(n) = \sum_{k=0}^\infty (-1)^k / (2k+1)^n$. We note that the
$\zeta$ function at even arguments is related to Bernoulli's numbers,
\begin{equation}
        \zeta(2k) = {\pi^{2k} 2^{2k-1} \over (2k)!} \, B_k \; ,
                \quad \zeta(0) = - {1 \over 2} \, , \; 
        \zeta(2) = {\pi^2 \over 6} \, , \; \zeta(4) = {\pi^4 \over 90}
        \, , \; \zeta(6) = {\pi^6 \over 945} \, \ldots \ ,
 \label{zetafn}
\end{equation}
while the $\xi$ function at odd arguments is connected with Euler's
numbers,
\begin{equation}
        \xi(2k+1) = {\pi^{2k+1} \over 4^{k+1} (2k)!} \, E_k \; ,
                        \quad \xi(1) = {\pi \over 4} \, , \; 
                        \xi(3) = {\pi^3 \over 32} \, \ldots \ ,
 \label{xifunc}
\end{equation}
and $\zeta(3) \approx 1.202$, $\xi(2) = G \approx 0.915965$ (Catalan's
constant \cite{graryz}).



\begin{table}
\setdec 0.00
\caption{Superfluid boson -- flux liquid analogy.}
\medskip

\begin{tabular}{lcccccccc}
Superfluid bosons & $m$ & $\hbar$ & $\beta\hbar$ & $V(r)$ & $n$ &
$\mu$ & $i{\bf v}$ & $\rho_s$ \\
\tableline
Vortex liquid & ${\tilde \epsilon}_1$ & $T$ & $L$ & 
$2 \epsilon_0 K_0(r/\lambda)$ & $B/\phi_0$ & $(H-H_{c_1}) \phi_0/4\pi$
& ${\bf h}/{\tilde \epsilon}_1$ & $\rho^2 c_{\rm 44}^{-1}$ \\
\end{tabular}
 \label{analog}
\end{table}

\bigskip \bigskip

\begin{table}
\setdec 0.00
\caption{Varieties of disorder.}
\medskip
\begin{tabular}{ll}
Vortex liquids in three dimensions & Superfluid Helium films \\
\tableline
Point disorder (e.g., oxygen vacancies) & 
        Space-- and time--dependent random potential \\
Columnar disorder (e.g., screw dislocations, & 
        Amorphous substrate modelled by uncorrelated \\
$\quad$ columnar damage tracks) &
        $\quad$ ``point--like'' random potential \\
Planar disorder (e.g., twin boundaries, &
        ``Smooth'' or crystalline substrate with line \\
$\quad$ grain boundaries) & 
        $\quad$ disorder deposited microlithographically \\
\end{tabular}
 \label{disvar}
\end{table}



\begin{figure}

FIG.~\ref{phases}. Schematic representation of the Abrikosov flux
        lattice (a), a disentangled flux liquid (b), and an entangled
        flux liquid (c). In the boson picture, the directed lines map
        onto the imaginary--time particle world lines, and these
        phases correspond to a crystalline solid (a), a normal liquid
        (b), and a superfluid (c), respectively.
\bigskip

FIG.~\ref{wavfun}. Schematic plot of the time--averaged cross section
        of a thermally fluctuating flux line in a slab with free
        boundary conditions. The bulk part away from the boundaries
        has a width determined by the square of the bosonic ``wave
        function'', which is a Gaussian for a free random walker. Near
        the top and bottom surfaces, thermal wandering is enhanced,
        reflected by a probability distribution determined just by the
        ``wave function'' (not its square).
\bigskip

FIG.~\ref{monpol}. Lowest--energy solid phase contribution to the
        correlation function ${\tilde G}({\bf r},z;{\bf r}',z')$,
        which inserts a flux head and tail into a crystalline vortex
        array. Dashed lines represent a row of vortices slightly
        behind the plane of the page. In (a), a vacancy is created at
        ``time'' $z$, which then propagates and is destroyed at
        ``time'' $z'$. Interstitial propagation from $z'$ to $z$ is
        shown in (b). The energy of the ``string'' defect connecting
        the head to the tail increases linearly with the separation in
        both cases and leads to the exponential decay of 
        ${\tilde G}({\bf r},z;{\bf r}',z')$. 
\bigskip

FIG.~\ref{cdtxty}. Critical temperatures $T_x(\alpha)/T_{\rm BG}$
        (solid line) and $T_y(\alpha)/T_{\rm BG}$ (dashed) where the
        tilt moduli $\overline{{c^v_{xx}}^{-1}}$ and
        $\overline{{c^v_{yy}}^{-1}}$ in a system with parallel
        columnar defects tilted along the $x$ direction vanish,
        respectively, as a function of the dimensionless parameter 
        $u = {\tilde \epsilon}_1^2 (\tan \alpha)^2 / 4 \pi n_0 a$.
        The curve $T_x(\alpha)/T_{\rm BG}$ may be viewed as an
        estimate for the phase boundary of the Bose glass, which
        remains stable towards external tilt as long as 
        $\alpha < \alpha_c$ (transverse Meissner effect). 
\bigskip

FIG.~\ref{sctxty}. Tilt instability temperatures 
        $T_x(\alpha)/T_{\rm BG}$ (thick solid line) and 
        $T_y(\alpha)/T_{\rm BG}$ (dashed) for a system with two
        families of symmetrically tilted columnar defects. For
        comparison, $T_x(\alpha)/T_{\rm BG}$ (thin solid line) for a  
        single family of tilted linear disorder is also depicted.
\bigskip

FIG.~\ref{pdsptx}. Tilt instability temperature 
        $T_x(\alpha)/T_{\rm BG}$ for both parallel planar defects
        uniformly tilted along the $x$ direction (solid line) and two
        families of symmetrically tilted planes (dashed). The critical
        tilt angle marking the instability of the Bose glass phase of
        localized flux lines is shifted upwards in the
        symmetric--splay case. 

\end{figure}


\newpage


\tighten

\centerline{--- FIG.~7 in Ref.~\cite{nelseu} ---}
\centerline{}
\vfill
\begin{figure}
\caption{Schematic representation of the Abrikosov flux
        lattice (a), a disentangled flux liquid (b), and an entangled
        flux liquid (c). In the boson picture, the directed lines map
        onto the imaginary--time particle world lines, and these
        phases correspond to a crystalline solid (a), a normal liquid
        (b), and a superfluid (c), respectively.}
 \label{phases}
\end{figure}

\bigskip \bigskip

\begin{figure}
\epsfxsize = 8.5 truein
\epsffile{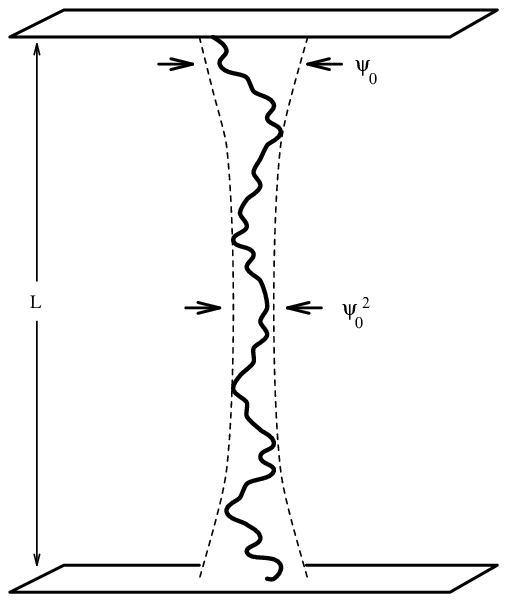}
\caption{Schematic plot of the time--averaged cross section
        of a thermally fluctuating flux line in a slab with free
        boundary conditions. The bulk part away from the boundaries
        has a width determined by the square of the bosonic ``wave
        function'', which is a Gaussian for a free random walker. Near
        the top and bottom surfaces, thermal wandering is enhanced,
        reflected by a probability distribution determined just by the
        ``wave function'' (not its square).}
 \label{wavfun}
\end{figure}

\newpage

\centerline{--- FIG.~12 in Ref.~\cite{nelala} ---}
\centerline{}
\vfill
\begin{figure}
\caption{Lowest--energy solid phase contribution to the correlation
        function ${\tilde G}({\bf r},z;{\bf r}',z')$, which inserts a
        flux head and tail into a crystalline vortex array. Dashed
        lines represent a row of vortices slightly behind the plane of
        the page. In (a), a vacancy is created at ``time'' $z$, which
        then propagates and is destroyed at ``time'' $z'$. 
        Interstitial propagation from $z'$ to $z$ is shown in (b). The 
        energy of the ``string'' defect connecting the head to the
        tail increases linearly with the separation in both cases and
        leads to the exponential decay of 
        ${\tilde G}({\bf r},z;{\bf r}',z')$.}  
 \label{monpol}
\end{figure}

\bigskip \bigskip

\begin{figure}
\epsfxsize = 6.5 truein
\epsffile{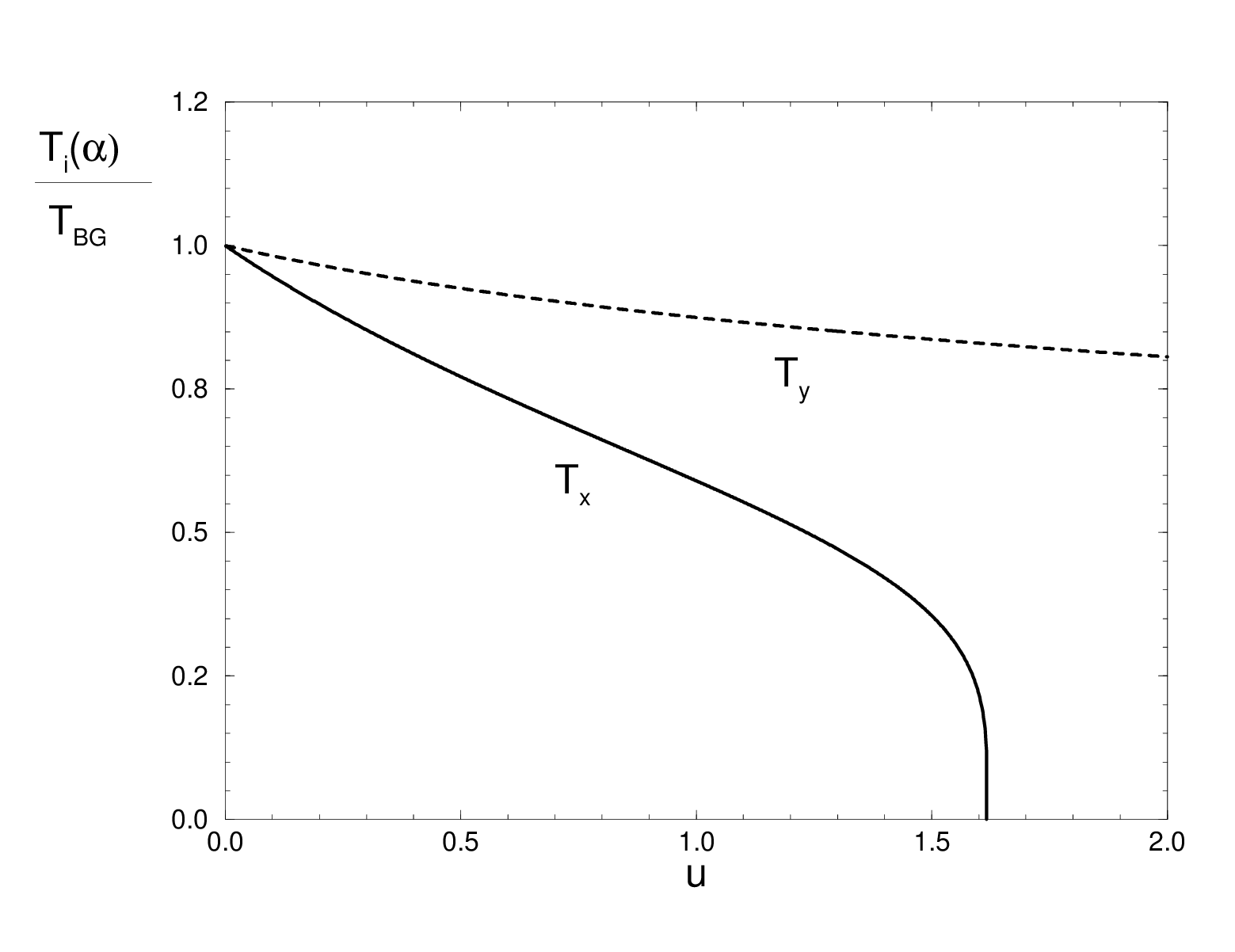}
\caption{Critical temperatures $T_x(\alpha)/T_{\rm BG}$
        (solid line) and $T_y(\alpha)/T_{\rm BG}$ (dashed) where the
        tilt moduli $\overline{{c^v_{xx}}^{-1}}$ and
        $\overline{{c^v_{yy}}^{-1}}$ in a system with parallel
        columnar defects tilted along the $x$ direction vanish,
        respectively, as a function of the dimensionless parameter 
        $u = {\tilde \epsilon}_1^2 (\tan \alpha)^2 / 4 \pi n_0 a$.
        The curve $T_x(\alpha)/T_{\rm BG}$ may be viewed as an
        estimate for the phase boundary of the Bose glass, which
        remains stable towards external tilt as long as 
        $\alpha < \alpha_c$ (transverse Meissner effect).}
 \label{cdtxty}
\end{figure}

\newpage

\begin{figure}
\epsfxsize = 6.5 truein
\epsffile{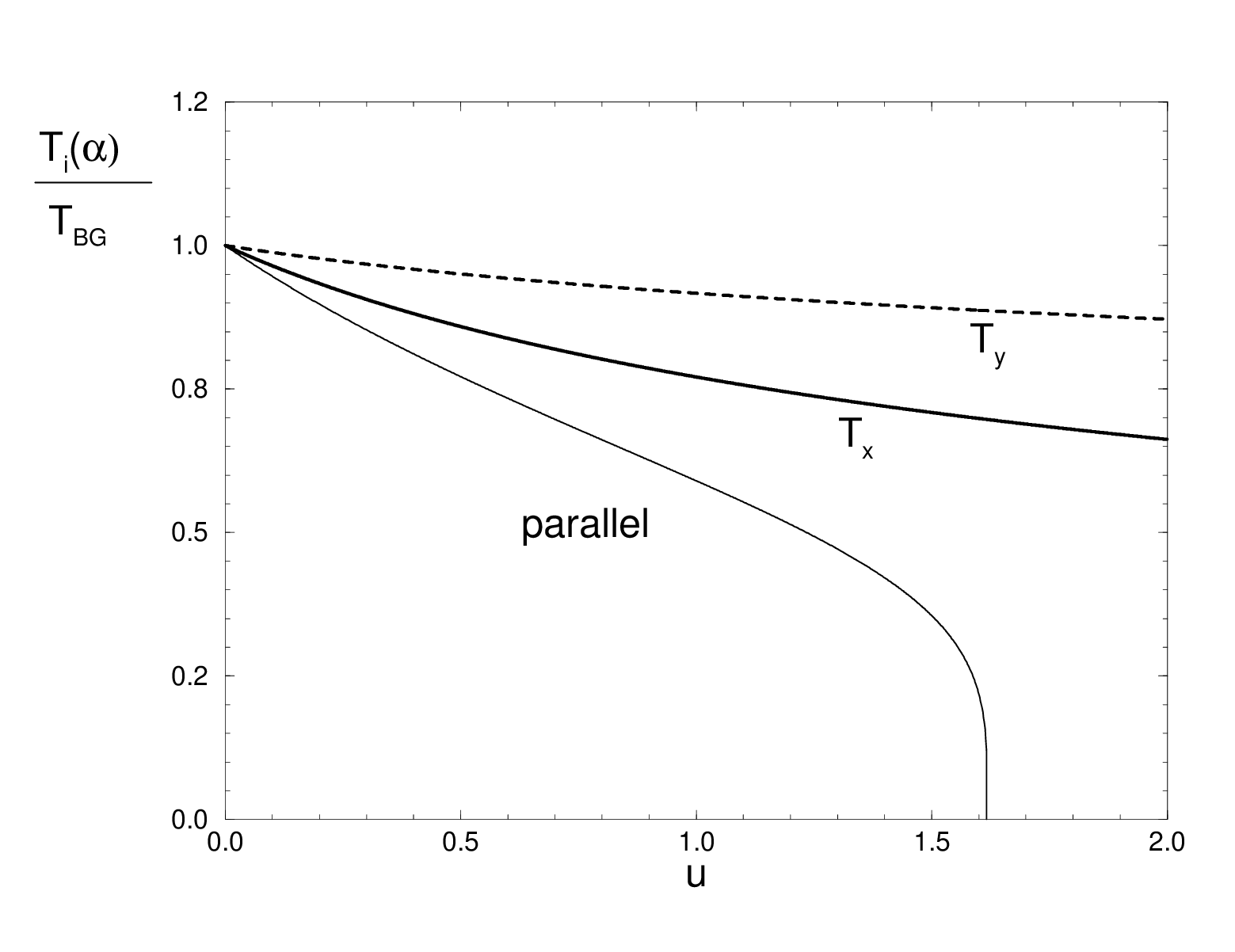}
\caption{Tilt instability temperatures 
        $T_x(\alpha)/T_{\rm BG}$ (thick solid line) and 
        $T_y(\alpha)/T_{\rm BG}$ (dashed) for a system with two
        families of symmetrically tilted columnar defects. For
        comparison, $T_x(\alpha)/T_{\rm BG}$ (thin solid line) for a  
        single family of tilted linear disorder is also depicted.}
 \label{sctxty}
\end{figure}

\newpage

\begin{figure}
\epsfxsize = 6.5 truein
\epsffile{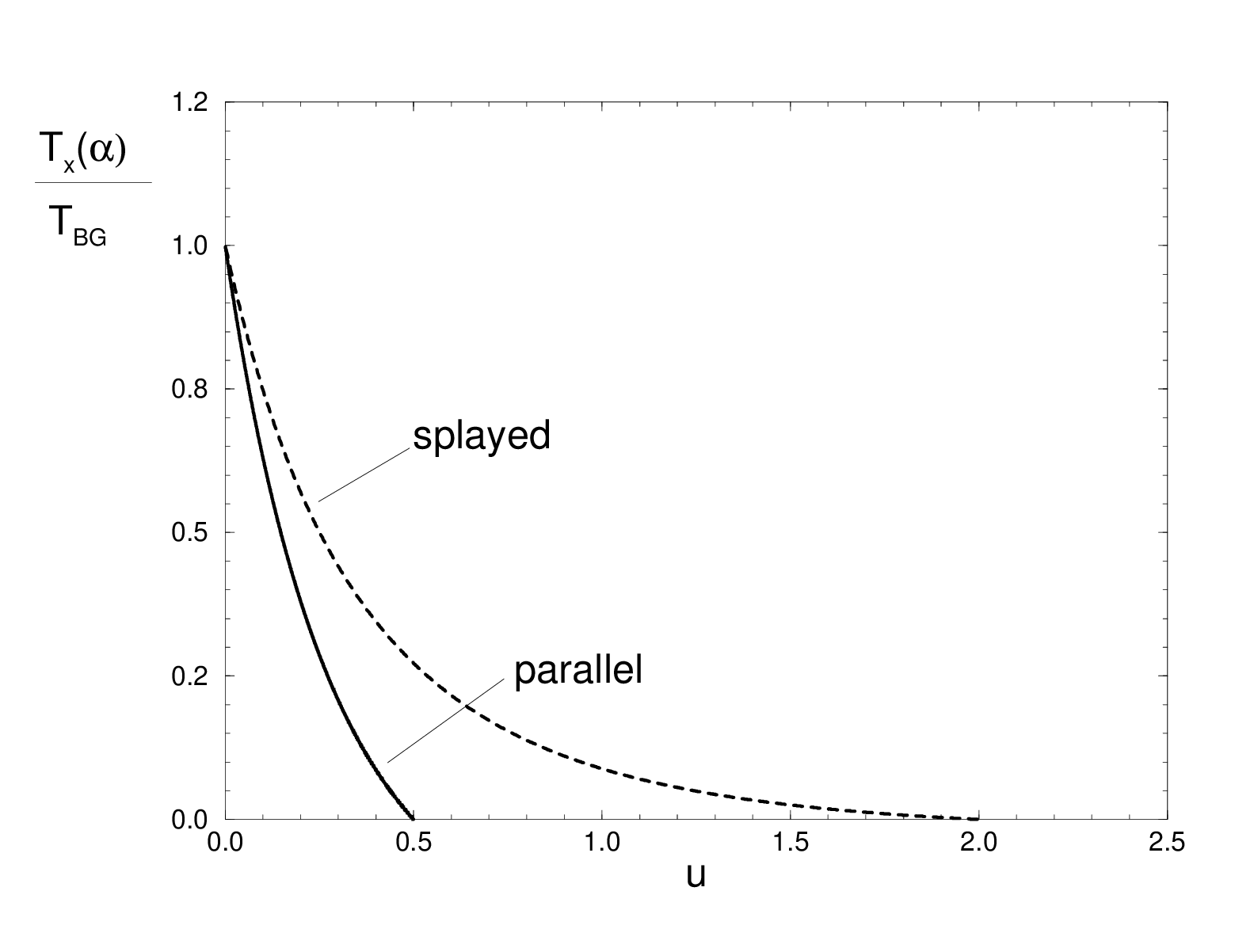}
\caption{Tilt instability temperature 
        $T_x(\alpha)/T_{\rm BG}$ for both parallel planar defects
        uniformly tilted along the $x$ direction (solid line) and two
        families of symmetrically tilted planes (dashed). The critical
        tilt angle marking the instability of the Bose glass phase of
        localized flux lines is shifted upwards in the
        symmetric--splay case.}
 \label{pdsptx}
\end{figure}


\end{document}